\documentclass[]{JFM-FLM_Au}
\usepackage{subcaption}
\usepackage{comment}
\newcolumntype{C}[1]{>{\centering\arraybackslash}p{#1}}

\DeclareMathOperator{\erf}{erf}
\DeclareMathOperator{\ran}{Range}

\newcommand{\Si}{S^{\star}}
\newcommand{\e}{\varepsilon}
\newcommand{\om}{\omega}

\newcommand{\pa}{\partial}
\newcommand{\al}{\alpha}
\newcommand{\bet}{\beta}

\newcommand{\kap}{\kappa}
\newcommand{\sig}{\sigma}
\newcommand{\de}{\delta}
\newcommand{\da}{\dagger}

\newcommand{\di}{\mathrm{d}}
\newcommand{\Di}{\mathrm{D}}
\newcommand{\mZ}{\mathcal{Z}}
\newcommand{\Nre}{N_{\text{re}}}

\newcommand{\hps}{\hat{\psi}}

\newcommand{\hch}{\hat{\chi}}
\newcommand{\hc}{\hat{c}}
\newcommand{\hw}{\hat{w}}
\newcommand{\tc}{\tilde{c}}
\newcommand{\Ks}{K}

\newcommand{\He}{H_e}
\newcommand{\Hea}{H_{e,1}}
\newcommand{\Heb}{H_{e,2}}
\newcommand{\deea}{\de_{e,1}}

\newcommand{\dee}{\delta_e}

\newcommand{\wnl}{\om_{nl}}

\newcommand{\pr}{\varpi}
\newcommand{\hr}{\hat{\pr}}

\newcommand{\hf}{\hat{f}}
\newcommand{\tq}{\tilde{q}}

\newcommand{\hn}{\hat{n}}
\newcommand{\hu}{\hat{u}}

\newcommand{\rA}{r_{\dA}}

\newcommand{\qp}{q^+}
\newcommand{\qm}{q^-}

\newcommand{\bLa}{\boldsymbol{\Lambda}}
\newcommand{\bLac}{\Lambda}
\newcommand{\badns}{\Lambda^{\text{DNS}}}

\newcommand{\bk}{\boldsymbol{k}}

\newcommand{\bka}{\boldsymbol{k}_A}
\newcommand{\bkb}{\boldsymbol{k}_B}

\newcommand{\bW}{\boldsymbol{W}}

\newcommand{\bx}{\boldsymbol{x}}
\newcommand{\bee}{\boldsymbol{e}}
\newcommand{\br}{\boldsymbol{r}}
\newcommand{\bi}{\boldsymbol{h}}
\newcommand{\bww}{\bi_1}

\newcommand{\bqdns}{\boldsymbol{q}^{\text{DNS}}}
\newcommand{\hbqdns}{\hat{\boldsymbol{q}}^{\text{DNS}}}

\newcommand{\bJ}{\boldsymbol{J}}

\newcommand{\bd}{\boldsymbol{d}}

\newcommand{\hbd}{\hat{\boldsymbol{d}}}

\newcommand{\bkn}{\bk \in \mathbb{Z}^2\backslash\set{\bz}}
\newcommand{\bp}{\boldsymbol{p}}
\newcommand{\ba}{\boldsymbol{a}}
\newcommand{\bn}{\boldsymbol{n}}
\newcommand{\bz}{\boldsymbol{0}}
\newcommand{\bu}{\boldsymbol{u}}

\newcommand{\bff}{\boldsymbol{f}}

\newcommand{\mm}{\boldsymbol{m}}

\newcommand{\hbu}{\hat{\boldsymbol{u}}}
\newcommand{\hbn}{\hat{\boldsymbol{n}}}
\newcommand{\tbn}{\tilde{\boldsymbol{n}}}
\newcommand{\hba}{\hat{\boldsymbol{a}}}
\newcommand{\hbq}{\hat{\boldsymbol{q}}}
\newcommand{\hbr}{\hat{\boldsymbol{r}}}
\newcommand{\tbq}{\tilde{\boldsymbol{q}}}
\newcommand{\bq}{\boldsymbol{q}}
\newcommand{\bqm}{\boldsymbol{q}_{\text{m}}}
\newcommand{\hbg}{\hat{\boldsymbol{g}}}
\newcommand{\bgg}{\boldsymbol{g}}

\newcommand{\hbf}{\boldsymbol{\hat{f}}}
\newcommand{\bH}{\boldsymbol{H}}

\newcommand{\nab}{\boldsymbol{\nabla}}
\newcommand{\hnab}{\hat{\boldsymbol{\nabla}}}

\newcommand{\TAS}{\overline{T}_{\dA\dS}}
\newcommand{\TSA}{\overline{T}_{\dS\dA}}
\newcommand{\TSB}{\overline{T}_{\dS\dB}}

\newcommand{\bM}{\mathsfbi{M}}
\newcommand{\bI}{\mathsfbi{I}}
\newcommand{\bP}{\mathsfbi{P}}

\newcommand{\bPo}{\mathsfbi{P}_{o}}
\newcommand{\bPs}{\mathsfbi{P}_{S}}
\newcommand{\bPoa}{\mathsfbi{P}_{o,\bka}}
\newcommand{\bPob}{\mathsfbi{P}_{o,\bkb}}

\newcommand{\bF}{\mathsfbi{F}}
\newcommand{\bN}{\mathsfbi{N}}
\newcommand{\bC}{\mathsfbi{C}}

\newcommand{\bxi}{\boldsymbol{\xi}}
\newcommand{\bzet}{\boldsymbol{\zeta}}

\newcommand{\bL}{\mathsfbi{L}}
\newcommand{\bSi}{\boldsymbol{\Sigma}}
\newcommand{\bPhi}{\boldsymbol{\Phi}}

\newcommand{\bQ}{\mathsfbi{D}}
\newcommand{\hQ}{\hat{D}}
\newcommand{\hbQ}{\hat{\bQ}}
\newcommand{\tbQ}{\tilde{\bQ}}

\newcommand{\bD}{\mathsfbi{H}}

\newcommand{\bR}{\mathsfbi{R}}
\newcommand{\hbR}{\hat{\mathsfbi{R}}}

\newcommand{\bS}{\mathsfbi{S}}

\newcommand{\bT}{\mathsfbi{T}}
\newcommand{\bU}{\mathsfbi{U}}

\newcommand{\bE}{\mathsfbi{E}}
\newcommand{\hbE}{\hat{\mathsfbi{E}}}

\newcommand{\bB}{\mathsfbi{B}}
\newcommand{\hbB}{\hat{\mathsfbi{B}}}

\newcommand{\bZ}{\mathsfbi{0}}
\newcommand{\bZz}{\mathsfbi{Z}}

\newcommand{\bA}{\partial\mathcal{A}}
\newcommand{\boB}{\partial\mathcal{B}}
\newcommand{\bO}{\partial\Omega}

\newcommand{\dA}{\mathcal{A}}
\newcommand{\dB}{\mathcal{B}}
\newcommand{\dS}{\mathcal{S}}

\newcommand{\ops}{\mathcal{S}_\mathcal{O}}
\newcommand{\opr}{\mathcal{R}_\mathcal{O}}

\newcommand{\ti}{\text{i}}
\newcommand{\cc}{\text{c.c.}}
\newcommand{\bdot}{\mathbf{\cdot}}

\newcommand{\Ap}{ A_{+} }
\newcommand{\Am}{ A_{-} }
\newcommand{\Bp}{ B_{+} }
\newcommand{\Bm}{ B_{-} }

\newcommand{\mA}{ |A| }
\newcommand{\mB}{ |B| }

\newcommand{\msC}{ \mathcal{C} }

\newcommand{\mAp}{ |A_{+}| }
\newcommand{\mAm}{ |A_{-}| }
\newcommand{\mBp}{ |B_{+}| }
\newcommand{\mBm}{ |B_{-}| }

\newcommand{\hbpa}{\hbd^{\phi}_{\bka}}
\newcommand{\hbpb}{\hbd^{\phi}_{\bkb}}
\newcommand{\hqpa}{\hbq^{\phi}_{\bka}}

\newcommand\pae[1]{\left( #1 \right)}
\newcommand\sae[1]{\left[ #1 \right]}
\newcommand\set[1]{\left\{ #1 \right\}}
\newcommand\bNc[1]{\hat{\mathsfbi{N}}[ #1 ]}

\newcommand\dt[2]{\hbg_{#1}^{(#2)}}

\newcommand\myeq{\mathrel{\stackrel{\makebox[0pt]{\mbox{\normalfont\tiny i.b.p.}}}{=}}}

\newcommand\nn[1]{||#1||}

\newcommand\ssp[2]{\left \langle #1 \middle| #2 \right \rangle_{u,p}}
\newcommand\ssd[2]{\left \langle #1 \middle| #2 \right \rangle}

\newcommand\ea[1]{\mathbb{E}\left[ #1 \right]}

\newcommand{\tr}[1]{\operatorname{tr}\left(#1\right)}
\newcommand{\vect}[1]{\operatorname{vec}\left(#1\right)}
\newcommand{\ve}[1]{\underline{#1}}

\newcommand\vem[3]{\begin{bmatrix}
#1 \\
#2 \\
#3
\end{bmatrix}}

\newcommand\vemo[5]{\begin{bmatrix}
#1 \\
#2 \\
#3 \\
#4 \\
#5 
\end{bmatrix}}

\newcommand{\spap}[1]{ \ssd{\tbq^{\Ap,\da}}{#1} }

\newcommand{\spam}[1]{ \ssd{\tbq^{\Am,\da}}{#1} }

\newcommand{\spbp}[1]{ \ssd{\tbq^{\Bp,\da}}{#1} } 

\newcommand{\spbm}[1]{ \ssd{\tbq^{\Bm,\da}}{#1}}  

\newcommand{\spaa}[1]{ \ssd{\tbq^{A,\da}}{#1} }  

\newcommand{\spbb}[1]{ \ssd{\tbq^{B,\da}}{#1} }  

\lefttitle{Y.-M. Ducimeti\`ere and M.J. Shelley }
\righttitle{A weakly nonlinear analysis of a stochastically forced active fluid model}

\title{Rare transitions between collective states in an active fluid via a weakly nonlinear reduction}

\author{Yves-Marie Ducimeti\`ere \aff{1}, \and Michael John Shelley \aff{1,2}}

\affiliation{\aff{1}{Courant Institute of Mathematical Sciences, New York University, New York, NY 10012, USA}
\aff{2}{Center for Computational Biology, Flatiron Institute, Simons Foundation, New York, New York, 10010, USA}}

\corresau{Yves-Marie Ducimeti\`ere, \email{yd3213@nyu.edu}}

\begin{document}
\maketitle

\begin{abstract}
We study a model for a dilute suspension of rod-like particles swimming at constant velocity in a Stokes flow. As the translational diffusivity of the particles decreases, a two-dimensional uniform concentration of randomly aligned particles undergoes either a codimension-$2$ pitchfork bifurcation or a codimension-$4$ Hopf bifurcation, depending on the particles' swimming speed. We use a weakly nonlinear expansion to reduce the system to a low-dimensional one for the amplitudes of the bifurcating eigenmodes. The originality of our calculations lies in incorporating spatio-temporal white noise forcing. The stochastic forcing terms in the amplitude equations are derived analytically, from the noise acting on the original system, and \textit{via} a generalized non-resonance condition in the variance.   

Past the onset of the bifurcations, the particles deterministically self-organize into steady or oscillating states of collective motion. For the Hopf bifurcation scenario, two stable periodic orbits are found to coexist, each corresponding to a distinct collective dynamics. The stochastic forcing induces rare transitions between them. Owing to the low dimensionality of amplitude equations, steady and dynamical statistics can be computed directly from the Fokker-Planck equation, or via the Adaptive Multilevel Splitting (AMS) rare-event algorithm. In particular, extremely long mean transition times and associated out-of-equilibrium paths between the metastable periodic orbits are obtained. These paths can be understood in light of the invariant manifolds of the low-dimensional system, which brings insights into the mechanism behind the transitions. 

We also performed fully nonlinear stochastic simulations and used the AMS algorithm directly on the full system. The statistics are in good quantitative agreement with those computed on the reduced systems, the latter being obtained at a considerably lower numerical cost. 
\end{abstract}



\section{Introduction}
\label{sec:intro}

Active matter systems consist of collections of discrete agents, for example, particles or micro-swimmers, each of which can convert a source of energy (usually chemical) into mechanical work. Examples include experimental works on fish schools \citep{Katz11} and bird flocks \citep{Ballerini08}, ensembles of biopolymers cross-linked by motor proteins \citep{Sumino12, Sanchez12}, and both experimental and theoretical works on colloidal suspensions of self-propelled spherical particles \citep{Buttinoni13, Cates15, Geyer19} and suspensions of motile bacteria \citep{Dombrowski04, Tuval05, Subramanian09}. 
Agents that constitute an active matter system typically interact on their own individual scale. This can occur either directly (e.g., dipolar magnetic interactions, excluded volume, etc.) or indirectly through the medium in which they move. These interactions, occurring at the individual scale of the agents, can sometimes translate into collective dynamics at much larger scales, as reported in \cite{Toner05, Marchetti13, Cavagna14, Zottl16, Zhang17}, among many others. Active matter systems are thus known to exhibit an extremely rich phenomenology, spanning many length and time scales. In the context of swimming bacteria, this includes chaos \citep{Mendelson99, Dombrowski04, Tuval05, Cisneros11, Dunkel13}, turbulence \citep{Dunkel13}, and complex pattern formation \citep{Sokolov09, Ohm22}.

The system of present interest is a coarse-grained version of the Doi-Saintillan-Shelley (DSS) model for a dilute suspension of active, rod-like particles \citep{Saintillan08, Saintillan08b}. The DSS model consists of a high-dimensional conservation equation for particle number, coupled with the Stokes equations for the surrounding fluid velocity. In the dilute limit (considered here), the particles interact with each other only hydrodynamically through an active stress within the fluid. Here, we remove the orientational dependency of the model by evolving only the first three moments of the density function with respect to the orientational coordinates. This results in a ``coarse-grained" model, depending only on space and time. However, such moment expansions are notorious for leading to a closure problem, since the equations for the lower moments require higher moments. Consequently, we adopt the generalized polar Bingham closure to express all the higher moments in terms of the first three, which was shown in \cite{Weady22} to preserve the thermodynamics structure of the system. 

For some fixed set of parameters, the DSS model was reported in \cite{Ezhilan13} and \cite{Ohm22} to exhibit multiple, distinct stable states of collective particle motion (see figures~8(c) and 10(a) therein, respectively). As we shall demonstrate, this is also true for the coarse-grained version studied here. This coexistence of distinct stable states, for a fixed set of parameters, is referred to as ``multistability''. It is the central motivation behind the present analysis. In the presence of noise, multistability implies that the system can exist in a specific state of collective motion for an extremely long time, but, from time to time, due to a rare fluctuation, it exits the basin of attraction of that state and transitions to another, e.g. \cite{Grafke17}. Noise-induced transitions typically are increasingly rare as the intensity of the fluctuations vanishes.

Characterizing the transition to collective behavior in an active matter system is of both historical \citep{Kramers40}, conceptual, and possibly practical value. In many situations, bacterial contamination manifests as biofilms, typically in domestic water systems \citep{Costerton05} or coronary implants \citep{Flemming02}. Biofilm growth is triggered when the local bacterial population density exceeds a certain threshold. Thus, it would be useful to elucidate how a uniform density evolves into a non-uniform one. More generally, a proper modeling of the transition to collective motion could help control active matter systems, with applications in liquid crystal displays.

In the context of equilibrium statistical mechanics, steady states of a system minimize a potential. Therefore, in simple situations, formulas for the transition rates can be derived analytically. Consider, for example, a bi-stable overdamped system driven by a stochastic noise $\epsilon \xi(t)$ with intensity $\epsilon$
\begin{equation}
\mathrm{d}_tx = -\mathrm{d}_xV(x)  + \epsilon \xi(t), \nonumber
\end{equation}
with $V$ a double-well potential with a potential barrier $\Delta V$. The Arrhenius law predicts the expected time $T$ between two transitions to go like the exponential of minus the potential barrier separating two attractors, divided by the square of the intensity of the fluctuations, i.e., $T \propto \exp(-2\Delta V/\epsilon^2)$.  
 
However, active matter systems are out of equilibrium, with energy constantly injected and dissipated at the individual scale. Due to their out-of-equilibrium nature, no potential exists, which complicates the computation of transition rates and trajectories. For instance, the absence of potential implies that the forward and backward routes from one basin of attraction to another are generically not the time-reversals of one another.

Another fundamental difficulty is that descriptions of active matter systems of interest usually involve a considerably large number of degrees of freedom. This applies to the coarse-grained model considered here. This means that direct numerical simulation is too costly a tool for the statistical study of extremely rare transition events.

Overall, the out-of-equilibrium nature of active matter systems and the large number of degrees of freedom involved make the statistical characterization of rare transitions a current scientific challenge. Some advances have been made based on the Freidlin-Wentzell's large deviation theory, valid for out-of-equilibrium systems subject to infinitesimal noise \citep{Freidlin98}. Their theory says that the most probable trajectory selected by the system for a rare transition from one basin to another, called an ``instanton path", can be computed \textit{a priori}, as it minimizes a certain action functional in the path integral representation of the system. The transition rates are then proportional to the exponential of the minimal action (i.e., the action along the instanton) divided by the squared noise intensity. This minimal action is called a ``quasipotential", as it is analogous to a potential barrier in the equilibrium case. Recent progress in numerical methods has made it possible to compute instanton paths in different contexts, ranging from fluid mechanics \citep{Bouchet11, Grafke13, Wan15, Schorlepp22} to reaction-diffusion equations \citep{Heymann08, Zakine23}, and canonical low-dimensional dynamical systems \citep{Kikuchi20}.

Another, more indirect, way to determine the instanton path is to use rare events algorithms, which aim to oversample the number of transition events and related trajectories. One such algorithm is the Adaptive Multilevel Splitting (AMS) algorithm. \citep{Cerou07, Brehier15, Rolland16}. 
In short, the AMS algorithm performs killing and cloning iterations to generate a larger number of trajectories, initiated in the basin of one attractor, and ending in the basin of another. This large number of transition paths is expected to concentrate around the instanton and reveal it. Moreover, the algorithm still applies for non-infinitesimal forcing, where the mean transition paths are sometimes found not to be the instantons \citep{Borner24, Rolland24}.

The present article aims to characterize the statistics of rare noise-induced transition events in a multistable, out-of-equilibrium, space-time-dependent coarse-grained version of the DSS model. For that purpose, and in the continuation of \cite{Ducimetiere24}, we propose the following approach: the dimensionality of the model, subject to a stochastic forcing, will first be reduced analytically. As in \cite{Ohm22}, we restrict the parameter space to be asymptotically close to a pitchfork or Hopf bifurcation point. This justifies using an asymptotic expansion to approximate the solution. The originality of the present calculations with respect to those in \cite{Ohm22} lies in the inclusion of stochastic forcing and in the treatment of the closure map inherent to coarse-graining procedures. The original model, a system of PDEs with a continuously infinite number of degrees of freedom, can thus be reduced to a system of $4$ or $8$ ODEs for the scalar amplitudes of the bifurcating eigenmodes. 

Critically, the reduced systems obtained here are substantially easier to study and physically interpret than the original equations. In particular, they are amenable to standard tools from statistical mechanics at low numerical cost, such as the Fokker-Planck equation or the AMS algorithm. We can therefore make predictions, for instance, about the average out-of-equilibrium transition rates and associated trajectories between the different attracting states of particle collective motion. These predictions align well with the results obtained by applying the AMS algorithm directly to the full DSS model, which requires considerably more numerical effort. 

We begin with a brief presentation of the DSS model and its coarse-grained version in §~\ref{sec:prdef}, followed by a review of the linear stability properties of the isotropic, uniform base flow in §~\ref{sec:linf}. The method for deriving a stochastically forced amplitude equation system past the onset of the Hopf and pitchfork bifurcations is outlined in §~\ref{sec:wnlf}. We present numerical and algorithmic techniques in §~\ref{sec:fnlm}. Sections \ref{sec:pitch} and \ref{sec:hopf} are dedicated to our weakly nonlinear results, which we systematically compare to those obtained from the full DSS model. We summarize our main findings and propose further work in §~\ref{sec:conc}. 

\section{Problem definition}
\label{sec:prdef}

In this section, we first briefly recall the DSS model for a dilute suspension of rod-like particles swimming in a Stokes fluid \citep{Saintillan08, Ezhilan13, Subramanian09}. We then present a coarse-grained version of this model, which is studied in this article \citep{Weady22}.

\subsection{The DSS model}
\label{sec:prdef_kin}

Consider a suspension of $N$ swimming rod-like particles, each of length $l$ and much smaller diameter $b \ll l$, so that the aspect ratio $b/l \ll 1$ of each particle is small. The particles are immersed in a fluid, and the whole system has a characteristic length scale $L \gg l$ (e.g., domain size) much larger than the particle length.

Let $\psi(\bx,\bp,t)$ denote the particle density function, with $\bx$ the spatial coordinate, $\bp$ with $\nn{\bp}=1$ the orientational vector of the particles, and $t$ the time. The function $\psi(\bx,\bp,t)$ should be interpreted in the sense of continuum mechanics. Specifically, $\psi(\bx,\bp,t)$ gives the number density of particles found within an elementary element centered at $(\bx,\bp)$, at a time $t$. Thereby, $\psi$ has the dimension of number of particles per (spatial and orientational) volume. We first assume that there are no particle sources/sinks. The ensuing conservation of the number of particles inside a material volume leads, via the transport theorem, to
\begin{eqnarray}
\frac{\pa \psi }{\pa t} = - \nab \cdot \pae{\dot{\bx}\psi} - \nab_p \cdot \pae{\dot{\bp}\psi}.
  \label{eq:nor}
\end{eqnarray}
In (\ref{eq:nor}), $\dot{\bx}$ and $\dot{\bp}$ denote the material derivatives of $\bx$ and $\bp$, respectively. We have also introduced the symbols $\nab$ for the standard spatial gradient operator and $\nab_p \coloneq \pae{\bI-\bp \bp^T}(\pa_{p_1}, ..., \pa_{p_{d}})^T$ with $d$ the spatial dimension of the problem (typically $d=2$ or $d=3$). The latter operator is a standard gradient along the orientational coordinates $\bp=(p_1, ..., p_{d})$, in which only the component orthogonal to $\bp$, i.e., tangent to the unit sphere $\nn{\bp}=1$, is kept by application of the projection matrix $\pae{\bI-\bp \bp^T}$. Indeed, variations normal to the unit sphere are not permitted by imposition of $\nn{\bp}=1$. 

We consider a periodic spatial domain $\Omega$, where the coordinate along each of the $d$ dimensions ranges from $0$ and $L$. The total number of particles within $\Omega$ is conserved and equal to $N$, i.e., $\int_{\Omega} \int_{\nn{\bp}=1} \psi(\bx,\bp,t) \di \bp \di \bx = N$, for all times.

In the DSS framework, all the particles have the same swimming (i.e., intrinsic) velocity $V\bp$; this is why $\psi$ does not depend on velocity coordinates: because the swimming velocity of a particle can be deduced from its $\bp$. The background fluid velocity is named $\bu(\bx,t)$. Based on the slender body approximation, the DSS model is given by
\begin{eqnarray}
\dot{\bx} &=& V\bp + \bu - D_T \nab(\log\psi), \quad \text{and} \label{eq:dx} \\
\dot{\bp} &=& \pae{\bI-\bp \bp^T} (\nab \bu)\bp - D_R \nab_p(\log\psi), \label{eq:dp}
\end{eqnarray}
with $(\nab \bu)\bp=(\bp \cdot \nab)\bu$ the matrix-vector product between $\nab \bu$ and $\bp$. Equation (\ref{eq:dx}) says the velocity of a particle is the sum of its intrinsic swimming speed $V\bp$ plus the advection velocity $\bu$ from the fluid. 
In (\ref{eq:dp}), the term 
$(\nab \bu)\bp = \lim_{l \rightarrow 0}\pae{\bu(\bx+l\bp,t)-\bu(\bx,t)}/l $
is the spatial derivative of the fluid velocity at $\bx$, and in the direction $\bp$ of the particle. Thereby, for an infinitely slender and infinitesimally small rod-like particle, the component of $(\nab \bu)\bp$ normal to $\bp$, i.e., $ \pae{\bI - \bp \bp^T} (\nab \bu)\bp$, naturally gives the rotation rate $\dot{\bp}$ of the particle under the action of the fluid \citep{Jeffery22}. By contrast, the component of $(\nab \bu)\bp$ along $\bp$ would give the stretching rate of a particle, which is prohibited by its assumed rigidity. Such resistance to stretching should translate into a stress exerted on the fluid, but it is formally second-order in concentration and thus is neglected in the model. 

Eventually, in (\ref{eq:dx})-(\ref{eq:dp}), the terms in $D_T$ and $D_R$ model random (thermal) and supposedly isotropic collisions of the fluid molecules with a particle. At a continuum level, this must translate into translational and rotational isotropic diffusion of the particle density function, respectively.

To close the system, we need an equation for the fluid velocity field $\bu(\bx,t)$. Because we consider small particles as compared to the characteristic macroscopic length scale ($L \gg l$), as well as a small velocity scale, we model the fluid by the incompressible Stokes equations 
\begin{eqnarray}
-\mu \Delta \bu + \nab \pr &=& \nab \cdot \bSi, \quad \text{with} \quad \bSi(\bx,t) = a_0 \int_{\nn{\bp}=1} \psi(\bx,\bp,t) \bp \bp^T \di \bp , \label{eq:us} \\
\nab \cdot \bu &=& 0. \label{eq:divu}
\end{eqnarray}
Above, $\mu$ denotes the fluid viscosity, $\pr(\bx,t)$ the pressure field, and $\bSi(\bx,t)$ the stress tensor exerted by the particles upon the fluid. The latter consists of the force dipole $a_0 \bp \bp^T$ that a single particle oriented along $\bp$ exerts upon the fluid, weighted by the density function and integrated over all possible orientations. For self-propelled particles, the dipole strength $a_0$ typically scales as $a_0 \propto \mu V l^2$, with a negative sign for pusher particles and a positive sign for puller particles. However, $a_0$ can be nonzero even when particles do not swim: ``shaker" particles, for instance, still exert the same stress on the fluid even though $V=0$ \citep{Ezhilan13, Stenhammar17}. Note that, in (\ref{eq:us}), one can equivalently replace $\bSi(\bx,t)$ by its traceless version $\bSi(\bx,t)-\bI/d$ and absorb the removed isotropic part in the tensor in the pressure. 

Overall, (\ref{eq:nor})-(\ref{eq:divu}) form a closed system of equations. The particles are advected by the fluid velocity and rotated by the associated shear, and feed back onto the fluid by exerting a stress on it due to their intrinsic movement (swimming and/or shaking). These phenomena are local in space, but (\ref{eq:us}) is non-local in orientation.  

While no modeling assumptions are made when deriving (\ref{eq:nor}) (other than each particle having intrinsic velocity $V\bp$), there are several in (\ref{eq:dx})-(\ref{eq:divu}). In particular, we have neglected direct (``contact'', or ``steric'') interactions between the particles. Such interactions could result, for instance, from the minimization of excluded volume between the particles. In the context of the Maier-Saupe theory \citep{Maier58}, or in the work of Doi and Edward \citep{Doi88}, contact interactions are modeled as an additional torque in (\ref{eq:dp}), deriving from an interaction potential that aligns neighboring particles. Furthermore, this tendency must create a flow, which would translate into an extra stress tensor in the fluid equations. 

We neglect contact interactions in this article, which is justified in the ``dilute suspension" limit where the mean number density is small, i.e., $N/L^d \ll 1$. We refer to \cite{Baskaran10, Ezhilan13} for a proper inclusion of contact interaction terms in (\ref{eq:dp}). 


The same gauges as in \cite{Ohm22} are chosen for nondimensionalizing the equations. That is, we choose (i) the mean density $\Psi_c = N/L^d$ for the density scale, (ii) the rescaled periodic box length $L_c = L/(2\upi)$ for the length scale, (iii) the inverse shear rate of the fluid under an active stress $\sim \Psi_c|a_0|$ for the time scale, $T_c = \mu/(\Psi_c|a_0|)$, and eventually (iv) the associated velocity difference across a distance $L_c$ for the velocity scale $U_c=L_c/T_c$. This gives the set of nondimensional equations
\begin{eqnarray}
\pa_t \psi &=& - \nab \cdot \pae{\dot{\bx}\psi} - \nab_p \cdot \pae{\dot{\bp}\psi}, \label{eq:psin}  \\
\dot{\bx} &=& \bet \bp + \bu - D_T \nab(\log\psi), \label{eq:xn}  \\
\dot{\bp} &=& \pae{\bI-\bp \bp^T} (\nab \bu)\bp - D_R \nab_p(\log\psi),  \label{eq:pn} \\
-\Delta \bu + \nab \pr &=& \pm \int_{\nn{\bp}=1} \bp \bp^T \nab \psi \di \bp, \label{eq:un}\\
\nab \cdot \bu &=& 0. \label{eq:divun}
\end{eqnarray}
In (\ref{eq:xn}) and (\ref{eq:pn}), the diffusion coefficients have been nondimensionalized according to $D_T\rightarrow T_c L_c^d D_T$ and $D_R \rightarrow T_c D_R$. Moreover, the nondimensional swimming speed $\bet = V/U_c$ was introduced in (\ref{eq:xn}).

\subsection{A coarse-grained version with Bingham closure}

System (\ref{eq:psin})-(\ref{eq:divun}) possesses $d$ dimensions in space and $d-1$ dimensions in orientation ($\bp$ is a $d$-component vector but recall the constraint $\nn{\bp}=1$). The numerical cost associated with this high dimensionality is prohibitive, as we anticipate that producing steady statistics requires long (and/or many) simulations. The numerical cost is particularly problematic in the statistical characterization of rare events, which is the main focus of this article. 

We will study a version of (\ref{eq:psin})-(\ref{eq:divun}) that is coarse-grained over the orientational degrees of freedom. Specifically, we will not solve for $\psi(\bx,\bp,t)$ directly but for its first orientational moments, which by definition depend only on space and thus reduce the domain to $d$ dimensions. 


Let us define the zeroth to fourth moments of $\psi$ as
\begin{align}
\begin{split}
&c(\bx,t)   \coloneq \int_{\nn{\bp}=1} \psi \di \bp, \quad \bn(\bx,t) \coloneq \int_{\nn{\bp}=1} \psi \bp \di \bp, \quad \bQ(\bx,t) \coloneq \int_{\nn{\bp}=1} \psi \bp \bp^T \di \bp, \nonumber  \\
&\bR(\bx,t) \coloneq \int_{\nn{\bp}=1} \psi  \bp^{(3)} \di \bp, \quad \text{and} \quad \bS(\bx,t) \coloneq \int_{\nn{\bp}=1} \psi \bp^{(4)} \di \bp, \nonumber
\end{split}
\end{align}
with $\bp^{(n)}$ the $n$th-order dyadic product of $\bp$ (e.g. $\bp^{(3)} = \bp \otimes \bp \otimes \bp$, etc...). The zeroth-moment, $c$, is the (spatial) concentration field of the particle. The first moment, $\bn$, is the polarization vector of the particles, not to be confused with their preferred direction of nematic alignment. The polarization vector is meaningful only when particles exhibit a head-tail asymmetry, arising, for example, from propulsive mechanisms as here \citep{Pedley92}, polarity sorting \citep{Gao15}, or asymmetric geometric effects \citep{Yamada03}. The vector $\bn(\bx,t)$ thus gives the local polar (i.e., signed) orientation of the particles. 

The second moment, $\bQ$, is such that $c^{-1}\bQ$ corresponds to the nematic tensor. The second moment obeys the conditions $\tr{\bQ(\bx,t)} = c(\bx,t)$ and $\bQ(\bx,t)^T=\bQ(\bx,t)$, where the trace condition follows from $\nn{\bp} = 1$. Due to its symmetry, $\bQ$ has orthogonal eigenvectors with real eigenvalues. 
The eigenvector associated with the largest eigenvalue is called the ``director'', and represents the preferred nematic (unsigned) alignment direction of the particles. The isotropic case, in which particles within the elementary volume are oriented randomly (and independently), corresponds to $\bQ(\bx,t) = \bI/d$. 

Note that particles can exhibit a clear preferred nematic alignment direction for a given $\bx$, say $\mm$, such that $\bQ(\bx,t) \approx \mm\mm^T$, while having a zero polarization vector $\bn(\bx,t)=0$ there. It suffices that, around $\bx$, there are as many particles along $-\mm$ as along $\mm$. 

By injecting (\ref{eq:xn}) and (\ref{eq:pn}) in (\ref{eq:psin}), then integrating the resulting equation for $\psi$ against $1$, $\bp$ and $\bp\bp^T$ over the unit sphere $\set{\bp : \nn{\bp}=1}$, we are left with evolution equations for the moments $c$, $\bn$, and $\bQ$, given by
\begin{align}
\begin{split}
\frac{\Di c}{\Di t} &= - \bet \nab \cdot \bn + D_T \Delta c ,  \\
\frac{\Di \bn}{\Di t} - (\nab \bu) \bn + \bR : \bE &= - \bet \nab \cdot \bQ + D_T \Delta \bn +(1-d)D_R\bn, \\
\frac{\Di \bQ}{\Di t} - (\nab \bu) \bQ - \bQ (\nab \bu)^T  + 2\bS : \bE &= - \bet \nab \cdot \bR + D_T \Delta \bQ - 2dD_R\pae{\bQ-\frac{c}{d}\bI}. 
\label{eq:c}
\end{split}
\end{align}
Above, $\Di(\bullet)/\Di t \coloneq \pa_t (\bullet) + (\bu \cdot \nab) (\bullet)$ is the material derivative. Terms on the left-hand side in (\ref{eq:c}) have purely kinematic origins and result from hydrodynamic advection and rotations. They involve the contractions $\sae{\bR : \bE}_{i} = R_{ijk}E_{jk}$ and $\sae{\bS : \bE}_{ij} = S_{ijkl}E_{kl}$, where $\bE \coloneq (\nab \bu + (\nab \bu)^T)/2$ is the symmetric strain-rate tensor. On the right-hand side of (\ref{eq:c}), all terms arise from translational and rotational diffusion, except those pre-multiplied by $\bet$, which result from particles' motility. In each equation, the term in $\bet$ acts as a conservative source term involving the next-order moment. 


System (\ref{eq:c}) is not closed, for it involves the unknown third and fourth moments $\bR$ and $\bS$. Proceeding similarly to deriving (\ref{eq:c}), the evolution equations for $\bR$ and $\bS$ can be obtained by integrating (\ref{eq:nor}) against $\bp^{(3)}$ and $\bp^{(4)}$, respectively. However, this will involve the fifth and sixth moments, for which equations must also be derived, and so on: the system will never close. A closure map, providing $\bR$ and $\bS$ as functions of $c$, $\bn$, and $\bQ$, is thus needed for the final system to have as many equations as unknowns. 

For that purpose, we proceed along the lines of \cite{Weady22}, Sec.~ II. D, and seek solutions for the density function $\psi$ under the generalized polar ``Bingham'' form
\begin{eqnarray}
\psi_B(\bx,\bp,t) = \frac{1}{Z(\bx,t)}\exp\pae{\bB(\bx,t):\bp\bp^T + \ba(\bx,t)\cdot\bp}. \label{eq:Bin}
\end{eqnarray}
Above, $\bB$ is a $d\times d$ symmetric and traceless tensor, $\ba$ is a $d$-dimensional vector, and $Z$ is a scalar normalization factor that ensures conservation of the total number of particles. Let us also introduce the ``relative'' entropy 
\begin{eqnarray}
S\sae{\psi || \psi_0 }(t)=\int_{\Omega}\int_{\nn{\bp}=1}\pae{\frac{\psi}{\psi_0}}\log\pae{\frac{\psi}{\psi_0}}\di \bp \di \bx, \nonumber
\end{eqnarray}
which, roughly speaking, provides a statistical measure of how different $\psi$ is from a given density function $\psi_0$ (hence the adjective ``relative''). In the following, $\psi_0$ is chosen as the uniform, isotropic density function $\psi_0=(\int_{\nn{\bp}=1}1\di \bp)^{-1}$, an equilibrium solution of (\ref{eq:psin})-(\ref{eq:divun}) corresponding in dimensional form to $N$ particles indeed. Then, the Bingham density function in (\ref{eq:Bin}) is special in that it minimizes the relative entropy between $\psi$ and $\psi_0$, under the constraints that the first three moments of $\psi$, i.e., $c$, $\bn$, and $\bQ$, are known and thus must be matched. 
The dependence of $\psi_B$ on $Z$, $\ba$, and $\bB$ purposely makes this matching possible. Indeed, integrating (\ref{eq:Bin}) against $1$, $\bp$ and $\bp\bp^T$, and equalizing with the known $c$, $\bn$ and $\bQ$, respectively, we obtain
\begin{eqnarray}
c(\bx,t)   = \int_{\nn{\bp}=1} \psi_B \di \bp, \ \bn(\bx,t) = \int_{\nn{\bp}=1} \psi_B \bp \di \bp, \ \text{and} \ \bQ(\bx,t) = \int_{\nn{\bp}=1} \psi_B \bp \bp^T \di \bp. \nonumber
\end{eqnarray}
Inverting the system above yields the proper $Z$, $\ba$, and $\bB$. Without matching constraints on any moments, $\psi_B$ would be equal to $\psi_0$. 

Seeking solutions in the form of $\psi_B$ effectively closes the system, for the knowledge of $Z$, $\ba$, and $\bB$ fully determines $\psi_B$, which, in turn, determines all the higher moments according to 
\begin{eqnarray}
\bR_B\sae{c,\bn,\bQ}(\bx,t) \coloneq \int_{\nn{\bp}=1} \psi_B \bp^{(3)} \di \bp,  \quad \bS_B\sae{c,\bn,\bQ}(\bx,t) \coloneq \int_{\nn{\bp}=1} \psi_B \bp^{(4)} \di \bp, \nonumber
\end{eqnarray}
etc. The square bracket denotes that, in effect, the higher moments become functionals of the first three only. The subscript $B$ emphasizes that the moments have been obtained through the Bingham closure.

That higher moments can be determined from the lower ones physically amounts to saying that the system remains in a quasi-equilibrium regime \citep{Levermore96, Levermore97, Abdelmalik16, Jiang21}. Indeed, $\psi_B$ is the density function that, conditioned on $c$, $\bn$, and $\bQ$, is the closest to $\psi_0$ under the entropic measure $S$ and at each time $t$. Thereby, enforcing the solution $\psi_B$ inherently assumes that all the moments higher than $\bQ$ tend to relax to the equilibrium $\psi_0$ much more rapidly than the typical time scales of $c$, $\bn$, and $\bQ$. This separation of time scales justifies slaving the higher moments to the lower ones, since then these former depart from $\psi_0$ only as a consequence of the matching constraints on these latter, and not because of their own dynamics.

By replacing $\bR$ and $\bS$ in (\ref{eq:c}) by $\bR_B\sae{c,\bn,\bQ}$ and $\bS_B\sae{c,\bn,\bQ}$, respectively, we arrive at a closed system of five equations
\begin{align}
\begin{split}
\Di_t c &= - \bet \nab \cdot \bn + D_T \Delta c , \\
\Di_t \bn - (\nab \bu) \bn + \bR_B : \bE &= - \bet \nab \cdot \bQ + D_T \Delta \bn +(1-d)D_R\bn,\\
\Di_t \bQ - (\nab \bu) \bQ - \bQ (\nab \bu)^T  + 2\bS_B : \bE &= - \bet \nab \cdot \bR_B + D_T \Delta \bQ - 2dD_R\pae{\bQ-\frac{c}{d}\bI} \\ 
-\Delta \bu + \nab \pr &= \pm \nab \cdot \bQ, \quad \text{and} \quad \nab \cdot \bu = 0, 
\label{eq:ecB}
\end{split}
\end{align}
for five unknown fields $(c,\bn,\bQ,\bu,\pr)(\bx,t)$. Compared to the DSS model, it was further shown in \cite{Weady22, Weady22b} that its coarse-grained version (\ref{eq:ecB}) yields the same temporal evolution equation for the relative entropy $S$. As we shall see, some linear stability properties are also in correct agreement. 

\section{Linear formulation}
\label{sec:linf}

In this section, we assess the linear stability properties of the coarse-grained model with Bingham closure (\ref{eq:ecB}). This will serve as a building block for the weakly nonlinear analysis in the next section. From now onward, the analysis is restricted to pusher particles. 

To lighten the notations, we concatenate the variables in the state vector $\bq$ such that
\begin{eqnarray}
\bq(\bx,t) \coloneq \pae{c,\bn,\vect{\bQ},\bu,\pr}^T(\bx,t), \nonumber
\end{eqnarray}
with $\vect{\bullet}$ the vectorization linear operator, sometimes abbreviated in $\ve{\bullet}$ and transforming a rank-$n$ tensor into a vector of dimension $d^n \times 1$. The vector $\bq$ thus has dimension $(1+d+d^2+d+1) \times 1$ (keeping all redundant entries for $\bQ$).

A linear stability analysis unravels the time-asymptotic response of a system to infinitesimal initial perturbations around some steady, or ``base'' state. The selected base state, solution to the governing equations (\ref{eq:ecB}), is
\begin{align}
c_0=1, \quad \bn_0 = \bz, \quad \bQ_0 = \bI/d, \quad \bu_0 = \bz, \quad \text{and} \quad \pr_0 = 0, \nonumber
\end{align}
all concatenated into the vector $\bq_0$. This corresponds to the uniform, isotropic density function $\psi_0=(\int_{\nn{\bp}=1}1\di \bp)^{-1}$ (e.g., $\psi_0=1/(2\upi)$ for $d=2$) for the particles, as well as zero base velocity and pressure.

We then consider infinitesimally small departures from $\psi_0$ and, accordingly, from $\bq_0$ and higher moments, according to
\begin{align}
\psi^{\e} & \coloneq \psi_0\pae{1 + \e \psi_1 + O(\e^2)} \equiv \psi_B = \exp\pae{\bB:\bp\bp^T + \ba\cdot\bp + w}, \label{eq:psi_e}\\
\bq^{\e}  & \coloneq \bq_0 + \e \bq_1 + O(\e^2), \ \bR_B^{\e}   \coloneq  \bR_0 + \e \bR_1 + O(\e^2), \ \bS_B^{\e}  = \bS_0 + \e \bS_1 + O(\e^2), \label{eq:bq_e}
\end{align}
with $\e \rightarrow 0$. In the coarse-grained framework, the density function $\psi^{\e}$ in (\ref{eq:psi_e}) must be under Bingham form $\psi_B$, where we have defined $w\coloneq \ln(Z^{-1})$ between (\ref{eq:Bin}) and (\ref{eq:psi_e}). Thereby, to be asymptotically consistent, in (\ref{eq:psi_e}) we must also expand 
\begin{align}
\begin{split}
w^{\e}  \coloneq w_0 + \e w_1 + O(\e^2), \ \ \ba^{\e} \coloneq \ba_0 + \e \ba_1 + O(\e^2), \ \ \text{and} \ \ \bB^{\e} \coloneq \bB_0 + \e \bB_1 + O(\e^2).\nonumber
\end{split}
\end{align}
Injecting these in (\ref{eq:psi_e}) (i.e., replacing each field by its version with the $\e$ superscript), making a Taylor expansion of the exponential in terms of powers of $\e$, then identifying each power of $\e$ between the right-hand side and the left-hand side of the equation, leads to,  
\begin{eqnarray}
\text{at $O(1)$} : \quad \psi_0 & = & \frac{1}{Z_0}\exp\pae{\bB_0:\bp\bp^T + \ba_0 \cdot \bp}, \label{eq:psi_0}\\
\text{at $O(\e)$}: \quad \psi_1 & = & \bB_1:\bp\bp^T + \ba_1 \cdot \bp + w_1, \label{eq:psi_1}
\end{eqnarray}
as well as higher order terms, irrelevant for the moment since $\e \rightarrow 0$. Equation (\ref{eq:psi_0}) simply leads to $Z_0=\int_{\nn{\bp}=1}1\di \bp$ (e.g., $Z_0=2\upi$ for $d=2$), $\ba_0 = \bz$, and $\bB_0 = \bZ$, thus confirming that the selected $\psi_0$ can indeed be written under Bingham form. By construction, (\ref{eq:psi_1}) is linear in $\bB_1$, $\ba_1$ and $w_1$. 

At the same time, taking the orientational moments of $\psi^{\e}$ in (\ref{eq:psi_e}), then identifying the powers of $\e$ with the moments expansions in (\ref{eq:bq_e}), brings about 
\begin{align}
\begin{split}
&c_j  = \int_{\nn{\bp}=1} \psi_0\psi_j \di \bp, \quad  \bn_j = \int_{\nn{\bp}=1} \psi_0\psi_j \bp \di \bp, \quad  \bQ_j = \int_{\nn{\bp}=1} \psi_0\psi_j \bp \bp^T\di \bp, \\
&\bR_j = \int_{\nn{\bp}=1} \psi_0\psi_j \bp^{(3)} \di \bp, \quad \bS_j = \int_{\nn{\bp}=1} \psi_0\psi_j \bp^{(4)} \di \bp, \nonumber
\end{split}
\end{align}
for each order $j=1,2,...$ and where the moments $c_j$, $\bn_j$, and $\bQ_j$ are those in the state vector $\bq_j$. Thereby, multiplying (\ref{eq:psi_1}) by $\psi_0$ and then taking the orientational moments up to the fourth one yields the series of linear systems 
\begin{eqnarray}
c_1 & = & \bQ_0:\bB_1 + \bn_0 \cdot \ba_1 + w_1c_0, \label{eq:c1}\\
\bn_1 & = & \bR_0:\bB_1 + \bQ_0: \ba_1 + w_1\bn_0, \label{eq:n1}\\
\bQ_1 & = & \bS_0:\bB_1 + \bR_0: \ba_1 + w_1\bQ_0, \label{eq:Q1}\\
\bR_1 & = & \bT_0:\bB_1 + \bS_0: \ba_1 + w_1\bR_0, \label{eq:R1}\\
\bS_1 & = & \bU_0:\bB_1 + \bT_0: \ba_1 + w_1\bS_0, \label{eq:S1} \quad \text{etc.}
\end{eqnarray}
where $\bR_0$, $\bS_0$, $\bT_0$ and $\bU_0$ are the third to sixth moments of the base density $\psi_0$. By linearity, (\ref{eq:c1})-(\ref{eq:Q1}) can be rewritten more synthetically under the form
\begin{eqnarray}
\vem{c_1}{\bn_1}{\ve{\bQ}_1} = \bM_0 \vem{w_1}{\ba_1}{\ve{\bB}_1},\label{eq:M0a}
\end{eqnarray}
where $\bM_0$ is the matrix such that (\ref{eq:M0a}) is equivalent to (\ref{eq:c1})-(\ref{eq:Q1}). The subscript ``$0$'' highlights its dependence on base moments (known). Similarly, we can rewrite $\bR_1$ and $\bS_1$ as
\begin{eqnarray}
\ve{\bR}_1 = \bM^{R}_0 \vem{w_1}{\ba_1}{\ve{\bB}_1} \quad \text{and} \quad \ve{\bS}_1 = \bM^{S}_0 \vem{w_1}{\ba_1}{\ve{\bB}_1} \label{eq:M0b},
\end{eqnarray}
respectively, where we have defined the matrices $\bM^{R}_0$ and $\bM^{S}_0$. The linear system (\ref{eq:M0a}) can be inverted after embedding the definitions $\tr{\bB_1}=0$ and $\bB^T_1=\bB_1$ (which must hold at each order $j=1,2,...$), for example by using projector operators or replacing $d(d-1)/2+1$ lines of the system by these constraint equations. Injecting the inverse of (\ref{eq:M0a}) in (\ref{eq:M0b}) then gives
\begin{eqnarray}
\ve{\bR}_1 = \bM^{R}_0 \bM^{-1}_0 \vem{c_1}{\bn_1}{\ve{\bQ}_1} \quad \text{and} \quad \ve{\bS}_1 = \bM^{S}_0 \bM^{-1}_0 \vem{c_1}{\bn_1}{\ve{\bQ}_1} \label{eq:M0c}.
\end{eqnarray}
Thereby, at $O(\e)$, the higher moments can be deduced as a linear (because linearized) combination of the first three. 

Furthermore, since the base solution does not vary over space or time, and since the spatial domain has periodic boundary conditions and the temporal one extends to positive infinity, perturbations can be sought in the form of Fourier modes
\begin{eqnarray}
\bq_1(\bx,t) & = & \hbq_1 \exp\pae{\ti\bk\cdot\bx + \sig t} + \cc, \label{eq:hbq1}\\
\psi_1(\bx,t) & = & \hps_1 \exp\pae{\ti\bk\cdot\bx + \sig t} + \cc.
\end{eqnarray}
The components of the wavenumber vector $\bk \in \mathbb{Z}^d$ must be integers because the spatial domain is finite and $2\pi$-periodic, which quantizes the admissible wavenumbers. On the other hand, $\sig \in \mathbb{C}$ is a complex-valued scalar. The real part of $\sig$, i.e., $\sig_r$, represents the growth rate of the mode, while the imaginary part, $\sig_i$, represents its frequency. The quantities $\hbq_1$ and $\hps_1$ do not depend on space or time. 

Injecting the expansion (\ref{eq:bq_e}) of the state vector into the governing equations (\ref{eq:ecB}), then using (\ref{eq:M0c}) and (\ref{eq:hbq1}), and eventually collecting then terms at $O(\e)$, results in the linear system
\begin{eqnarray}
\sigma \bD \hbq_1 = \bL_{\bk} \hbq_1, \ \text{under the constraints: $\tr{\hbQ_1}=\hc_1$ and $\hbQ_1^T=\hbQ_1$}.
\label{eq:lin1}
\end{eqnarray}
We have defined the singular mass matrix 
\begin{eqnarray}
\bD \coloneq  \begin{bmatrix}
1 & \bZ & \bZ & \bZ & 0 \\
0 & \bI_{d\times d} & \bZ & \bZ & 0 \\
0 & \bZ & \bI_{d^2 \times d^2} & \bZ & 0 \\
0 & \bZ & \bZ & \bZ &  0\\
0 & \bZ & \bZ & \bZ & 0 \\
\end{bmatrix},
\nonumber
\end{eqnarray}
with $\bI_{d\times d}$ the identity matrix of dimension $d\times d$. We have also defined the linear operator $\bL_{\bk}$, acting over $\hbq_1$ according to 
\begin{align}
\begin{split}
&\bL_{\bk} \hbq_1 \coloneq \\
&\begin{bmatrix}
-\ti \bet \bk^T\hbn_1 - D_T\nn{\bk}^2 \hc_1 \\
-\ti \bet \hbQ_1 \bk - \pae{D_T\nn{\bk}^2 + (d-1)D_R }\hbn_1\\
\vect{-\ti \bet \hbR_1 : \bk + \ti (\hbu_1\bk^T)\bQ_0 + \ti \bQ_0(\bk\hbu_1^T) - 2\bS_0:\hbE_1 - \pae{D_T\nn{\bk}^2 + 2dD_R }\hbQ_1 + 2D_R \bI \hc_1}\\
-\ti \hbQ_1 \bk - \nn{\bk}^2 \hbu_1 - \ti \bk \hr_1 \\
\ti \bk^T \hbu_1
\end{bmatrix},
\label{eq:Ldef}
\end{split}
\end{align}
with $\hbE_1 = \ti(\hbu_1\bk^T + \bk\hbu_1^T)/2$ and where it is implicit that $\hbR_1$ depends linearly on $\hc_1$, $\hbn_1$ and $\hbQ_1$ by application of the matrix $\bM_0^R \bM_0^{-1}$ in (\ref{eq:M0c}). In (\ref{eq:Ldef}), the subscript ``$\bk$" emphasizes the dependence of the operator on the wavenumber vector $\bk$.

Equation (\ref{eq:lin1}) is a generalized eigenvalue problem for the eigenvalues $\sig$ and corresponding eigenvectors. Since the trace and symmetry constraints correspond to linear algebraic equations ($(d-1)d/2+1$ in number), they can be enforced by means of a projection operator. This idea is further developed in Appendix~\ref{app:lind}, as it is important for the practical implementation of (\ref{eq:lin1}). In the remainder, we ignore the eigenvectors contained in the kernel of $\bD$, i.e., those with nonzero components only in pressure and/or velocity, as they are associated with infinite eigenvalues. Consequently, for a given $\bk$, we only retain $1+d+d^2$ eigenvectors. We assume they constitute a complete basis for the concentration, orientation vector, and second moment (but incomplete in velocity and pressure because of the unconsidered kernel).

Although this is not done in what follows, we note that, for such a simple base solution, the analytical developments could be taken further. For instance, taking $d=2$, it is possible to show that
\begin{align}
\begin{split}
\lim_{D_R \rightarrow 0} \sig(k) = \frac{1}{8} \pm \frac{1}{8}\sae{1-16 (\bet k)^2}^{1/2} - D_T k^2,
\label{eq:ana}
\end{split}
\end{align}
where we have defined $k \coloneq \nn{\bk}$. By isotropy of the base solution, the wavefront direction does not matter for the stability properties, such that the eigenvalues depend only on $k$. We refer to \cite{Weady22} for further details regarding the derivation of (\ref{eq:ana}), although a different nondimensionalization was chosen therein. 

From (\ref{eq:hbq1}), it is clear that an eigenmode with wavenumber $k$ is (linearly) unstable if the real part of the corresponding eigenvalue $\sig(k)$ is strictly positive. By contrast, an eigenmode is strictly stable if the real part of the associated eigenvalue is strictly negative, and neutral (sometimes also called ``marginally stable") if the real part of the eigenvalue is null. By abuse of language, a wavenumber $k$ is said to be unstable if it exhibits at least one unstable eigenmode, and is stable in the opposite scenario where all its eigenmodes are stable. The special case $\bk=0$ requires a special treatment, detailed in Appendix~\ref{app:lind}. Indeed, it must be solved in a reduced space free of velocity and pressure. It is then shown in Appendix~\ref{app:lind} that, by virtue of the conservation of total particle number, $\bk=0$ yields only strictly stable eigenmodes if $D_R>0$.  

\begin{figure}
            \begin{subfigure}{0.499\textwidth}
            \centering
            \scalebox{0.475}{\includegraphics[trim=0cm 0cm 0cm 0cm]{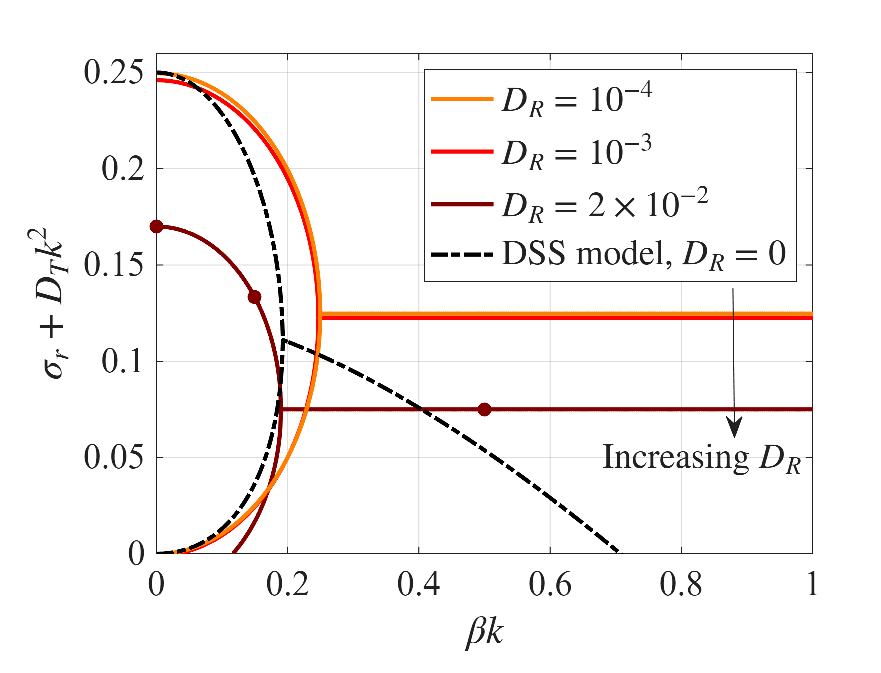}}
             \subcaption{}\label{fig:lindr_a}
            \end{subfigure}
            \begin{subfigure}{0.499\textwidth}
            \centering
            \scalebox{0.475}{\includegraphics[trim=0cm 0cm 0cm 0cm]{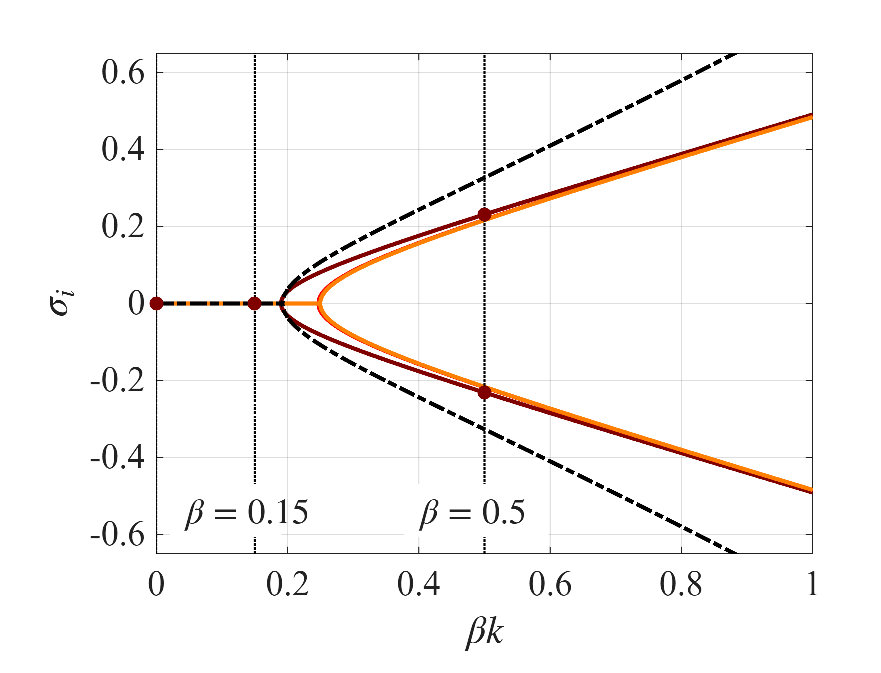}}
             \subcaption{}\label{fig:lindr_b}
            \end{subfigure}
     \caption{Linear dispersion relation for the DSS model (dashed-dotted line, for $D_R=0$) and its coarse-grained version (continuous lines, each for a different $D_R \in \set{10^{-4},10^{-3},0.02}$, darker shades for larger $D_R$). 
     The symbol $k\coloneq\nn{\bk}$ designates the norm of the wavenumber vector. Quantities are shown as a function of $k>0$ rescaled by the swimming speed $\bet$.
     (a) Growth rate $\sig_r\coloneq \Re(\sig)$ of the most unstable branch; the stabilizing contribution from translational diffusion, $-D_T k^2$, is subtracted to bring insight to the cut-off mechanisms. The critical $D_{T,c}$ below which the $k=1$ eigenmodes become unstable is defined as that for which $\sig_r(k=1)=0$. The maximum value of $\sig_r+D_T k^2$ over $\bet k$ thus corresponds to $D_{T,c}$ over $\bet$. Three dot markers highlight the values of $D_{T,c}$ at the parameters selected for further analysis: $D_R=0.02$ and $\bet \in \set{0,0.15,0.5}$.
     (b) Corresponding frequency $\sig_i \coloneq \Im(\sig)$ of the most unstable branch. It also corresponds to the dependence of $\om$ on $\bet$, with $\om$ the frequency of the $k=1$ neutral eigenmodes at $D_T=D_{T,c}$; if $|\om| \neq 0$ (resp. $\om=0$), the $k=1$ wavenumbers experience a Hopf (resp. pitchfork) bifurcation by decreasing $D_T$ below $D_{T,c}$} \label{fig:lindr}
\end{figure}

For all the other $\bk$, such that $k = \nn{\bk}>0$, we show in figure~\ref{fig:lindr_a} the real parts of the eigenvalues $\sigma(k)$ solving (\ref{eq:lin1}). Part of these results were already presented in \cite{Weady22} (figure~$1$ therein). 
The results are shown as a function of $\bet k$, and the stabilizing contribution from diffusion, $-D_Tk^2$, is subtracted from the growth rates to highlight potential other stabilizing mechanisms. Only the eigenvalue branches for which $\sigma_r(k)+D_Tk^2$ becomes positive over some $k$ range are shown. The corresponding imaginary parts of the eigenvalues, i.e., the oscillation frequency of the corresponding eigenmodes, are shown in figure~\ref{fig:lindr_b}. Three different values of $D_R$ are considered. For the sake of comparison, we also reproduce from \cite{Ohm22} the result from the DSS model for $D_R=0$ (figure~$3$ therein).

For the coarse-grained model, the dispersion relation (\ref{eq:ana}) is recovered in the limit $D_R \rightarrow 0$, such that $\sig_r+D_T k^2$ becomes constant and equal to $1/8$ for $\bet k \geq 1/4$. Still in that limit, for $\bet k < 1/4$ there are two distinct branches with $\sig_r+D_T k^2 \geq 0$, each of which is associated with a zero frequency. By contrast, for $\bet k \geq 1/4$, the growth rates of these two branches merge, and the frequencies become nonzero and equal and opposite.  
For $\bet=0$, i.e., immotile particles, the agreement with the dispersion relation from the DSS model is exact as the third moment $\hbR_1$ vanishes in (\ref{eq:Ldef}), and thus the closure model is unimportant. The agreement between the DSS model and its coarse-grained version then progressively degrades as $\bet k$ increases. In particular, the threshold value of $\bet k$ at which the two branches merge seems to be slightly overestimated by the coarse-grained version. Moreover, above that threshold, the DSS model presents a decreasing $\sig_r+D_T k^2$. This suggests that there are stabilizing effects that are not solely due to translational diffusion and that are not captured in the coarse-grained version. 

Increasing the rotational diffusion has a stabilizing effect for all $\bet k$. Whether the branches represented in figure~\ref{fig:lindr_a} correspond to stable or unstable eigenmodes depends on the value of $D_T$. However, in all cases, for a fixed value of $\bet$, the growth rate $\sig_r$ decreases monotonically over $k>0$, such that among all $\bk \in \mathbb{Z}^{d}\setminus\set{\bz}$, the wavenumber vector amplitude $k=1$ systematically is the most unstable or the least stable. Recall that $\bk$ must have integer components and that $k=0$ is a special case that is always strictly stable.

Let us now define $D_{T,c}$ as the critical/threshold value of the translational diffusivity $D_T$ below which the $k=1$ wavenumbers become unstable. In other words, at $D_T=D_{T,c}$, the growth rate of the $k=1$ wavenumbers is zero by definition. Decreasing (resp. increasing) $D_T$ below $D_{T,c}$ renders them unstable (resp. strictly stable). We also define $\om$ as the frequencies of the neutral eigenmode(s) at $D_T=D_{T,c}$.

Precisely because $\sig_r(k=1)=0$ for $D_T=D_{T,c}$, the dependence of $D_{T,c}$ (resp. $\om$) on $\bet$, for a given $D_R$, is the same as that of the maximum value of $\sig_r+D_T k^2$ over $\bet k$ in figure~\ref{fig:lindr_a} (resp. figure~\ref{fig:lindr_b}). In particular, there is a threshold swimming speed $\bet$ above which $D_{T,c}$ becomes independent of it. As said, this threshold tends to $\bet=1/4$ in the limit $D_R\rightarrow 0$, and is found to be $\bet \approx 0.19$ for $D_R = 0.02$. Below this threshold $\bet$, there is only one neutral eigenmode per $\bk$ (with $k=1$), with $\om=0$. Above the threshold, there are a pair of neutral eigenmodes per $\bk$ with equal and opposite nonzero frequencies. Accordingly, below this threshold $\beta$, the $k=1$ wavenumbers undergo a static pitchfork bifurcation as $D_T$ decreases below $D_{T,c}$, whereas they experience an oscillatory Hopf bifurcation above it. We shall study the weakly nonlinear dynamics past the onset of these bifurcations in $D_T$ only.

In what follows, for a given $\bk$, the corresponding family of eigenvectors (or ``eigenbasis'') is denoted $\set{\tbq_j}_{j \geq 1}$, where each $\tbq_j$ is a solution of (\ref{eq:lin1}) with eigenvalue $\sig_j$. They are sometimes referred to as the ``direct'' eigenmodes, in opposition to the ``adjoint" ones, which we introduce in a moment. Within the family, the eigenvectors are ordered such that $\sig_{r,1} \geq \sig_{r,2} \geq ...$.

In the following, it is instructive 
to extract the component of the state vector along the bifurcating eigenmode(s) at $k=1$. For that purpose, we shall define another basis, biorthogonal to the eigenbasis under a predefined inner product. Since the fields do not depend on space anymore after isolating the Fourier component along $\bk$, the inner product is chosen as the Hermitian dot product $\ssp{\hbq_a}{\hbq_b} \coloneq \hbq_a^H \hbq_b$, where $\hbq_a$ and $\hbq_b$ are two arbitrary vectors that do not depend on space (but can depend on time) and where the superscript $H$ designates the Hermitian transpose. The subscript ``$(u,p)$" emphasizes that the inner product includes velocity and pressure, even though the latter are not differentiated with respect to time due to the presence of $\bD$. From the choice of the inner product, it is possible to construct an operator $\bL_{\bk}^{\da}$, ``adjoint" to $\bL_{\bk}$, and defined as 
$\ssp{\bL_{\bk}\hbq}{\hbq^{\da}} \coloneq \ssp{\hbq}{\bL^{\da}_{\bk}\hbq^{\da}}$, for all $\hbq \in \mathcal{D}(\bL_{\bk}), \hbq^{\da} \in \mathcal{D}(\bL^{\da}_{\bk})$
where $\mathcal{D}(\bullet)$ is the domain of the operator. By definition of the second moment, the adjoint fields must obey the same trace and symmetry conditions as the direct one, i.e., $\tr{\bQ^{\da}} = c^{\da}$ and $\bQ^{\da,T}=\bQ^{\da}$. The two domains $\mathcal{D}(\bL_{\bk})$ and $\mathcal{D}(\bL^{\da}_k)$ each embed these conditions and are the same. For a given $\bk$, the eigenmodes of the adjoint system, constituting the adjoint eigenbasis $\set{\tbq^{\da}_j}_{j \geq 1}$. The adjoint eigenmodes satisfy $\sig_j^* \bD \tbq^{\da}_j = \bL^{\da}_{\bk} \tbq^{\da}_j$, under the trace and symmetry constraints, and for $j=1,2,...$. We have used the straightforward result $\bD^{\da} = \bD$, and $\sig_j^*$ is the complex conjugate of the eigenvalue $\sig_j$ of the corresponding direct problem. We discuss the practical construction of the adjoint eigenbasis in Appendix~\ref{app:lind}.

Crucially, for a given $\bk$, the corresponding adjoint and direct families are biorthogonal under the inner product 
\begin{eqnarray}
\ssd{\hbq_a}{\hbq_b} \coloneq \ssp{\hbq_a}{\bD \hbq_b} = \hbq_a^H\bD\hbq_b.
\nonumber
\end{eqnarray}
In words, $\ssd{\hbq_a}{\hbq_b}$ is the Hermitian dot product that includes the mass matrix, $\bD$, which removes contributions from velocity and pressure. The biorthogonality property is such that
\begin{eqnarray}
\ssd{\tbq^{\da}_{j}}{\tbq_{i}} = \de_{ij}, 
\label{eq:biorth}
\end{eqnarray}
with $\de_{ij}$ the Kronecker symbol.

We restrict the subsequent analysis to two spatial dimensions, i.e., $d=2$, with $\bx=(x,y)^T$. Since $\bkn$, then the $k=1$ wavenumbers of interest only include $\bk = \pm \bka $ with $\bka \coloneq (1,0)^T$ and $\bk = \pm \bkb$ with $\bkb \coloneq  (0,1)^T$. 
At $D_T = D_{T,c}$, each of these wavenumbers is associated with either one or two neutral eigenmodes, depending on the value of $\bet$. For values of $\bet$ corresponding to $\om=0$, i.e., in the pitchfork bifurcation region, such as for $\bet \leq 0.19$ for $D_R=0.02$ (see figure~\ref{fig:lindr_b}), there is only one. The neutral eigenmode associated with the wavenumber $\bka$ (and $\om=0$) is denoted $\tbq^A$, as emphasized by the superscript. Accordingly, the neutral eigenmode associated with $-\bka$ is the complex conjugate $\tbq^{A,*}$. We also define the neutral eigenmode $\tbq^B$ for $\bkb$ and $\om=0$ (and is complex conjugate for $-\bkb$). The adjoint eigenmodes are denoted $\tbq^{A,\da}$ and $\tbq^{B,\da}$, respectively. 

By contrast, for $\bet > 0.19$ there is a pair of equal but opposite nonzero frequencies (see figure~\ref{fig:lindr_b}), and thus there are two neutral eigenmodes per wavenumber. They consist of $\tbq^{A_{\pm}}$ for the eigenmodes oscillating along the wavenumber $\bka$ with frequencies $\pm \om$ (with $\om >0$), as well as $\tbq^{B_{\pm}}$ along $\bkb$ with frequencies $\pm \om$. Mathematically, 
\begin{align}
\begin{split}
(\ti \om \bD - \bL_{\bka})\tbq^{\Ap} = (-\ti \om \bD-\bL_{\bka})\tbq^{\Am} =   (\ti \om \bD-\bL_{\bkb})\tbq^{\Bp} = (-\ti \om\bD -\bL_{\bkb})\tbq^{\Bm} = \bz.
  \label{eq:deig}
\end{split}
\end{align}
To facilitate the comparison of our results with those in \cite{Ohm22}, all direct eigenmodes for all $\bk$ (including $\tbq^A$, $\tbq^B$, $\tbq^{A_{\pm}}$ and $\tbq^{B_{\pm}}$) are normalized according to 
\begin{eqnarray}
\ssd{\tbq_j}{\tbq_j} = \frac{1}{32}.
\label{eq:normal}
\end{eqnarray}
The normalization of the adjoint eigenmodes follows from (\ref{eq:biorth}). We have now completely characterized the neutral eigenmodes spanning the slow manifold of (\ref{eq:ecB}). 

\section{Stochastic and weakly nonlinear formulation}
\label{sec:wnlf}

We now introduce the stochastically forced version of the system (\ref{eq:ecB}), as this is the central focus of our analysis. We then proceed to reduce the dimensionality of the subsequent system to its slow manifold. 

Let us write $\bp=(\cos(p),\sin(p))^T$ with $0\leq p\leq 2\pi$. Before coarse-graining, the stochastic forcing is introduced solely in the density conservation equation (\ref{eq:nor}), as
\begin{eqnarray}
\frac{\pa \psi }{\pa t} = - \nab \cdot \pae{\dot{\bx}\psi} - \nab_p \cdot \pae{\dot{\bp}\psi} + F f_{\psi},
  \label{eq:nor_st}
\end{eqnarray}
where $f_{\psi}(\bx,p,t)$ is an additive Gaussian noise, white in space, orientation, and time, and with zero average and unit intensity, such that,
\begin{eqnarray}
\ea{f_{\psi}(\bx,p,t)f_{\psi}(\bx',p',t')}=\de(\bx-\bx')\de(p-p')\de(t-t'),
  \label{eq:dcor}
\end{eqnarray}
The symbol $\ea{\bullet}$ denotes the ensemble average over noise realizations. In (\ref{eq:nor_st}) we have also introduced the free scalar parameter $F \in \mathbb{R}$, so that the intensity of $F f_{\psi}$ is directly given by $F^2$. 

Importantly, we require the stochastic forcing to have zero spatial average for every orientation and time, i.e., $\int_{\Omega}f_{\psi}(\bx,p,t)\di \bx=0$ for each $p$ and $t$. This guarantees that the total number of particles is conserved for each time $t$, despite the presence of forcing.  

The stochastic forcing $F f_{\psi}$ represents the effect of neglected terms in the model, for instance, due to intrinsic particle dynamics, which would modify the conservation equation (see Sec.~2.7.2 in \cite{Saintillan07}). It could also model the effect of unresolved scales. In the absence of additional experimental information, the noise is chosen to be as non-specific as possible, that is, white. Note that it would perhaps be more physically relevant to write $f_{\psi}$ under conservative form, i.e., as the divergence of a tensor. Nonetheless, our paper primarily focuses on methodology, and we expect our developments to be easily adaptable to conservative noise.

The stochastic forcing $F f_{\psi}$ translates into the coarse-grained model as
\begin{align}
\begin{split}
\Di_t c &= - \bet \nab \cdot \bn + D_T \Delta c + F f_c,  \\
\Di_t \bn - (\nab \bu) \bn + \bR_B : \bE &= - \bet \nab \cdot \bQ + D_T \Delta \bn +(1-d)D_R\bn + F \bff_n, \\
\Di_t \bQ - (\nab \bu) \bQ - \bQ (\nab \bu)^T  + 2\bS_B : \bE &= - \bet \nab \cdot \bR_B + D_T \Delta \bQ - 2dD_R\pae{\bQ-\frac{c}{d}\bI} +F \bF_D,\\ 
-\Delta \bu + \nab \pr &= \nab \cdot \bQ, \quad \text{and} \quad \nab \cdot \bu = 0 .
\end{split}
\label{eq:ec_st}
\end{align}
Equations for the moments differ from their deterministic counterparts in (\ref{eq:ecB}) by the inclusion of the stochastic forcing terms in $f_c(\bx,t)$, $\bff_n(\bx,t)$, and $\bF_D(\bx,t)$, defined as
\begin{align}
\begin{split}
&f_c \coloneq \int_{\nn{\bp}=1} f_{\psi} \di \bp, \quad  \bff_n \coloneq \int_{\nn{\bp}=1} f_{\psi} \bp \di \bp, \quad  \bF_D \coloneq \int_{\nn{\bp}=1} f_{\psi} \bp \bp^T\di \bp.
 \label{eq:fdef}
\end{split}
\end{align}
The equations for the velocity, however, remain unforced and unchanged. The forcing fields $f_c$, $\bff_n$, and $\bF_D$ inherit from $\bff$ the property of having zero spatial average at each time $t$. Furthermore, their definitions (\ref{eq:fdef}) imply the trace and symmetry conditions, $\tr{\bF_D}=f_c$, and $\bF_D^T = \bF_D$.

The variance and covariance between each component of $f_c$, $\bff_n$, and $\bF_D$ can be characterized directly from (\ref{eq:fdef}) and (\ref{eq:dcor}). For example, 
\begin{equation}
\begin{split}
\ea{f_c(\bx,t) F_{D,xx}(\bx',t')} &= \int_{0}^{2\pi} \int_{0}^{2\pi} \cos^2(p') \ea{f_{\psi}(\bx,p,t) f_{\psi}(\bx',p',t')} \di p \di p' \\
&= \de(\bx-\bx')\de(t-t')\int_{0}^{2\pi} \cos^2(p) \di p = \de(\bx-\bx')\de(t-t')\pi. \nonumber
\end{split}    
\end{equation}
Proceeding similarly with all the other components, the associated variance and covariance are summarized in table~\ref{tab:cov_f}.
\begin{table}
  \begin{center}
\def~{\hphantom{0}}
  \begin{tabular}{lccccc}
         & $f_c$   &   $f_{n,x}$  & $f_{n,y}$ & $F_{D,xx}$ & $F_{D,xy}$ \\[3pt]
       $f_c$   & 2$\upi$ & 0 & 0 & $\upi$ & 0 \\
       $f_{n,x}$  &  & $\upi$ & 0 & 0 & 0 \\
       $f_{n,y}$  &  &  & $\upi$ & 0 & 0\\
       $F_{D,xx}$ &  &  &  & $3\upi/4$ & 0 \\
       $F_{D,xy}$ &  &  &  &  & $\upi/4$ \\
  \end{tabular}
  \caption{Covariance amplitudes, multiplying $\de(\bx-\bx')\de(t-t')$, between different components of the stochastic forcing at the coarse-grained level. The table is symmetric, so its lower-triangular part is not written explicitly. }
  \label{tab:cov_f}
  \end{center}
\end{table}
All forcing components acting at the coarse-grained level are also white in time and space.

Let us move to the weakly nonlinear reduction of (\ref{eq:ec_st}). We consider values of $D_T$ smaller but asymptotically close to the threshold $D_{T,c}$, by taking
\begin{equation}
\begin{split}
D_T = D_{T,c}-\e^2,
\label{eq:edef}
\end{split}    
\end{equation}
with $0\leq \e \ll 1$. 

For $D_T$ as in (\ref{eq:edef}), the neutral eigenmodes at $k=1$ computed in the previous section become unstable with $O(\e^2)$ growth rates. Consequently, in the deterministic regime (i.e., without stochastic forcing), these eigenmodes would slowly grow in amplitude until nonlinear effects become important. All the other eigenmodes remain strictly stable with $O(1)$ damping rates and thus, in the absence of sustained excitation, would rapidly vanish over time. In other words, in the deterministic regime, there is a separation of time scales (or ``spectral gap") between the strictly stable, ``fast", eigenmodes and the bifurcating, ``slow", ones. The fast eigenmodes tend to relax to the base state much more rapidly than do the slow ones; therefore, it is these latter that are expected to dominate the deterministic, linear, and weakly nonlinear dynamics near the threshold. In the following, ``fast" and ``slow" will always refer to the deterministic dynamics

We will maintain this hierarchy between the slow and fast eigenmodes in the stochastically forced case, by assuming a weak stochastic forcing, i.e., $F \ll 1$. In particular, $F$ is scaled in terms of $\e$ as
\begin{equation}
\begin{split}
F = \e^2 \phi \ll 1,
\label{eq:fstocdef}
\end{split}    
\end{equation}
where $\phi=O(1)$ is a free and real-valued scalar. Given that all eigenmodes are excited indiscriminately and in a sustained manner by the stochastic forcing, it is perhaps not obvious ``\textit{a priori}" that the slow eigenmodes still dominate the dynamics. However, under (\ref{eq:fstocdef}), the slow eigenmodes will show a much larger response to the forcing than the fast ones, because they are, by definition, closer to resonance. Intuitively, the slow eigenmodes have made less progress in relaxing to equilibrium between successive noise impulses than do the fast ones, and thus can sustain a larger variance.  

The solution of the system (\ref{eq:ec_st}) is approximated through an asymptotic expansion with powers of $\e$, similar to (\ref{eq:psi_e})-(\ref{eq:bq_e}), though retaining some higher-order terms as
\begin{eqnarray}
\psi^{\e} &=& \psi_0\pae{1 + \e \psi_1 + \e^2 \psi_2 + \e^3 \psi_3 + O(\e^4)}  \equiv \psi_B, \label{eq:psi_ew}\\
\bq^{\e}  &=& \bq_0 + \e \bq_1 + \e^2 \bq_2 + \e^3 \bq_3 + O(\e^4), \label{eq:qq_ew}
\end{eqnarray}
and expanding similarly $\bR_B^{\e} =  \bR_0 + \e \bR_1 + \e^2 \bR_2 + \e^3 \bR_3 + O(\e^4)$, $\bS_B^{\e}$, $w^{\e}$, $\ba^{\e}$, and $\bB^{\e}$. Again, the density function (\ref{eq:psi_ew}) is sought under Bingham form. Consequently, proceeding similarly to the previous section, we obtain
\begin{align}
\psi_j = \bB_j : \bp\bp^T + \ba_j \bdot \bp + w_j + \chi_j, \quad j = 1,2,...
\label{eq:psi_j}
\end{align}
where each $\chi_j$ stems from nonlinear interactions of the $\psi_j$s obtained at previous orders. In particular, 
\begin{align}
\chi_1=0, \quad \chi_2 = \frac{1}{2}\psi_1^2, \quad \chi_3 = -\frac{1}{3}\psi_1^3 + \psi_1\psi_2, \quad \text{etc.}
\label{eq:chi_j}
\end{align}
Again, multiplying (\ref{eq:psi_j}) by $\psi_0$ and taking the zeroith to second moments gives
\begin{align}
\vem{c_j}{\bn_j}{\ve{\bQ}_j} = \bM_0 \vem{w_j}{\ba_j}{\ve{\bB}_j}  + \vem{c_{\chi,j}}{\bn_{\chi,j}}{\ve{\bQ}_{\chi,j}}, \quad j = 1,2,...\label{eq:M0_w}
\end{align}
where the subscript ``$(\chi,j)$" denotes the corresponding moment of $\psi_0 \chi_j$. That is, 
\begin{align}
\begin{split}
&c_{\chi,j}  \coloneq \int_{\nn{\bp}=1} \psi_0\chi_j \di \bp, \quad  \bn_{\chi,j} \coloneq \int_{\nn{\bp}=1} \psi_0\chi_j \bp \di \bp, \quad  \text{etc.} \nonumber
\end{split}
\end{align}
Thereafter, multiplying (\ref{eq:psi_j}) by $\psi_0$, taking the third and fourth moments, and applying $\bM^{-1}_0$ in (\ref{eq:M0_w}), which again embeds the constraints $\tr{\bB_j}=0$ and $\bB^T_j=\bB_j$, leads to 
\begin{align}
\ve{\bR}_j = \bM^{R}_0 \bM^{-1}_0 \vem{c_j}{\bn_j}{\ve{\bQ}_j} + \ve{\bR}_{f,j}, \ \ \text{where} \ \ \ve{\bR}_{f,j} \coloneq - \bM^{R}_0 \bM^{-1}_0 \vem{c_{\chi,j}}{\bn_{\chi,j}}{\ve{\bQ}_{\chi,j}} + \ve{\bR}_{\chi,j},
\label{eq:R_w}
\end{align}
for the third moment, as well as 
\begin{align}
\ve{\bS}_j = \bM^{S}_0 \bM^{-1}_0 \vem{c_j}{\bn_j}{\ve{\bQ}_j} + \ve{\bS}_{f,j}, \ \ \text{where} \ \ \ve{\bS}_{f,j} \coloneq - \bM^{S}_0 \bM^{-1}_0 \vem{c_{\chi,j}}{\bn_{\chi,j}}{\ve{\bQ}_{\chi,j}} + \ve{\bS}_{\chi,j},
\label{eq:S_w}
\end{align}
for the fourth, and where $j=1,2,...$. The third and fourth moments (and all the higher ones) are expressed as the sum of two terms. The first term is linear in the first three moments (zeroth to second). In contrast, the second term acts as a forcing term (hence the subscript ``$f$)`` and arises purely from the moments of nonlinear interactions among the $\psi_j$ determined at previous orders. Thus, at the stage of solving for $j$th order, equations (\ref{eq:M0_w}), (\ref{eq:R_w}), and (\ref{eq:S_w}) are indeed all linear (because linearized), and the first three moments are the only unknowns. Again, all higher moments can be determined from the first three.   
Equation (\ref{eq:M0_w}) is important in what follows, for it is the one that, together with (\ref{eq:psi_j}), links the $j^{th}$ correction in the state vector to the $j$th correction in the density. 

Keeping this in mind, let us now examine the expansion (\ref{eq:qq_ew}) of the state vector. As motivated above, the leading-order solution $\bq_1$ is chosen to be contained in the slow manifold, such that  
\begin{align}
\begin{split}
\bq_1(\bx,t) = &\Ap(t) \tbq^{\Ap} e^{\ti \bka \bdot \bx + \ti \om t} + \Am(t) \tbq^{\Am} e^{\ti \bka \bdot \bx - \ti \om t } \\
& + \Bp(t) \tbq^{\Bp} e^{\ti \bkb \bdot \bx + \ti \om t } + \Bm(t) \tbq^{\Bm} e^{\ti \bkb \bdot \bx - \ti \om t} +\cc, 
  \label{eq:q1c4H}
  \end{split}
\end{align}
for the codimension-$4$ Hopf bifurcation, and 
\begin{align}
\begin{split}
\bq_1(\bx,t) = A(t) \tbq^{A} e^{\ti \bka \bdot \bx} + B(t)\tbq^{B} e^{\ti \bkb \bdot \bx} + \cc, 
  \label{eq:q1c2P}
  \end{split}
\end{align}
for the codimension-$2$ pitchfork bifurcation. In (\ref{eq:q1c4H}) and (\ref{eq:q1c2P}), the eigenmodes are pre-multiplied by $O(1)$ complex-valued scalar amplitudes, depending on time only, and for the moment unknown. The purpose of the expansion procedure is to derive the evolution equations of these amplitudes. 

Readers should be aware that the following calculations are intricate. While they may appeal to those interested in the article's methodological and mathematical aspects, those interested in the physical results are invited to accept the amplitude equations (\ref{eq:ampeq_Pi}) and (\ref{eq:ampeq_H}) and jump directly to Sec. \ref{sec:pitch_det} for the results. 

Without further ado, and no loss of generality, we will first restrict our calculations to the codimension-$4$ Hopf bifurcation case. The codimension-$2$ pitchfork bifurcation case will then be treated simply by dropping particular terms in the final results.

For (\ref{eq:q1c4H}), the evolution equations for the amplitudes are sought in the form  
\begin{align}
\begin{split}
&\frac{\di \bLac_j}{\di t} = \e^2 \pae{ f_{\bLac_j}(\bLa) + \e f^{(2)}_{\bLac_j}(\bLa) + \hdots }+ \e \phi \pae{ \xi_{\bLac_j}(t) + \e \xi^{(2)}_{\bLac_j}(\bLa,t) + \hdots }, \\
&\text{with} \quad \bLa \doteq (\Ap,\Am,\Bp,\Bm,\Ap^*,\Am^*,\Bp^*,\Bm^*),
  \label{eq:nf}
  \end{split}
\end{align}
for $j=1,2,...,8$. Each function $f_{\bLac_j}$ and its higher-order corrections, $f^{(2)}_{\bLac_j}$, $f^{(3)}_{\bLac_j}$, ..., have a purely deterministic origin and contain both linear (for the leading-order $f_{\bLac_j}$) and nonlinear polynomial terms in the amplitudes. The $\e^2$ prefactor follows from the scaling of the distance to criticality in (\ref{eq:edef}).

An additive white noise process $\xi_{\bLac_j}$ is included in (\ref{eq:nf}), as well as multiplicative higher-order corrections $\xi^{(2)}_{\bLac_j}$, $\xi^{(3)}_{\bLac_j}$, etc. The $\e \phi$ scaling follows from the fact that, according to (\ref{eq:fstocdef}), the forcing is introduced at $O(\e^2)$ in the original equations, with prefactor $\phi$, whereas the amplitudes appear at $O(\e)$. Both $f_{\bLac_j}$ and $\xi_{\bLac_j}$ are unknowns for the moment, but will be determined by the weakly nonlinear procedure, thus validating \textit{a posteriori} the Ansatz in (\ref{eq:nf}). Note that $\xi_{\bLac_j}$ could have been assumed to also be multiplicative (or not necessarily white), without changing its final expression.

From now on, it would be possible to make the amplitudes depend on a slow timescale $\tau \coloneq \e^2 t$, as $\bLac_j = \bLac_j(\tau)$, and proceed with the multiple-scale method. However, while this slow-time dependence is justified in the deterministic regime \citep{Ohm22}, we choose not to enforce it explicitly here. That is because the amplitudes are stochastically forced by white noise, and thus their temporal derivatives are all diverging (at least in the $L^2$ space of interest here). Therefore, we found it mathematically questionable to stipulate that they vary slowly over time. Nevertheless, we argue in Appendix~\ref{app:mms} that the multiple-scale method (or at least a certain interpretation of it) leads to the same final systems at leading-order.  
The method proposed below, on the other hand, relies on basic algebraic manipulations and, in that sense, may appear less arbitrary than the multiple-scale method. 


A first-order correction to the state vector (\ref{eq:q1c4H}) corresponds to a first-order correction to the density function. Using (\ref{eq:psi_j}) and (\ref{eq:M0_w}) with $\chi_1=0$, we can derive
\begin{align}
\begin{split}
\psi_1(\bx,\bp,t) =& \Ap(t) \hps^{\Ap}(\bp) e^{\ti \bka \bdot \bx + \ti \om t} + \Am(t) \hps^{\Am}(\bp) e^{\ti \bka \bdot \bx - \ti \om t } \\
& + \Bp(t) \hps^{\Bp}(\bp) e^{\ti \bkb \bdot \bx + \ti \om t } + \Bm(t) \hps^{\Bm}(\bp) e^{\ti \bkb \bdot \bx - \ti \om t} + \cc,
  \label{eq:psi1_g}
 \end{split}
\end{align}
with
\begin{align}
\begin{split}
\hps^{\bLac}(\bp) \coloneq \hbB^{\bLac} : \bp\bp^T + \hba^{\bLac} \bdot \bp + \hw^{\bLac}, \quad \text{and} \quad \vem{\hw^{\bLac}}{\hba^{\bLac}}{\ve{\hbB}^{\bLac}} = \bM^{-1}_0 \vem{\tc^{\bLac}}{\tbn^{\bLac}}{\ve{\tbQ}^{\bLac}}, 
  \nonumber
 \end{split}
\end{align}
where the symbol $\bLac$ denotes any one of the components of $\bLa$.

Injecting the scalings (\ref{eq:edef}) and (\ref{eq:fstocdef}), as well as the expansion (\ref{eq:qq_ew}), into the system (\ref{eq:ec_st}) results in the new expansion
\begin{align}
\begin{split}
& \e  \sae{\pae{\bD\pa_t-\bL}\bq_1 } + \e^2 \sae{\pae{\bD\pa_t-\bL}\bq_2 - \frac{1}{2}\bN[\bq_1,\bq_1] - \bet \br_2 - \phi \bff} \\ 
& + \e^3 \sae{\pae{\bD\pa_t-\bL}\bq_3 + \bD \Delta \bq_1 - \bN[\bq_1,\bq_2]- \bet \br_3 }  + O(\e^4) = \bz.
  \label{eq:wo1}
 \end{split}
\end{align}
In (\ref{eq:wo1}), we have used that the base state $\bq_0$ is an exact solution of (\ref{eq:ec_st}), thus there are no terms at $O(1)$. In addition, $\bL$ results from the linearization of (\ref{eq:ec_st}) around $\bq_0$ and at $D_{T,c}$. The fact that we consider $D_T$ smaller than $D_{T,c}$, with a $\e^2$ difference, naturally results in the $\bD \Delta \bq_1$ term at $O(\e^3)$ in (\ref{eq:wo1}). By acting over some $\bq_j$, the linear operator $\bL$ encompasses the contribution to $\bR_j$ that is linear in the first three moments in (\ref{eq:R_w}) (it produces a divergence term pre-multiplied by $\bet$ in the linear operator). On the other hand, the contributions to $\bR_j$ with the subscript $f$ result from nonlinear interactions of previous order terms, thus result as forcing terms denoted by $\br_j$ in (\ref{eq:wo1}), and which is such that
\begin{align}
\begin{split}
\br_j \coloneq \pae{0,\bz,-\vect{\nab \bdot \bR_{f,j}},\bz,0}^T, \quad \text{with} \quad j=1,2,... .
  \label{eq:dpq}
\end{split}
\end{align}
In (\ref{eq:wo1}), we have also defined the nonlinear, quadratic operator
\begin{align}
\begin{split}
\bN[\bq_i,\bq_j] \coloneq \bC[\bq_i,\bq_j] + \bC[\bq_j,\bq_i], \nonumber
\end{split}
\end{align}
with
\begin{align}
\begin{split}
\bC[\bq_i,\bq_j] \coloneq \vemo{-(\bu_j\bdot\nab)c_i}{-(\bu_j\bdot\nab)\bn_i +(\nab\bu_j)\bn_i - \bR_i:\bE_j}{\vect{-(\bu_j\bdot\nab)\bQ_i + (\nab\bu_j)\bQ_i + \bQ_i(\nab\bu_j)^T -2\bS_i:\bE_j} }{\bz}{0}, \nonumber
\end{split}
\end{align}
embedding the nonlinear interactions between the velocity field $\bu_j$ and the moments $c_i$,$\bn_i$, and $\bQ_i$. Note that $\bN[\bq_i,\bq_j] = \bN[\bq_j,\bq_i]$. Eventually, the symbol $\bff$ denotes
\begin{align}
\begin{split}
\bff(\bx,t) \coloneq \pae{f_c,\bff_n,\vect{\bF_D},\bz,0}^T(\bx,t).\label{eq:fcgd}
\end{split}
\end{align}

Applying $\pae{\bD\pa_t-\bL}$ to the expression (\ref{eq:q1c4H}) of $\bq_1$, we obtain 
\begin{align}
\begin{split}
&\pae{\bD\pa_t-\bL}\bq_1 \\
= &\e \phi \bD \sae{e^{\ti \bka \bdot \bx}\pae{\xi_{\Ap} \tbq^{\Ap} e^{\ti \om t} + \xi_{\Am} \tbq^{\Am} e^{- \ti \om t}} +  e^{\ti \bkb \bdot \bx}\pae{\xi_{\Bp} \tbq^{\Bp} e^{\ti \om t} + \xi_{\Bm} \tbq^{\Bm} e^{- \ti \om t}}}\\
&+\e^2 \bD \Bigg[ e^{\ti \bka \bdot \bx}\sae{\pae{ f_{\Ap} + \phi \xi^{(2)}_{\Ap}} \tbq^{\Ap} e^{\ti \om t} + \pae{ f_{\Am} + \phi \xi^{(2)}_{\Am}}\tbq^{\Am} e^{- \ti \om t}} \\
&+ e^{\ti \bkb \bdot \bx} \sae{   \pae{f_{\Bp} + \phi \xi^{(2)}_{\Bp}} \tbq^{\Bp} e^{\ti \om t} + \pae{ f_{\Bm} + \phi\xi^{(2)}_{\Bm} } \tbq^{\Bm} e^{- \ti \om t}} \Bigg] + \cc +O(\e^3),
  \label{eq:q1in}
\end{split}
\end{align}
where we have used (\ref{eq:nf}) and the definition of the eigenmodes in (\ref{eq:deig}). Indeed, we recall that all linear operators in the expansion are constructed at $D_{T,c}$. 

On the other hand, assuming the statistics of the desired solution to be uniform in space, each higher-order solution $\bq_j$, with $j=2,3,...$, as well as the stochastic forcing, are represented as a Fourier series
\begin{align}
\begin{split}
\bq_j(\bx,t) = \bq_{j,\bz}(t) + \sum_{\bkn}\hbq_{j,\bk}(t) e^{\ti \bk \bdot \bx}, \quad \text{and} \quad \bff(\bx,t) = \sum_{\bkn}\hbf_{\bk}(t) e^{\ti \bk \bdot \bx},
  \label{eq:defq}
\end{split}
\end{align}
where $\bq_{j,0}(t)$ corresponds to the spatial average of $\bq_j(\bx,t)$, and where recall that the spatial average of the forcing is null by construction, i.e., $\hbf_{\bz}(t)=\bz, \forall t$. From this point forward, the index $\bk$ denotes the Fourier component oscillating in space at wavenumber $\bk$ of the corresponding field.

Introducing (\ref{eq:q1c4H}), (\ref{eq:q1in}) and (\ref{eq:defq}) into (\ref{eq:wo1}) produces a new expansion for each of the spatial Fourier components. The expansion for the component oscillating at $\bka$, in particular, reads  
\begin{align}
\begin{split}
&\e^2 \sae{\pae{\bD\di_t-\bL_{\bka}}\hbq_{2,\bka} - \phi \hbf_{\bka} + \phi \xi_{\Ap} \bD \tbq^{\Ap} e^{\ti \om t} + \phi \xi_{\Am}\bD \tbq^{\Am} e^{-\ti \om t} - \hbd_{2,\bka} } \\ 
& +  \e^3 \Big [ \pae{\bD\di_t-\bL_{\bka}}\hbq_{3,\bka} + \pae{ f_{\Ap} - \Ap + \phi \xi^{(2)}_{\Ap} }\bD\tbq^{\Ap} e^{\ti \om t} \\
& +  \pae{ f_{\Am} - \Am + \phi \xi^{(2)}_{\Am}}\bD \tbq^{\Am} e^{-\ti \om t} - \hbd_{3,\bka}   \Big]  + O(\e^4) = \bz.
  \label{eq:o3ka}
 \end{split}
\end{align}
The symbol $\bd_j$ denotes the forcing term, appearing at order $j$, and stemming from nonlinear interactions of fields determined at previous orders. Its sole purpose is to lighten the notations. From (\ref{eq:wo1}), it corresponds to 
\begin{eqnarray}
\bd_2 \coloneq  \frac{1}{2}\bN[\bq_1,\bq_1] + \bet \br_2,  \quad \bd_3 \coloneq  \bN[\bq_1,\bq_2] +  \bet \br_3, \quad \text{etc}.\label{eq:bd3}
\end{eqnarray}
In deriving the expansion (\ref{eq:o3ka}), we have also used that the Laplacian $\Delta$ becomes $-\nn{\bk}^2$ when applied to a Fourier $\bk$-component, but $\nn{\bka}=\nn{\bkb}=1$. 

The expansion for the component oscillating at $\bkb$ is similar to (\ref{eq:o3ka}), but $\Ap$, $\Am$, and $\bka$, are replaced by $\Bp$, $\Bm$ and $\bkb$, respectively (including within the superscripts and subscripts) 


Let us now characterize the forcing term $\bd_2$, induced by nonlinear interactions of the first-order with itself. For this, we first inject the expression (\ref{eq:psi1_g}) for $\psi_1$, into that for $\chi_2$ given in (\ref{eq:chi_j}). We then use (\ref{eq:R_w}) to compute the corresponding forcing in the fourth moment, $\bR_{f,2}$, from which we can deduce $\br_2$ in line with (\ref{eq:dpq}). In addition, the expression (\ref{eq:q1c4H}) for $\bq_1$ is introduced in $\bN[\bq_1,\bq_1]$. Since the first-order solution only includes the wavenumbers $\pm \bka$ and $\pm \bkb$, and since both $\chi_2$ and $\bN[\bq_1,\bq_1]$ are quadratic nonlinearities, $\bd_2$ only includes the wavenumbers contained in the set 
\begin{align}
\begin{split}
\Ks \coloneq \set{\bz, \pm 2\bka, \pm 2\bkb, \pm(\bka+\bkb), \pm(\bka-\bkb)}. \nonumber
\end{split}
\end{align}
This implies in particular that $\hbd_{2,\bka} = \hbd_{2,\bkb} = \bz$, which should accounted for in the expansion (\ref{eq:o3ka}) for $\bka$ and its equivalent for $\bkb$. After tedious but elementary algebra, it is possible to disentangle the following expression
\begin{align}
\begin{split}
\bd_2(t) =& \bd_{2,\bz}(t) + \Big (  \hbd_{2,2\bka}(t) e^{2 \ti \bka\bdot\bx} + \hbd_{2,2\bkb}(t) e^{2\ti \bkb\bdot\bx} \\
& + \hbd_{2,\bka+\bkb}(t) e^{\ti(\bka+\bkb)\bdot\bx} + \hbd_{2,\bka-\bkb}(t) e^{\ti(\bka-\bkb)\bdot\bx} + \cc \Big ). 
\label{eq:d2full} 
\end{split}
\end{align}
The contribution from the wavenumber $\bk = \bz$ itself can be further detailed as 
\begin{align}
\begin{split}
\bd_{2,\bz}(t) =& \mAp^2 \hbd^{\Ap \Ap^*} + \mAm^2 \hbd^{\Am \Am^*} + \mBp^2 \hbd^{\Bp \Bp^*} + \mBm^2  \hbd^{\Bm \Bm^*} \\
& + \pae{\Ap\Am^* \hbd^{\Ap\Am^*} e^{2\ti \om t} + \cc} + \pae{ \Bp\Bm^*\hbd^{\Bp\Bm^*}e^{2\ti \om t} + \cc },  \nonumber
\end{split}
\end{align}
where the dependence of the amplitudes on time is implied. The pair of amplitudes in the superscript of each $\hbd^{\bLac_i\bLac_j}$ emphasizes that the latter is pre-multiplied by $\bLac_i\bLac_j$. The order of appearance of the amplitudes in the superscript does not matter, and $\hbd^{\bLac_i\bLac_j}=\hbd^{\bLac_j\bLac_i}$. Each $\hbd^{\bLac_i\bLac_j}$ is time-independent and reads
\begin{align}
\begin{split}
\hbd^{\bLac_i\bLac_j} \coloneq \gamma  \bNc{\tbq^{\bLac_i},\tbq^{\bLac_j}} + \bet \hbr^{\bLac_i\bLac_j}, \quad \text{with} \quad \gamma = \begin{cases}
1 & \text{if} \; i\neq j \\
\frac{1}{2} & \text{if} \; i= j
\end{cases}
\nonumber
\end{split}
\end{align}
Because $\bq_1$ is real-valued, $\tbq^{\bLac_i} = \tbq^{\bLac_{i-4},*}$ for $i=5,...,8$, e.g., $\tbq^{\Ap^*} = \tbq^{\Ap,*}$, etc. The operator $\bNc{\bullet,\bullet}$ is similar to $\bN[\bullet,\bullet]$, except that the gradient $\nab$ is replaced by $\hnab \doteq \ti \bk$. If it applies to $\tbq^{\Ap}$, then $\hnab = \ti \bka$, etc. The term $\hbr^{\bLac_i\bLac_j} = (0,\bz,-\vect{\hnab \bdot \hbR^{\bLac_i\bLac_j}_{f}},\bz,0)^T$ requires the knowledge of $\hbR^{\bLac_i\bLac_j}_{f}$, which is computed from (\ref{eq:R_w}) by replacing $\chi_2$ by its contribution $\hch^{\bLac_i\bLac_j}$, where
\begin{align}
\begin{split}
\hch^{\bLac_i\bLac_j}(\bp) = \gamma \hps^{\bLac_i}(\bp)\hps^{\bLac_j}(\bp). \nonumber
\end{split}
\end{align}
Again, because $\psi_1$ is real-valued, $\hps^{\bLac_i}=\hps^{\bLac_{i-4},*}$ for $i\geq 5$. For example, we evaluate $\hch^{\Ap\Ap^*} = |\hps^{\Ap}|^2$, $\hch^{\Ap\Ap} = \hps^{\Ap\Ap}/2$, etc. 

The contributions of the other wavenumbers to $\bd_2$ in (\ref{eq:d2full}) are detailed as
\begin{align}
\begin{split}
\hbd_{2,2\bka}(t) =& \Ap\Am \hbd^{\Ap\Am} + \Ap^2 \hbd^{\Ap\Ap} e^{2\ti \om t} + \Am^2\hbd^{\Am\Am}e^{-2\ti \om t}, \label{eq:d2ka}
\end{split}
\end{align}
followed by,
\begin{align}
\begin{split}
\hbd_{2,2\bkb}(t) =& \Bp\Bm \hbd^{\Bp\Bm} + \Bp^2 \hbd^{\Bp\Bp} e^{2\ti \om t} + \Bm^2\hbd^{\Bm\Bm}e^{-2\ti \om t}, \nonumber
\end{split}
\end{align}
then,
\begin{align}
\begin{split}
\hbd_{2,\bka+\bkb}(t) =& \Ap \Bm \hbd^{\Ap\Bm} + \Am \Bp \hbd^{\Am\Bp}  + \Ap \Bp \hbd^{\Ap\Bp}e^{2\ti \om t} + \Am \Bm \hbd^{\Am\Bm}e^{-2\ti \om t}, \nonumber
\end{split}
\end{align}
and, eventually,
\begin{align}
\begin{split}
\hbd_{2,\bka-\bkb}(t) =& \Ap \Bp^* \hbd^{\Ap \Bp^*} + \Am \Bm^*  \hbd^{\Am \Bm^*}  + \Ap \Bm^* \hbd^{\Ap \Bm^*} e^{2\ti \om t}  + \Am \Bp^* \hbd^{\Am \Bp^*} e^{-2\ti \om t}.
\nonumber
\end{split}
\end{align}

Meanwhile, the expansion associated with each wavenumber in the set $\Ks$ (the only ones contributing to $\bd_2$), reads
\begin{align}
\e^2 \sae{\pae{\bD\di_t-\bL_{\bk}}\hbq_{2,\bk} - \phi \hbf_{\bk} - \hbd_{2,\bk}} +O(\e^3)  = \bz, \quad \text{for} \quad \bk \in \Ks,
\label{eq:o2exps}
\end{align}
and remembering in particular that $\hbf_0=\bz$. By contrast with the expansions for $\bka$ in (\ref{eq:o3ka}) and $\bkb$, we do not show the $O(\e^3)$ term in (\ref{eq:o2exps}) as they do not play a role in what follows. 

We now rewrite each expansion in (\ref{eq:o2exps}) in integral form by applying $\pae{\bD\di_t-\bL_{\bk}}^{-1}$, which amounts to inverting for each $\hbq_{j,\bk}(t)$, with $j=2,3.,..$. For reasons that will become clear, under the Ansatz (\ref{eq:nf}), inverting for $\hbq_{j,\bk}(t)$ and then collecting terms at each power of $\e$ is asymptotically consistent, while doing the opposite, i.e., collecting terms first and then inverting, is not. 

In the system (\ref{eq:ec_st}), the Stokes equations are never differentiated with respect to time. Therefore, they act as algebraic constraints while the entire system evolves. Accordingly, we introduce the projector $\bPs$ such that $\bPs\hbq(t)$ automatically satisfies the Stokes equations in (\ref{eq:ec_st}), even if the state vector $\hbq(t)$ does not. In the remaining, we restrict the state vectors to evolve in the range of $\bPs$, meaning that the velocity and pressure are slaved to the moments for all time $t\geq 0$. Furthermore, the Stokes equations are such that, for given moments, there must correspond a unique velocity and pressure field. 

Note that requiring $\hbq(t) \in \ran(\bPs)$ for $t\geq 0$ constitutes a loss of generality, because (\ref{eq:ec_st}) need not be true for $t=0$, and thus the initial condition need not be in the range of $\bPs$ (i.e., it has no reason to be ``compatible'' with the equations). This loss of generality, however, is unimportant.  

It is then useful to introduce the propagator $\bPhi_{\bk}$, such that the operator $(\bD\di_t-\bL_{\bk})$ can be rewritten as 
\begin{align}
(\bD\di_t-\bL_{\bk})\hbq(t) = \bPhi_{\bk}(t,0)\di_t\pae{\bPhi_{\bk}(0,t)\hbq(t)},\quad \text{for all} \quad \hbq(t) \in \ran(\bPs), \label{eq:prpp}
\end{align}
where $\bD \di_t \bPhi_{\bk}(t,0) = \bL_{\bk} \bPhi_{\bk}(t,0)$ and $\bPhi_{\bk}$ acting in $\ran(\bPs)$. In the absence of the singular mass matrix $\bD$, the propagator $\bPhi_{\bk}(t,s)$ would simply be the matrix exponential $e^{\bL_{\bk}(t-s)}$. The propagator obeys the semi-group composition properties $\bPhi_{\bk}(t,s) = \bPhi_{\bk}(t,r)\bPhi_{\bk}(r,s)$, and $\bPhi_{\bk}(t,t)=\bI_{\ran(\bPs)}$. The system $(\bD\di_t-\bL_{\bk})=\hbg(t)$ can be easily inverted for an arbitrary $\hbg(t)$ using (\ref{eq:prpp}) by successively: (i) multiplying by $\bPhi_{\bk}(0,t)$, (ii) integrating between $0$ and $t$, and (iii) multiplying by $\bPhi_{\bk}(t,0)$. This gives $\hbq(t) = \bPhi(t,0)\hbq(0) + \bPhi(t,0)\int_{0}^{t} \bPhi(0,s) \hbg(s)\di s$.

Crucially, the propagator yields the dyadic decomposition 
\begin{align}
\bPhi_{\bk}(t,0) (\bullet)  = \sum_{j\geq 1} \ssd{\tbq^{\da}_j}{(\bullet)}\tbq_j e^{\sig_j t},\nonumber 
\end{align}
where we recall $\set{\tbq_j}_{j\geq 1}$ to be the set of $1+d+d^2$ eigenmodes solving (\ref{eq:lin1}) with finite eigenvalues and for a given $\bk$, and $\set{\tbq^{\da}_j}_{j\geq 1}$ to be the set of their respective adjoints. 
Since the inner product in the dyadic representation contains the mass matrix, a zero initial condition in $c$, $\bn$, and $\bQ$, implies that the linear response is zero in all components for all $t\geq0$ regardless of the initial velocity or pressure. That is precisely because we have restricted the state vector to be within $\ran(\bPs)$ also at $t=0$. This implies that the initial condition must be compatible with the equation, and that the velocity and pressure adapt to the first three moments: if the latter are zero, then the former must be zero as well by inverting the Stokes equations, and the linear solution remains zero for all times. This is consistent with our numerical time-stepping procedure in Sec.~\ref{sec:dns}.   

Rewriting the expansions in (\ref{eq:o2exps}) in inverted/integral form results in
\begin{align}
\begin{split}
& \e^2 \pae{\hbq_{2,\bk}(t) - \phi\int_{0}^{t}\bPhi_{\bk}(t,s)\hbf_{\bk}(s)\di s - \int_{0}^{t}\bPhi_{\bk}(t,s)\hbd_{2,\bk}(s)\di s} + O(\e^3)  = \bz, \label{eq:o21} 
\end{split}
\end{align}
for each $\bk \in \Ks $. Without loss of generality, we set $\hbq_{2,\bk}(0)=\bz$ when deriving (\ref{eq:o21}), and the initial conditions of each higher-order field are similarly set to zero in the following. Equivalently, all transients will be ignored because they vanish over time, and only the statistically steady regime will be considered. The integral term $\int_{0}^{t}\bPhi_{\bk}(t,s)\hbd_{2,\bk}(s)\di s$ involves monomials in the amplitudes. For example, if $\bk=2\bka\in \Ks$, then
\begin{align}
\begin{split}
& \int_{0}^{t}\bPhi_{2\bka}(t,s)\hbd_{2,2\bka}(s)\di s \\
& = \int_{0}^{t}\bPhi_{2\bka}(t,s)\pae{\Ap\Am \hbd^{\Ap\Am} + \Ap^2 \hbd^{\Ap\Ap} e^{2\ti \om s} + \Am^2\hbd^{\Am\Am}e^{-2\ti \om s}}\di s,
\label{eq:o2ka_1} 
\end{split}
\end{align}
where (\ref{eq:d2ka}) was used. As a consequence of the amplitudes having an $O(\e^2)$ deterministic dynamics and being forced with a noise of amplitude $O(\e)$ in (\ref{eq:nf}), the integrals involving functions of the amplitudes can (and thus must) be expanded asymptotically by performing successive integration by parts. This procedure extracts the amplitudes out of the temporal integrals up to any order in $\e$. For example, isolating the first of the three integral terms in (\ref{eq:o2ka_1}) and defining $\bi(s)\coloneq \int_{0}^{s}\bPhi_{2\bka}(t,x) \hbd^{\Ap\Am} \di x$, we develop
\begin{align}
&\int_{0}^{t} \Ap\Am \underbrace{\bPhi_{2\bka}(t,s) \hbd^{\Ap\Am}}_{= \di \bi(s) / \di s} \di s  \myeq \sae{\Ap\Am \bi}_{s=0}^{s=t} - \int_{0}^{t} (\Ap \di_s \Am + \Am \di_s \Ap) \bi \di s \nonumber \\
=& \Ap(t)\Am(t) \bi(t) - \e \phi \int_{0}^{t} (\Ap \xi_{\Am} +\Am \xi_{\Ap}) \bi \di s + O(\e^2) . \nonumber \\
\myeq & \Ap(t)\Am(t) \bi(t) - \e \phi \pae{\Ap(t) \int_{0}^{t} \xi_{\Am} \bi \di s  + \Am(t) \int_{0}^{t} \xi_{\Ap} \bi \di s} + O(\e^2),
\label{eq:o2ka_3}
\end{align}
An integration by parts was performed between the first and second line, and (\ref{eq:nf}) was used to express $\di_t \Am$ and $\di_t \Ap$. This resulted in the appearance of integral terms in $\e$, $\e^2$, $\e^3$, etc. In turn, the amplitudes and functions $f_{\bLac_j}$ can be extracted from these new integral terms by performing additional integration by parts, as was done between the second and third line.

Importantly, the term $\bi(t) = \int_{0}^{t}\bPhi_{2\bka}(t,x) \hbd^{\Ap\Am} \di x$ is identified as the solution to $(\bD\di_t-\bL_{2\bka})\bi(t) = \hbd^{\Ap\Am}$ at time $t$. As we have seen in Sec.~\ref{sec:linf}, the linear system $2\bka$ possesses only strictly stable eigenmodes. Therefore, we expect $\bi(t)$ to remain bounded for all times and, in particular, to converge to the time-independent vector $\hbq^{\Ap\Am}$, solving $-\bL_{2\bka}\hbq^{\Ap\Am} = \hbd^{\Ap\Am}$. We further show in Appendix~\ref{App:hotn} that the integral terms at $O(\e)$ in (\ref{eq:o2ka_3}) also yield a bounded variance despite the presence of white noise in the integrand, and for this reason can rigorously be scaled asymptotically. Overall, under (\ref{eq:nf}), the integral (\ref{eq:o2ka_3}) becomes 
\begin{align}
\begin{split}
\int_{0}^{t} \Ap\Am \bPhi_{2\bka}(t,s) \hbd^{\Ap\Am} \di s  = \Ap\Am\hbq^{\Ap\Am} + O(\e). \nonumber
\end{split}
\end{align}
Proceeding along the same lines of calculations for each of the other terms in (\ref{eq:o2ka_1}) permits us to transform the expansion (\ref{eq:o21}) into 
\begin{align}
\begin{split}
& \e^2 \bigg( \hbq_{2,2\bka}(t) - \phi\hbq^{\phi}_{2\bka}(t) - \Ap\Am \hbq^{\Ap\Am} - \Ap^2\hbq^{\Ap\Ap}e^{2\ti\om t} - \Am^2\hbq^{\Am\Am}e^{-2\ti\om t} \bigg) \\ 
& + O(\e^3)  = \bz, \label{eq:o21_bis} 
\end{split}
\end{align}
for $\bk=2\bka$, and where
\begin{align}
\begin{split}
\hbq^{\phi}_{\bk}(t) \coloneq \int_{0}^{t}\bPhi_{\bk}(t,s)\hbf_{\bk}(s)\di s, \quad \text{for} \quad \bk \in \Ks, \nonumber
\end{split}
\end{align}
is the linear response to the stochastic forcing $\hbf_{\bk}(t)$ (hence the superscript ``$\phi$''). Although $\hbq^{\phi}_{2\bka}(t)$ has a bounded root mean square because each $\bk \in \Ks$ is strictly stable, it does not converge to any specific value in the long-time limit, and thus has to be kept time-dependent.

In (\ref{eq:o21_bis}), the higher-order terms produced by the integration by parts have been absorbed at $O(\e^3)$. This is possible because the terms have not yet been collected. We now understand perhaps more clearly why it is asymptotically more consistent to first rewrite the expansions in (\ref{eq:o2exps}) in inverted form and then collect the terms, rather than doing the opposite. Because rewriting an expansion in inverted form reveals that contributions to certain integral terms must be prioritized asymptotically. Therefore, they must appear in their respective orders. For instance, the integral term pre-multiplied by $\e$ in (\ref{eq:o2ka_3}) must contribute to $\hbq_{3,2\bka}(t)$, the one at $O(\e^2)$ must contribute to $\hbq_{4,2\bka}(t)$, etc... Inverting the system only after having collected the terms would force the contributions at each order to all be absorbed within $\hbq_{2,\bka}(t)$. 

Another reason to collect the terms only after rewriting the expansion in inverted/integral form is that the white noise forcings in expansions (\ref{eq:o2exps}) have an infinite variance. However, as shown in Appendix~\ref{App:hotn}, their linear responses have finite variances. Thus, it is these latter responses that can be rigorously measured and collected accordingly at each power of $\e$. 

Now collecting terms at $O(\e^2)$ in (\ref{eq:o21_bis}) gives the $2\bka$-component of the second-order solution   
\begin{align}
\begin{split}
\hbq_{2,2\bka}(t) = \phi\hbq^{\phi}_{2\bka}(t) + \Ap\Am \hbq^{\Ap\Am} + \Ap^2\hbq^{\Ap\Ap}e^{2 \ti \om t} + \Am^2\hbq^{\Am\Am}e^{-2 \ti \om t}.
\label{eq:o21_bisbis} 
\end{split}
\end{align}

By applying for each of the others $\bk \in \Ks$ the established calculation pattern, that is: (i) writing the expansion under inverted/integral form, then (ii) expanding the integral involving the amplitudes asymptotically by integration by parts, and (iii) prioritizing the terms thus produced at their respective orders and only then collecting the terms at $O(\e^2)$, we gather,
\begin{align}
\begin{split}
\bq_{2,\bz}(t) =& \mAp^2 \hbq^{\Ap \Ap^*} + \mAm^2 \hbq^{\Am \Am^*} + \mBp^2 \hbq^{\Bp \Bp^*} + \mBm^2  \hbq^{\Bm \Bm^*} \\
& + \pae{\Ap\Am^* \hbq^{\Ap\Am^*} e^{2\ti \om t} + \cc} + \pae{ \Bp\Bm^*\hbq^{\Bp\Bm^*}e^{2\ti \om t} + \cc },  \nonumber
\end{split}
\end{align}
then, 
\begin{align}
\begin{split}
\hbq_{2,2\bkb}(t) =& \phi\hbq^{\phi}_{2\bkb}(t) + \Bp\Bm \hbq^{\Bp\Bm} + \Bp^2 \hbq^{\Bp\Bp} e^{2\ti \om t} + \Bm^2\hbq^{\Bm\Bm}e^{-2\ti \om t}, \label{eq:q2d2kb} 
\end{split}
\end{align}
followed by,
\begin{align}
\begin{split}
\hbq_{2,\bka+\bkb}(t) =& \phi\hbq^{\phi}_{\bka+\bkb}(t) + \Ap \Bm \hbq^{\Ap\Bm} + \Am \Bp \hbq^{\Am\Bp}  + \Ap \Bp \hbq^{\Ap\Bp}e^{2\ti \om t} \\
& + \Am \Bm \hbq^{\Am\Bm}e^{-2\ti \om t}, \label{eq:q2dkapkb} 
\end{split}
\end{align}
and, eventually,
\begin{align}
\begin{split}
\hbq_{2,\bka-\bkb}(t) =& \phi\hbq^{\phi}_{\bka-\bkb}(t) + \Ap \Bp^* \hbq^{\Ap \Bp^*} + \Am \Bm^*  \hbq^{\Am \Bm^*}  + \Ap \Bm^* \hbq^{\Ap \Bm^*} e^{2\ti \om t} \\
& + \Am \Bp^* \hbq^{\Am \Bp^*} e^{-2\ti \om t}.
 \label{eq:q2dkamkb} 
\end{split}
\end{align}
In the expressions above, each $\hbq^{\bLac_i\bLac_j}$ is time-independent and solves
\begin{align}
\begin{split}
\sae{ \ti (\om_{\bLac_i}+\om_{\bLac_j})\bD - \bL_{\bk_{\bLac_i}+\bk_{\bLac_j}}}\hbq^{\bLac_i\bLac_j} = \hbd^{\bLac_i\bLac_j}, \quad \text{where} \quad \bk_{\bLac_i}+\bk_{\bLac_j} \in \Ks
\label{eq:q2XY}
\end{split}
\end{align}
where $\om_{\bLac_j}$ and $\bk_{\bLac_j}$ are the frequency and wavenumber, respectively, of the eigenmode $\tbq^{\bLac_j}$ pre-multiplied by the amplitude $\bLac_j$, with $j=1,...,8$. For instance, if in (\ref{eq:q2XY}) we choose $i=1$ (recall that $\bLac_1=\Ap$) and $j=8$ (recall that $\bLac_8=\Bm^*$), then $\om_{\bLac_1} = \om$, $\bk_{\bLac_1} = \bka$, $\om_{\bLac_8} = \om$ and $\bk_{\bLac_8} = -\bkb$, such that $\pae{  2\ti\om \bD - \bL_{\bka-\bkb}}\hbq^{\Ap \Bm^*} = \hbd^{\Ap \Bm^*}$.

The linear systems in (\ref{eq:q2XY}) are invertible only because $\bk_{\bLac_i}+\bk_{\bLac_j} \in \Ks$, such that they are associated with strictly stable eigenvalues only. For the reasons mentioned in Sec.~\ref{sec:linf} (see (\ref{eq:Ldef_k0}) and the discussion below), the system for $\bk_{\bLac_i}+\bk_{\bLac_j}=\bz$ requires a special treatment before being invertible. In practice, it is solved in a projected space free of the velocity and pressure degrees of freedom, although it is also possible to solve for the pseudo-inverse in the full space using the command \texttt{pinv} of \texttt{Matlab}.

All the wavenumbers not contained in the set $\Ks \cup \set{\pm \bka,\pm \bkb}$ are only excited by the stochastic forcing at $O(\e^2)$, such that
\begin{align}
\e^2 \sae{\pae{\bD\di_t-\bL_{\bk}}\hbq_{2,\bk} - \phi \hbf_{\bk}} +O(\e^3)  = \bz, \quad \text{for} \quad \bk \notin \Ks \cup \set{\pm \bka,\pm \bkb},
\label{eq:o2ff}
\end{align}
which gives $\hbq_{2,\bk}(t) = \phi \hbq_{2,\bk}^{\phi}(t)$.

Now that we have determined the second-order corrections for all wavenumbers except $\bka$ and $\bkb$, let us focus on the expansions for these two latter wavenumbers. In the expansion for $\bka$, given in (\ref{eq:q1in}), again applying $(\bD \di_t - \bL_{\bk})^{-1}$ and using the propagator formalism brings about
\begin{align}
\begin{split}
&\e^2 \bigg [ \hbq_{2,\bka}(t) - \phi \int_{0}^{t}\bPhi(t,s)\hbf_{\bka}\di s + \phi \int_{0}^{t}\bPhi(t,s)\bD \pae{\xi_{\Ap} \tbq^{\Ap}  e^{\ti \om s} + \xi_{\Am} \tbq^{\Am}  e^{-\ti \om s}} \di s \bigg ] \\
& +  \e^3 \bigg [ \hbq_{3,\bka}(t) + \int_{0}^{t} \bPhi(t,s) \pae{ f_{\Ap} - \Ap + \phi \xi^{(2)}_{\Ap} }\bD\tbq^{\Ap} e^{\ti \om s}\di s \\
& + \int_{0}^{t} \bPhi(t,s) \pae{ f_{\Am} - \Am + \phi \xi^{(2)}_{\Am}}\bD \tbq^{\Am} e^{-\ti \om s} \di s - \int_{0}^{t} \bPhi(t,s)\hbd_{3,\bka}\di s \bigg ] + O(\e^4) = \bz.
\label{eq:o3ka_bis}
 \end{split}
\end{align}
where $\bPhi$ is the shortened symbol for $\bPhi_{\bka}$, and we have used that $\hbd_{2,\bka}=\bz$. The temporal integrals at $O(\e^2)$ cannot be further expanded asymptotically by integrating by parts, because they do not involve the amplitudes. Therefore, it is legitimate to directly collect the term at $O(\e^2)$, leading to
\begin{align}
\begin{split}
\hbq_{2,\bka}(t) = \phi \int_{0}^{t}\bPhi(t,s)\hbf_{\bka}\di s - \phi \int_{0}^{t}\bPhi(t,s) \bD \pae{\xi_{\Ap} \tbq^{\Ap}  e^{\ti \om s} + \xi_{\Am} \tbq^{\Am}  e^{-\ti \om s}} \di s.
\label{eq:o3ka_bb}
 \end{split}
\end{align}
By contrast with the cases treated above, the temporal integrals in (\ref{eq:o3ka_bb}) are problematic, for they produce diverging (or ``secular'') terms. That is because the linear system at $\bka$ possesses neutral eigenmodes. More precisely, the dyadic representation of the propagator $\bPhi(t,s) (=\bPhi_{\bka}(t,s))$ is 
\begin{align}
\begin{split}
\bPhi(t,s) (\bullet) =& \spap{(\bullet)} \tbq^{\Ap} e^{\ti \om (t-s)}  + \spam{(\bullet)} \tbq^{\Am} e^{-\ti \om (t-s)} \\
&+ \underbrace{\sum_{j\geq3} \ssd{\tbq^{\da}_j}{(\bullet)}\tbq_j e^{\sig_j (t-s)}}_{=\bPhi(t,s)\bPo (\bullet)},
\label{eq:dyad_ka} 
\end{split}
\end{align}
where $\tbq_1 = \tbq^{\Ap}$ and $\tbq_2 = \tbq^{\Am}$, such that all the eigenmodes involved in the sum from $j\geq 3$ are strictly stable. Accordingly, let us define the projector operator
\begin{align}
\bPo (\bullet) \coloneq \sum_{j\geq 1} \ssd{\tbq^{\da}_j}{(\bullet)}\tbq_j - \spap{(\bullet)}\tbq^{\Ap} -  \spam{(\bullet)}\tbq^{\Am} , 
\label{eq:bpo} 
\end{align}
which removes the component of an arbitrary state vector along the neutral eigenmodes $\tbq^{\Ap}$ and $\tbq^{\Am}$, and with respect to the adjoint basis. This implies that $\bPo$ is an oblique projector (hence the subscript ``$o$''). While $\bPo \tbq^{\Ap} = \bPo \tbq^{\Am} =\bz$, all the other, strictly stable eigenmodes remain untouched under the application of $\bPo$, i.e., $\bPo\tbq_j = \tbq_j$, for $j \geq 3$, since by bi-orthogonality $\ssd{\tbq^{\Ap,\da}}{\tbq_j}=\ssd{\tbq^{\Am,\da}}{\tbq_j} = 0$ for $j\geq 3$.   

In other words, (\ref{eq:dyad_ka}) says the action of the propagator can be decomposed as a non-decaying part contained in the neutral (slow) eigen-subspace, spanned by $\tbq^{\Ap}$ and $\tbq^{\Am}$, and a part $\bPhi(t,s)\bPo (\bullet)(=\bPo\bPhi(t,s)(\bullet))$ contained in the strictly stable (fast) eigen-subspace, and whose contribution thus decays exponentially with time $t$.

Expression (\ref{eq:dyad_ka}) can then be used in (\ref{eq:o3ka_bb}) to show that
\begin{align}
\begin{split}
\hbq_{2,\bka}(t) =& \phi e^{\ti\om t} \tbq^{\Ap} \underbrace{\int_{0}^{t}\spap{\hbf_{\bka}(s)}e^{-\ti\om s} - \xi_{\Ap}(s)\di s}_{\text{root mean square diverging $\propto \sqrt{t}$}}  \\
& + \phi e^{-\ti\om t} \tbq^{\Am} \underbrace{\int_{0}^{t} \spam{\hbf_{\bka}(s)}e^{\ti\om s} - \xi_{\Am}(s) \di s}_{\text{root mean square diverging $\propto \sqrt{t}$}} \\
&  + \phi\underbrace{ \int_{0}^{t} \bPhi(t,s) \bPo \pae{\hbf_{\bka}(s) -\xi_{\Ap}(s)\bD\tbq^{\Ap}e^{\ti\om s} - \xi_{\Am}(s)\bD\tbq^{\Am}e^{-\ti\om s}}\di s}_{\text{remains bounded as evolving in the strictly stable (fast) eigen-subspace}}.
  \label{eq:o3ka_o2_bis}
 \end{split}
\end{align}
where the bi-orthogonality property $\ssd{\tbq_1^{\Am,\da}}{\tbq^{\Ap}} =\ssd{\tbq_1^{\Ap,\da}}{\tbq^{\Am}}=0 $ was again used. The integral term containing $\bPo$ evolves into the strictly stable eigen-subspace. For that reason, it remains bounded and poses no threat to the asymptotic hierarchy. However, the two other terms, proportional to $\tbq^{\Ap}$ or $\tbq^{\Am}$, respectively, may be problematic. Indeed, say that $\xi_{\Ap}=\xi_{\Am}=0$. Then, the two terms proportional to $\tbq^{\Ap}$ or $\tbq^{\Am}$ integrate white noise processes without any stabilizing effect. Therefore, they are Wiener processes, whose root mean square $\propto \sqrt{t}$ diverges with time. This can easily be shown using the It\^{o} isometry. This ruins the asymptotic hierarchy (at least in the root mean square sense), as it implies in particular that $\hbq_{2,\bka}(t)$ should in fact be scaled at $O(\e)$ after a time $t \sim 1/\e^2$. Consequently, to preserve the asymptotics, the processes $\xi_{\Ap}$ and $\xi_{\Am}$ are chosen in such a way as to cancel the associated integrands in (\ref{eq:o3ka_o2_bis}), i.e.,
\begin{align}
\begin{split}
\xi_{\Ap}(t) &\coloneq \spap{\hbf_{\bka}(t)} e^{-\ti \om t}, \quad \xi_{\Am}(t) \coloneq \spam{\hbf_{\bka}(t)} e^{\ti \om t}.
  \label{eq:xika}
 \end{split}
\end{align}
This way, the two integral terms proportional to $\tbq^{\Ap}$ or $\tbq^{\Am}$ vanish for all times, and $\xi_{\Ap}$ and $\xi_{\Am}$ are white noise indeed, consistently with our Ansatz in (\ref{eq:nf}).

Note that, to preserve the asymptotic for all times larger than $t \sim 1/\e^2$, it would be sufficient to choose $\xi_{\Ap}$ and $\xi_{\Am}$ to cancel out only the low frequencies $|\om| \leq O(\e^2)$ of the integrand. In the latter case, $\xi_{\Ap}$ and $\xi_{\Am}$ would be low-pass filtered versions of their expressions in (\ref{eq:xika}). The choice made in (\ref{eq:xika}) is thereby conservative, because it cancels the components along $\tbq^{\Ap}$ and $\tbq^{\Am}$ for all frequencies. It has the advantage of avoiding introducing a non-specific low-pass filter.  


By proceeding similarly for the expansion associated with the wavenumber $\bkb$, we obtain 
\begin{align}
\begin{split}
\xi_{\Bp}(t) &\coloneq \spbp{\hbf_{\bkb}(t)} e^{-\ti \om t}, \quad \xi_{\Bm}(t) \coloneq \spbm{\hbf_{\bkb}(t)} e^{\ti \om t}.
  \label{eq:xikb}
 \end{split}
\end{align}

By combining the results for all wavenumbers, we eventually obtain that the second-order correction in the state vector is
\begin{align}
\begin{split}
\bq_2(\bx,t) =& \bq_{2,0}(t) + \Big (  \hbq_{2,2\bka}(t)e^{2\ti \bka\bdot\bx} + \hbq_{2,2\bkb}(t)e^{2\ti \bkb\bdot\bx} + \hbq_{2,\bka+\bkb}(t)e^{\ti(\bka+\bkb)\bdot\bx} \\
& + \hbq_{2,\bka-\bkb}(t)e^{\ti(\bka-\bkb)\bdot\bx} + \hbq_{2,\bk \notin \Ks}(\bx,t) + \cc \Big ), 
\label{eq:q2full} 
\end{split}
\end{align}
where the expression of each Fourier component was given above, and where we have regrouped all components $\bk \notin \Ks$ in the terms $\hbq_{2,\bk \notin \Ks}$.
It follows from the relations (\ref{eq:psi_j}) and (\ref{eq:M0_w}) that, to the second-order correction $\bq_2$ of the state vector, must correspond a second-order correction $\psi_2$ of the density of the same form  
\begin{align}
\begin{split}
\psi_2 =& \psi_{2,0} + \Big (  \hps_{2,2\bka}e^{2\ti \bka\bdot\bx} + \hps_{2,2\bkb}e^{2\ti \bkb\bdot\bx} + \hps_{2,\bka+\bkb}e^{\ti(\bka+\bkb)\bdot\bx} \\
& + \hps_{2,\bka-\bkb}e^{\ti(\bka-\bkb)\bdot\bx} + \hps_{2,\bk \notin \Ks} + \cc \Big ). 
\label{eq:ps2full} 
\end{split}
\end{align}
Each Fourier component of the density also has the same detailed shape as its counterpart in the state vector. For example, it follows from (\ref{eq:o21_bisbis}) that 
\begin{align}
\begin{split}
\hps_{2,2\bka}(\bp,t) =& \phi \hps^{\phi}_{2\bka}(\bp,t) + \Ap\Am \hps^{\Ap\Am}(\bp) + \Ap^2  \hps^{\Ap\Ap}(\bp)e^{2 \ti \om t}   \\
&+ \Am^2 \hps^{\Am\Am}(\bp)e^{-2\ti\om t} ,
\end{split}
\end{align}
where each $\hps^{\bLac_i\bLac_j}$ is such that
\begin{align}
\begin{split}
&\hps^{\bLac_i\bLac_j}(\bp) \coloneq \hbB^{\bLac_i\bLac_j} : \bp\bp^T + \hba^{\bLac_i\bLac_j} \bdot \bp + \hw^{\bLac_i\bLac_j} + \hch^{\bLac_i\bLac_j}, \quad  \text{and} \\
&\vem{\hw^{\bLac_i\bLac_j}}{\hba^{\bLac_i\bLac_j}}{\ve{\hbB}^{\bLac_i\bLac_j}} = \bM^{-1}_0 \pae{ \vem{\hc^{\bLac_i\bLac_j}}{\hn^{\bLac_i\bLac_j}}{\ve{\hbQ}^{\bLac_i\bLac_j}} - \vem{\hc_{\hch^{\bLac_i\bLac_j}}}{\hn_{\hch^{\bLac_i\bLac_j}}}{\ve{\hbQ_{\hch^{\bLac_i\bLac_j}}}} }. 
\nonumber
 \end{split}
\end{align}

Returning to the expansion for $\bka$ in (\ref{eq:o3ka_bis}), the nonlinearly-induced forcing term arising at third order, $\bd_3$, given in (\ref{eq:bd3}), involves a quadratic interaction between the first- and second-order solutions \textit{via} the operator $\bN[\bullet,\bullet]$. Through the term $\br_3$, which arises from the moments of $\chi_3$ in (\ref{eq:chi_j}), $\bd_3$ also involves a cubic nonlinearity in the first-order density correction. Overall, the components of $\bd_3$ oscillating at $\bka$ and $\bkb$ can only be produced via the interactions
\begin{align}
\begin{split}
\bka = \left\{\begin{matrix}
\bka+\bz \\
-\bka+2\bka \\
-\bkb+(\bka+\bkb) \\
\bkb+(\bka-\bkb)\\
\end{matrix}\right. , \quad \text{and} \quad \bkb = \left\{\begin{matrix}
\bkb+\bz \\
-\bkb+2\bkb \\
-\bka+(\bka+\bkb) \\
\bka-(\bka-\bkb)\\
\end{matrix}\right. ,
\end{split}
\nonumber
\end{align}
respectively. Therefore, in the second-order corrections (\ref{eq:q2full}) and (\ref{eq:ps2full}), the terms with $\bk \in \Ks$ are the only ones needed to compute $\hbd_{3,\bka}$ and $\hbd_{3,\bkb}$. By contrast, the terms $\hbq_{2,\bk \notin \Ks}$ and $\hps_{2,\bk \notin \Ks}$ have no importance. After long calculations, the forcing terms $\hbd_{3,\bka}$ and $\hbd_{3,\bkb}$ are found to be
\begin{align}
\begin{split}
\hbd_{3,\bka}(t) = & \phi\hbpa(\bLa,t) + \pae{\Am \Bp \Bm^* \dt{1}{1} + \Ap \sum_{j=1}^{4} |\bLac_j|^2 \dt{1,j}{2}} e^{\ti \om t} \\
& + \pae{\Ap \Bp^* \Bm \dt{2}{1} + \Am \sum_{j=1}^{4} |\bLac_j|^2 \dt{2,j}{2} }e^{-\ti \om t} + N.R.T.,  \label{eq:hbd3ka} 
\end{split}
\end{align}
and
\begin{align}
\begin{split}
\hbd_{3,\bkb}(t) = & \phi\hbpb(\bLa,t) + \pae{\Ap \Am^* \Bm \dt{3}{1} + \Bp \sum_{j=1}^{4} |\bLac_j|^2 \dt{3,j}{2} } e^{\ti \om t} \\
& + \pae{\Ap^* \Am \Bp \dt{4}{1} + \Bm \sum_{j=1}^{4} |\bLac_j|^2 \dt{4,j}{2} } e^{-\ti \om t} + N.R.T.,  \label{eq:hbd3kb} 
\end{split}
\end{align}
respectively. To lighten the notation, we have introduced the time-independent vectors $\hbg^{(1)}_{j}$ and $\hbg^{(2)}_{j,k}$, whose detailed expressions can be found in Appendix~\ref{app:o3t}. We have also defined the multiplicative stochastic processes $\hbpa$ and $\hbpb$, which result from the quadratic interactions between the first-order terms and the second-order ones of the form $\phi \hbq^{\phi}_{\bk \in \Ks}$, appearing in (\ref{eq:o21_bisbis}) and (\ref{eq:q2d2kb})-(\ref{eq:q2dkamkb}). The latter second-order terms are responses of the linearized system to the external noise, and thus $\hbpa$ and $\hbpb$ involve a convolution integral over the history of the external noise, i.e., they are memory-dependent. Their detailed expressions are given in Appendix~\ref{app:xi2}. The acronym ``N. R. T.'' stands for ``non-resonant'' terms, meaning harmonic terms that do not oscillate at the neutral frequencies $\om$ or $-\om$. We shall see in a moment that these ``N. R. T'' terms are unimportant, and thus their full expressions are not shown. 

We insert (\ref{eq:hbd3ka}) into the expansion (\ref{eq:o3ka_bis}), and, as previously developed, move the amplitudes (and their functions) outside of the temporal integrals by integrating by parts. For instance,  
\begin{align}
\begin{split}
&\int_{0}^{t} \bPhi(t,s) f_{\Ap}\bD \tbq^{\Ap} e^{\ti \om s}\di s \\
&\myeq f_{\Ap}\int_{0}^{t} \bPhi(t,s) \bD \tbq^{\Ap} e^{\ti \om s}\di s - \int_{0}^{t}\pae{\sum_{j=1}^{8}\frac{\di\bLac_j}{\di t}\frac{\pa f_{\Ap}}{\pa \bLac_j}} \bPhi(t,s) \bD \tbq^{\Ap} e^{\ti \om s}\di s \\
&= f_{\Ap}\int_{0}^{t} \bPhi(t,s) \bD \tbq^{\Ap} e^{\ti \om s}\di s + O(\e),
  \nonumber
 \end{split}
\end{align}
by virtue of (\ref{eq:nf}). All higher-order terms produced by integrating by parts are absorbed at $O(\e^4)$ in (\ref{eq:o3ka_bis}), and, by then collecting terms at $O(\e^3)$, we obtain 
\begin{align}
\begin{split}
\hbq_{3,\bka}(t) =& \phi \hqpa(\bLa,t) + \Ap \int_{0}^{t} \bPhi(t,s)\bD \tbq^{\Ap}  e^{\ti \om s} \di s + \Am \int_{0}^{t}\bPhi(t,s)\bD \tbq^{\Am}  e^{-\ti \om s} \di s \\
& + \Am \Bp \Bm^* \int_{0}^{t} \bPhi(t,s)\dt{1}{1}e^{\ti \om s} \di s  + \Ap \sum_{j=1}^{4} |\bLac_j|^2 \int_{0}^{t} \bPhi(t,s) \dt{1,j}{2} e^{\ti \om s} \di s\\
& + \Ap \Bp^*\Bm  \int_{0}^{t} \bPhi(t,s)\dt{2}{1}e^{-\ti \om s} \di s + \Am \sum_{j=1}^{4} |\bLac_j|^2 \int_{0}^{t} \bPhi(t,s) \dt{2,j}{2} e^{-\ti \om s} \di s \\
& -  f_{\Ap}\int_{0}^{t} \bPhi(t,s)\bD \tbq^{\Ap} e^{\ti \om s}\di s - f_{\Am} \int_{0}^{t} \bPhi(t,s) \bD \tbq^{\Am} e^{-\ti \om s} \di s + N.R.T. 
  \label{eq:o3ka_oe3}
 \end{split}
\end{align}
The term $\hqpa$ stems from the multiplicative stochastic forcing $\hbpa$, as well as the the higher-order noise corrections $\xi^{(2)}_{\Ap}$ and $\xi^{(2)}_{\Am}$ acting on the amplitudes. We show in Appendix~\ref{app:xi2} that the latter two processes are determined by canceling the components of $\hqpa$ along $\tbq^{\Ap}$ and $\tbq^{\Am}$, respectively. This choice conservatively preserves the asymptotic hierarchy at all times, since these components typically yield a diverging root mean square $\propto \sqrt{t}$, resulting from the integration of stochastic processes with low-frequency content. This conservative argument to avoid a secular growth in the root mean square is similar to that already advanced below (\ref{eq:o3ka_o2_bis}). The processes $\xi^{(2)}_{\Ap}$ and $\xi^{(2)}_{\Am}$ are multiplicative, history-dependent, higher-order noise terms acting over the amplitudes, comparable to those derived in \cite{Xu96} and \cite{Blomker07}. 

Using the dyadic decomposition of $\bPhi$, given in (\ref{eq:dyad_ka}), reveals that $\hbq_{3,\bka}(t)$ also contain terms that diverge algebraically $\propto t e^{\pm \ti \om t}$ along the neutral eigenmodes. Unlike the previous order, however, the divergence this time is purely deterministic and holds not only in the root-mean-square sense. This ensues from the fact that, in the integrands of (\ref{eq:o3ka_oe3}), the $e^{\pm \ti \om s}$ cancel out with the $e^{\mp \ti \om s}$ coming from the dyadic decomposition of the propagator. The resulting integrand is constant over time, and its integral is $\propto t$. For instance,   
\begin{align}
\begin{split}
\int_{0}^{t} \bPhi(t,s)\dt{1}{1} e^{\ti \om s} \di s =& \spap{\dt{1}{1}}\tbq^{\Ap} t e^{\ti \om t}  + \spam{\dt{1}{1}}\tbq^{\Am} e^{-\ti \om t}\int_{0}^{t}e^{2 \ti \om s} \di s \\
& +\int_{0}^{t} \bPhi(t,s) \bPo \dt{1}{1} e^{\ti \om s} \di s. 
\label{eq:hyb}
 \end{split}
\end{align}
While the last two terms correspond to bounded integrals, the first is proportional to $t e^{\ti \om t}$ and diverges with time. In particular, it should not be scaled at $O(\e^3)$ but at $O(\e)$ as early as $t \sim 1/\e^2$; this contradicts the asymptotic hierarchy. Such a diverging response of a system forced at its neutral (or ``natural") frequencies is a well-known ``resonance" effect. By decomposing each temporal integral in (\ref{eq:o3ka_oe3}) in a similar fashion as above, we then exploit the freedom afforded by the deterministic functions $f_{\Ap}$ and $f_{\Am}$ to cancel the terms diverging $\propto t$ along $\tbq^{\Ap}$ and $\tbq^{\Am}$, respectively. In doing so, the asymptotic hierarchy is preserved. This non-resonance condition results in 
\begin{align}
\begin{split}
f_{\Ap} &= \Ap + \mu \Ap \mAp^2 + \eta \Ap \mAm^2 + \nu \Ap \pae{\mBp^2 + \mBm^2} + \kap \Am \Bp \Bm^*, \\
f_{\Am} &= \Am + \mu^* \Am \mAm^2 + \eta^* \Am \mAp^2 + \nu^* \Am \pae{\mBp^2 + \mBm^2} + \kap^* \Ap \Bp^* \Bm,
\label{eq:ffka}
 \end{split}
\end{align}
where we have defined the complex-valued, scalar coefficients 
\begin{align}
\begin{split}
\mu  & \coloneq \spap{\dt{1,1}{2}} = \spam{\dt{2,2}{2}}^*, \quad \eta \coloneq \spap{\dt{1,2}{2}} = \spam{\dt{2,1}{2}}^*, \\
\nu  & \coloneq \spap{\dt{1,3}{2}} = \spap{\dt{1,4}{2}} = \spam{\dt{2,3}{2}}^*= \spam{\dt{2,4}{2}}^*, \\
\kap & \coloneq \spap{\dt{1}{1}} = \spam{\dt{2}{1}}^*.
\nonumber
 \end{split}
\end{align}
From the expansion at the wavenumber $\bkb$, we obtain accordingly 
\begin{align}
\begin{split}
f_{\Bp} &= \Bp + \mu \Bp \mBp^2 + \eta \Bp \mBm^2 + \nu \Bp \pae{\mAp^2 + \mAm^2} - \kap \Ap \Am^* \Bm, \ \text{and} \\
f_{\Bm} &= \Bm + \mu^* \Bm \mBm^2 + \eta^* \Bm \mBp^2 + \nu^* \Bm \pae{\mAp^2 + \mAm^2} - \kap^* \Ap^* \Am \Bp.
\label{eq:ffkb}
 \end{split}
\end{align}

The reduced-order system (\ref{eq:nf}) for $\Ap$, $\Am$, $\Bp$, and $\Bm$ is now fully characterized. It is further possible to make it $\e$-free by using the rescaled time $\tau = \e^2 t$, physically interpreted as a slow time scale, and using the scaling invariance of white noise, i.e., $\xi_{\bLac_j}(t) = \e \xi_{\bLac_j}(\e^2 t)$ (and similarly for the $\xi^{(2)}_{\bLac_j}(\bLa,t)$ terms given in Appendix~\ref{app:noise}). By then only keeping the leading-order deterministic and stochastic terms, we obtain
\begin{align}
\begin{split}
\left\{\begin{matrix}
\di_{\tau} \Ap &= f_{\Ap}(\bLa) + \phi \xi_{\Ap}(\tau), \\
\di_{\tau} \Am &= f_{\Am}(\bLa) + \phi \xi_{\Am}(\tau), \\
\di_{\tau} \Bp&= f_{\Bp}(\bLa) + \phi \xi_{\Bp}(\tau), \\
\di_{\tau} \Bm &= f_{\Bm}(\bLa) + \phi \xi_{\Bm}(\tau),
\end{matrix}\right.
\label{eq:ampeq_H}
 \end{split}
\end{align}
where $f_{\Ap}$ and $f_{\Am}$ are given in (\ref{eq:ffka}), and $f_{\Bp}$ and $f_{\Bm}$ in (\ref{eq:ffkb}). The white noises processes $\xi_{\Ap}$ and $\xi_{\Am}$ derive from the spatio-temporal forcing $\bff(\bx,t)$ applied to the full system, according to (\ref{eq:xika}), and $\xi_{\Bp}$ and $\xi_{\Bm}$ according to (\ref{eq:xikb}). Note that $\xi_{\Ap}$ and $\xi_{\Am}$ are generally correlated, and similarly for $\xi_{\Bp}$ and $\xi_{\Bm}$; however, the former pair is uncorrelated with the latter pair because the noise is white in space. 

The higher-order noise terms $\xi^{(2)}_{\bLac}$ only appear at $O(\e)$ in (\ref{eq:ampeq_H}), and thus are neglected. This may appear inconsistent since they are multiplied by $\e^2$ in (\ref{eq:nf}) in the same way that the terms $f_{\bLac}$ are. In addition, the terms $\xi^{(2)}_{\bLac}$ and $f_{\bLac}$ are all determined at the same $O(\e^3)$. One way to understand this is to notice that the deterministic functions $f_{\bLac}$ cancel secular terms growing $\propto t$, whereas the stochastic processes $\xi^{(2)}_{\bLac}$ cancel terms growing $\propto \sqrt{t}$. Therefore, the former terms prevent secular terms from emerging at $O(\e)$ after a time $\tau\sim 1$ (and thus must appear in (\ref{eq:ampeq_H})), while the latter terms prevent secular terms from emerging at $O(\e)$ after a much longer time, $\tau \sim 1/\e^2$. That is why $\xi^{(2)}_{\bLac}$ do not appear at the leading-order in (\ref{eq:ampeq_H}), as it is written over $\tau$. 

In the codimension-2 pitchfork bifurcation case, the frequency $\om = 0$, such that only one eigenmode per wavenumber bifurcates. Therefore, the system (\ref{eq:ampeq_H}) reduces to
\begin{align}
\begin{split}
\left\{\begin{matrix}
\di_{\tau} A &=  A + \mu A\mA^2 + \nu A \mB^2 + \phi \xi_{A}(\tau), \\
\di_{\tau} B &=  B + \mu B\mB^2 + \nu B \mA^2 + \phi \xi_{B}(\tau),
\end{matrix}\right.
\label{eq:ampeq_Pi}
 \end{split}
\end{align}
where we have renamed $\Ap$ as $A$ and set $\Am$ to zero. The reverse could as easily have been done (rename $\Am$ as $A$ and set $\Ap$ to zero) since the coefficients become real-valued as soon as $\om = 0$. We have also renamed $\Bp$ in $B$ and set $\Bm$ to zero. The definitions of the coefficients and white noise processes remain the same as for the codimension-4 Hopf bifurcation, and $\om$ should simply be set to $0$ in (\ref{eq:xika}) and (\ref{eq:xikb}). This gives, in particular 
\begin{align}
\begin{split}
\xi_{A}(t) &\coloneq \spaa{\hbf_{\bka}(t)}, \quad \xi_{B}(t) \coloneq \spbb{\hbf_{\bkb}(t)}.
  \label{eq:xis_Pi}
 \end{split}
\end{align}
Note that the deterministic version of (\ref{eq:ampeq_Pi}) was already derived in \cite{Ohm22}.

Overall, the stochastic terms $\xi_{\bLac_j}$, $\xi^{(2)}_{\bLac_j}$,... in (\ref{eq:nf}) are chosen so that each stochastic higher-order correction in the state vector is fully contained within the strictly stable eigen-subspace. This is a conservative way to avoid secular stochastic terms from emerging in the expansion. On the other hand, the deterministic terms $f_{\bLac_j}$, $f^{(2)}_{\bLac_j}$,... prevent deterministic secular growth while still allowing for the deterministic higher-order corrections to have (bounded) component in the neutral eigen-subspace (e.g., the second term in the right-hand-side in (\ref{eq:hyb})). While the former approach to deriving the stochastic terms adopts the center-manifold perspective, the latter approach to deriving the deterministic terms adopts the normal-form or multiple-scale perspectives.  

\section{Fully nonlinear numerical methods}
\label{sec:fnlm}

The predictions from the low-dimensional systems of amplitude equations, derived above, are compared with the results obtained by directly simulating the full system (\ref{eq:ec_st}). We say a word on the direct numerical simulations (DNS) of (\ref{eq:ec_st}) in Sec.~\ref{sec:dns} and briefly recall the working principle of the rare event Adaptive Multilevel Splitting (AMS) algorithm in Sec.~\ref{sec:AMS}. 

\subsection{Direct numerical simulations}
\label{sec:dns}

We denote $\bqm \coloneq(c,\bn,\vect{\bQ})^T$ the reduced state vector, free of the velocity and pressure components. System (\ref{eq:ec_st}) can be written in compact form as
\begin{align}
\begin{split}
\frac{\pa \bqm}{\pa t} = D_T \Delta \bqm + \bgg(\bqm,\bu) + F \bff, 
\label{eq:dtbqm}
 \end{split}
\end{align}
where $\bgg$ includes the nonlinear terms in (\ref{eq:ec_st}). We evolve (\ref{eq:dtbqm}) by proceeding as follows
\begin{enumerate}
    \item From known $\bqm^{(n)}$ at the $n$th time step, the corresponding velocity $\bu^{(n)}$ and pressure $\pr^{(n)}$ can be found by inverting the linear Stokes equations, which is done in the Fourier space.
    \item From $\bqm^{(n)}$, $\bu^{(n)}$, and $\pr^{(n)}$ (and the same fields at $n-1$), $\bqm^{(n+1)}$ is computed by using the stochastic generalization of the SBDF2 time-stepper (It\^{o} convention). It is implicit in the Laplacian term in (\ref{eq:dtbqm}), and explicit in $\bgg$ and the stochastic forcing terms (although the latter two are discretized differently), and is given by
    \begin{align}
    \begin{split}
    \pae{\frac{3}{2\di t}\bI-D_T \Delta} \bqm^{(n+1)} = &\frac{4\bqm^{(n)}-\bqm^{(n-1)}}{2\di t} +  2\bgg^{(n)}-\bgg^{(n-1)} + F \frac{\bW^{(n+1)}}{\sqrt{\di t}\sqrt{(\di x)^2}},
    \label{eq:sbdf2}
    \end{split}
    \end{align}
    where $\bgg^{(n)} \coloneq \bgg(\bqm^{(n)},\bu^{(n)})$ and $\di t$ is the time increment. We use $N$ spatial discretisation points, uniformly spaced by $\di x$ along each spatial dimension. The exponent of two in $\sqrt{(\di x)^2}$ arises because the noise is white in two spatial dimensions. Each component of the vector $\bW=\pae{W_c,\bW_{n},\vect{\bW_D}}^T$ is a Gaussian-distributed random variable with zero mean and finite variance, which can be deduced from Table~\ref {tab:cov_f}. More precisely, at each time step, the components of $\bW$ are drawn randomly according to
    \begin{align}
    \begin{split}
    & W_c  \sim \mathcal{N}(0,2\upi), \quad  W_{n,x} \sim \mathcal{N}(0,\upi), \quad  W_{n,y} \sim \mathcal{N}(0,\upi) \\
    & W_{D,{xx}} =  \frac{W_c}{2} +  Y, \ \text{with} \ Y \sim \mathcal{N}\pae{0,\frac{\upi}{4}}, \quad W_{D,{xy}} \sim \mathcal{N}\pae{0,\frac{\upi}{4}},
    \label{eq:Wp}
    \end{split}
    \end{align}
    as well as $W_{D,{yx}} = W_{D,{xy}}$ and $W_{D,{yy}} = W_c - W_{D,xx}$. The random variables in (\ref{eq:Wp}) are drawn independently, but, importantly, $W_{D,{xx}}$ depends on $W_c$. Any spatial average is removed from each field at each time step.

    System (\ref{eq:sbdf2}) is also solved in Fourier space. Since it requires the two previous time steps, it is replaced by the semi-implicit Euler scheme for $n=1$. We refer to \cite{Delong13} for different examples of integrators for stochastic PDEs. 
    \item Iterate to $1.$
\end{enumerate}

After computing a fully nonlinear trajectory, $\bqdns(\bx,t)$, the amplitudes $\Ap$, $\Am$, $\Bp$, and $\Bm$, or $A$ and $B$, are extracted for comparison with the weakly nonlinear predictions. The first step is to extract the Fourier component oscillating at wavenumber $\bk$, i.e., $\hbqdns_{\bk}(t)$, with $\bk=\bka$ and $\bk=\bkb$.
Each $\hbqdns_{\bk}$ is further decomposed into the eigenbasis at the corresponding wavenumber, $\set{\tbq_j}_{j\geq 1}$, obtained by solving (\ref{eq:lin1}). This gives $\hbqdns_{\bk}(t) = \sum_j a_j(t) \tbq_j$, with $a_j(t)$ the amplitude along the eigenmode $\tbq_j$. Eventually, by using the bi-orthogonality property (\ref{eq:biorth}) between the direct and adjoint eigenbasis, it follows that the amplitudes along the bifurcating eigenmodes are
\begin{align}
\begin{split}
&\badns(t) = \ssd{\tbq^{\bLac,\da}}{\hbqdns_{\bk}(t)}  \quad  \text{where} \quad  \begin{cases}
\bk = \bka & \text{if}\; \bLac \in \set{A,\Ap,\Am} \\
\bk = \bkb & \text{if}\; \bLac \in \set{B,\Bp,\Bm}. \\
\end{cases}
\label{eq:Aex}
\end{split}
\end{align}

\subsection{The Adaptive Multilevel Splitting (AMS) algorithm}
\label{sec:AMS}

We now briefly recall the working principle of the Adaptive Multilevel Splitting algorithm, which we employ to determine rare events statistics. We closely follow the procedure originally developed in \cite{Cerou07}. We also refer to \cite{Gome22}, Section~2(b), for a complete description of the algorithm.

\subsubsection{Mean transition time}
\label{sec:AMS_tt}

Two slightly different versions of the algorithm are implemented. The first, described in this subsection, aims to estimate the ``mean transition time". It is the average time for a trajectory, initiated within a given basin of attraction of the system, to reach a distinct basin. This version of the AMS algorithm was applied to the Ginzburg-Landau partial differential equation in \cite{Rolland16}, and shown to accurately compute, as compared to analytical results, the mean transition time between two attractors. It was also successful in computing the transition path in the dynamics of multistable turbulent jets \citep{Bouchet19} or turbulent plane shear flows \citep{Gome22, Rolland24}. 

To start, let us define $\dA$ and $\dB$ as two non-overlapping regions in the phase space of the system. Each region is contained within a distinct (and deterministic) basin of attraction and corresponds to a particular stable state of interest. The goal of the AMS algorithm is to produce a large number of ``reactive'' trajectories, initiated within $\dA$ and ending within $\dB$. Thereby, each trajectory is associated with a noise-induced transition. 

For this, the first and perhaps the most sensitive step is to construct a scalar real-valued ``cost function'' $\varphi[\bq](t)$ that, for each time $t$, quantifies how close $\bq(\bx,t)$ is to $\dB$. Specifically, $\varphi$ is taken so that there exist two real scalars $h_{\dA}$ and $h_{\dB}$, with $h_{\dA} < h_{\dB}$, such that $\varphi[\bq] < h_{\dA}$ implies $\bq \in \dA$ and $\varphi[\bq] > h_{\dB}$ implies $\bq \in \dB$. 
Note that the isosurface $\varphi=h_{\dA}$ need not coincide with the boundary of $\dA$, nor does $\varphi=h_{\dB}$ with the boundary of $\dB$. The regions $\dA$ and $\dB$ are characterized prior to the cost function. The latter is chosen in consequence so that its sublevel set $\set{\varphi[\bq] < h_{\dA}}$ lies within $\dA$, and its superlevel set $\set{\varphi[\bq] > h_{\dB}}$ lies within $\dB$. Larger values of $\varphi$ are intended to indicate a greater progress of a trajectory toward $\dB$.
We also require $\varphi$ to have a non-zero gradient, at least in the region of interest. The algorithm also requires a scalar $h_{\dS}$, associated with an isosurface $\dS$, such that $h_{\dA} < h_{\dS} < h_{\dB}$ but $h_{\dS}$ is chosen close to $h_{\dA}$ (i.e., $\dS$ is closely enclosing $\dA$). We refer to \cite{Gome22}, figure~$4$ therein for an illustration (we purposefully use the same notation). For high-dimensional systems, the cost function $\varphi$ is usually constructed based on heuristic considerations, since the theoretical method for constructing involves computing the committor functions between the attractors, which quickly becomes impractical.  

The AMS algorithm then proceeds as follows:

\begin{enumerate}
    \item The initialization step requires computing $N_t$ stochastic trajectories, i.e., $\set{\bq_j(\bx,t)}_{1 \leq j \leq N_t}$. Each trajectory is initiated randomly within $\dA$ and is computed until $\varphi[\bq_j](t) > h_{\dB}$, or, much more likely, until $\varphi[\bq_j](t) < h_{\dA}$ after it has accomplished $\varphi[\bq_j](t) > h_{\dS}$ (i.e., until it re-enters $\dA$ after going as far as $h_{\dS}$). Therefore, the initial $N_t$ trajectories all have generally different final times. The latter are generally not too long since $h_{\dS}$ is close to $h_{\dA}$. Note that each trajectory is computed using a different noise realization.

    The algorithm requires storing all trajectories, which can lead to excessive CPU and memory usage. To mitigate this effect, we perform a two-dimensional spatial Fourier transform of each trajectory and store only the first $12$ Fourier modes in each direction. That is, we only save a low-pass spatially filtered version of the trajectories. We have checked that the final results rapidly converge with the number of saved Fourier modes, since, as seen in Sec.~\ref{sec:linf}, modes with increasing $\nn{\bk}>1$ are increasingly damped by the system.   
    \item At the $n$th iteration, the maximum value reached by the cost function along the $j$th trajectory, say $\varphi^{(n)}_j$, is determined. These maximum values are then used to sort the trajectories, from the lowest ones (i.e., the trajectory that has made the smallest excursion towards $h_{\dB}$) to the largest (i.e., the trajectory that has made the largest excursion). Namely, the trajectories at iteration $n$ are re-indexed such that
    \begin{align}
    \begin{split}
    \varphi^{(n)}_1 \leq \varphi^{(n)}_2 \leq ... \leq \varphi^{(n)}_K \leq ... \leq \varphi^{(n)}_{N_t}.
    \nonumber
    \end{split}
    \end{align}
    The idea is then to discard all trajectories with indices from $j=1$ to $j=K$ (included), corresponding to the $K$ trajectories whose maximal value of the cost function is the lowest. We note that these maxima are possibly equal, and thus in practice $K=K^{(n)}$ depends on the iteration index. For instance, if we choose \textit{a priori} $K=1$ but $\varphi^{(n)}_1 =\varphi^{(n)}_2 < \varphi^{(n)}_3 \leq ... $, then these are the first two trajectories that must be discarded, and $K^{(n)}=2$. All remaining trajectories, with index $j \in \set{K^{(n)}+1,...,N_t}$ are retained.

    A new trajectory replaces each of the discarded ones. A new trajectory is constructed to be equal to one of the retained trajectories, with index $l$ chosen at random in the retained set $\set{K^{(n)}+1,...,N_t}$, from the initial time $t=0$ and until the time $t^{\text{clone}}$ where $\varphi[\bq_l](t^{\text{clone}}) =  \varphi^{(n)}_K$. The remainder of the new trajectory from $\bq_l(\bx,t^{\text{clone}})$, which serves as an ``initial'' condition for the time stepper, is then computed by using a new noise realization, until it has reached either $\dA$ or $\dB$.

    Replacing the discarded $K^{(n)}$ trajectories accordingly, yields a new set of trajectories such that the maximum cost function value for each index $j$ is greater than or equal to $\varphi^{(n)}_K$, and thus greater than or equal to that of the previous set. In that sense, the new set of trajectories has made more progress towards $\dB$, at least as measured by $\varphi$, than the previous set. The value of $n$ is incremented to $n+1$, and this step is repeated as many times as necessary until the stopping condition of the algorithm is met. 
    \item The iterations stop as soon as all trajectories reach $\dB$, i.e., as soon as $\varphi^{(n)}_1 > h_{\dB}$. We have thus computed a large number $N_t$ of reactive trajectories from $\dA$ to $\dB$, which can be used to generate converged statistics on the $\dA \rightarrow \dB$ noise-induced transition. In particular, the probability that a trajectory initially on the surface $\dS$ ends up in $\dB$ is estimated by 
    \begin{align}
    \begin{split}
    \hat{p} \coloneq \prod_{n=1}^{M}\pae{1-\frac{K^{(n)}}{N_t}},
    \label{eq:pSB}
    \end{split}
    \end{align}
    where $M$ is the total number of iterations needed for the algorithm to reach its stopping condition. Accordingly, the probability that a trajectory goes from $\dS$ to $\dA$ is $1-\hat{p}$, and, by construction of the initial set, the probability that it goes from $\dA$ to $\dS$ is $1$. From this initialization step, it is possible to estimate $\TAS$, the average duration of a trajectory conditioned to start within $\dA$ and to end on the surface $\dS$. Similarly, an estimate for $\TSA$, the average duration of a trajectory conditioned to start on $\dS$ and to end by re-entering $\dA$ can also be computed from the initial set of trajectories. Because $\dS$ is chosen close to $\dA$, neither $\TAS$ nor $\TSA$ are expected to be large. From the converged set of $N_t$ reactive trajectories, we obtain an estimate for $\TSB$, the average duration of a trajectory conditioned to start on $\dS$ and to end by entering $\dB$. Although such a trajectory has a very low probability $\hat{p} \ll 1$ of occurring, the conditioned time $\TSB$ need not be large either. From all these quantities, and by describing the $\dA$ to $\dB$ transition dynamics as a Markov chain (see figure~$5$ in \cite{Gome22}), follows an estimate for the mean transition time $T$ from $\dA$ to $\dB$, as
    \begin{align}
    \begin{split}
    T \coloneq \pae{\TAS +\TSA}\frac{1-\hat{p}}{\hat{p}}+\pae{\TAS +\TSB} \nonumber
    \end{split}
    \end{align}
    This time is not conditioned on the trajectory starting in $\dA$ and ending in $\dB$. It really is an estimate of the average time required for a trajectory to reach $\dB$ while started within $\dA$, and as a consequence of rare external fluctuations. Because $\hat{p}$ is very small in the weak forcing limit considered in this article, $T$ is expected to be extremely large. 
    
\end{enumerate}

\subsubsection{Mean return time}
\label{sec:AMS_rt}

The second version of the AMS algorithm we employ does not require multi-stability. It is used to estimate the average waiting time until a scalar observable of the trajectory exceeds a given threshold in a statistically steady regime. Threshold values of interest are typically much larger than the standard deviation. This average waiting time is called the ``mean return time'' in the rest of the article. This version of the algorithm was presented in \cite{Lestang18} and used in \cite{Lestang20} to compute the mean return time for the drag force acting on a square, under the action of an impinging turbulent flow, to become extremely large.

The working principle of this version of the AMS algorithm is similar to that used to estimate the mean transition time. However, because it does not seek to characterize the transition between different basins of attraction, it is here unnecessary to define $\dA$, $\dS$ and $\dB$, and related thresholds $h_{\dA}$, $h_{\dS}$ and $h_{\dB}$. 

For a stochastic trajectory $\bq(\bx,t)$ in a statistically steady regime, the algorithm estimates the mean return time of the event $\varphi[\bq](t) \geq 1$. The cost function $\varphi$ here represents the observable of interest, and we scale it so that the exceedance threshold is always $1$. The algorithm inherently assumes that the rare event of interest follows a Poisson process. This holds if $1$ is much larger than the standard deviation of the observable, making the exceedance events independent. Consequently, the mean return times can be obtained from the probability $\tilde{p}$ of observing the event over a period of time $t_{\text{max}}$. The time $t_{\text{max}}$ must be much larger than the typical correlation time ($\tau \sim 1$, i.e., $t \sim 1/\e^2$) of the system, but, crucially, can be shorter than the mean return time. 

The probability $\tilde{p}$ is obtained by using the same killing-and-cloning iterative procedure over the set of $N_t$ trajectories, as presented above. One difference is that, at each iteration, the $N_t$ trajectories all have the same maximum duration $t_{\text{max}}$. The trajectories are then prioritized based on the maximum value reached by $\varphi[\bq](t)$ between $t=0$ and $t=t_{\text{max}}$. Within this interval, the computation of a trajectory is stopped as soon as $\varphi[\bq](t) \geq 1$, and the exceedance threshold $1$ thus plays the role of $h_{\dB}$ in the previous version of the algorithm. Iterations stop as soon as all $N_t$ trajectories have reached the exceedance threshold. If $M$ again denotes the final number of iterations, the probability $\tilde{p}$ is the same expression as in (\ref{eq:pSB}). Because of the Poisson process approximation, the mean return time, say $r$, ensues from $\tilde{p}$ as $r = - t_{\text{max}}/\ln(1-\tilde{p})$.

For the two implemented versions of the algorithm, the final estimate of the mean transition time $T$, or return time $r$, is itself a random variable. Thereby, $\Nre>1$ realizations of the algorithm are performed, and the resulting estimates are ensemble-averaged. In simple cases, the associated standard deviation can be shown to scale as $1/\sqrt{K N_t}$. Thus, a larger $N_t$ and/or a smaller $K$ result in a more reliable average estimate, although coming at a higher computational cost. The rate of convergence of the average estimate was also shown in \cite{Brehier15, Brehier16} to depend on the choice of the function $\varphi$ (see figure~$4$ in \cite{Brehier16}). 

\section{Results in the pitchfork bifurcation region}
\label{sec:pitch}

In this section, we present our results for parameters corresponding to the codimension-2 pitchfork bifurcation. Both the weakly (low-dimensional) and fully (high-dimensional) nonlinear approaches are systematically compared. First, in Sec.~\ref{sec:pitch_det}, we report our results in the purely deterministic regime, followed by comparison of some steady statistics in Sec.~\ref{sec:pitch_sts}, and eventually dynamical statistics in Sec.~\ref{sec:pitch_dys}. 

While we also study rare events in this section, readers specifically interested in noise-induced transitions between two distinct collective states are invited to jump directly to Sec.~\ref{sec:hopf}.

\subsection{Deterministic results ($\phi=0$)}
\label{sec:pitch_det}

The deterministic regime is recovered by setting $\phi=0$. The system (\ref{eq:ampeq_Pi}) of amplitude equations then predicts the trajectories to converge to the unique attractor in the magnitude phase space, given by
\begin{align}
\begin{split}
(\mA,\mB) = (\He, \He), \quad \text{where} \quad \He \coloneq \frac{1}{\sqrt{-(\mu+\nu)}},
\label{eq:He_Pi}
 \end{split}
\end{align}
and arbitrary phases for both $A$ and $B$ (independently). Indeed, the unforced version of (\ref{eq:ampeq_Pi}) remains invariant under a phase-shift of $A$ and $B$. For the stable equilibria in (\ref{eq:He_Pi}) to exist, implying that the bifurcation is ``supercritical", the sum $\mu+\nu$ must be negative. In the opposite case where the sum $\mu+\nu$ is strictly positive, $\He$ is not defined, and the weakly nonlinear expansion needs to be pursued at higher orders. The bifurcation is then said to be ``subcritical".  

\begin{figure}
            \begin{subfigure}{0.495\textwidth}
            \centering
            \scalebox{0.44}{\includegraphics[trim=0cm 0cm 0cm 0cm]{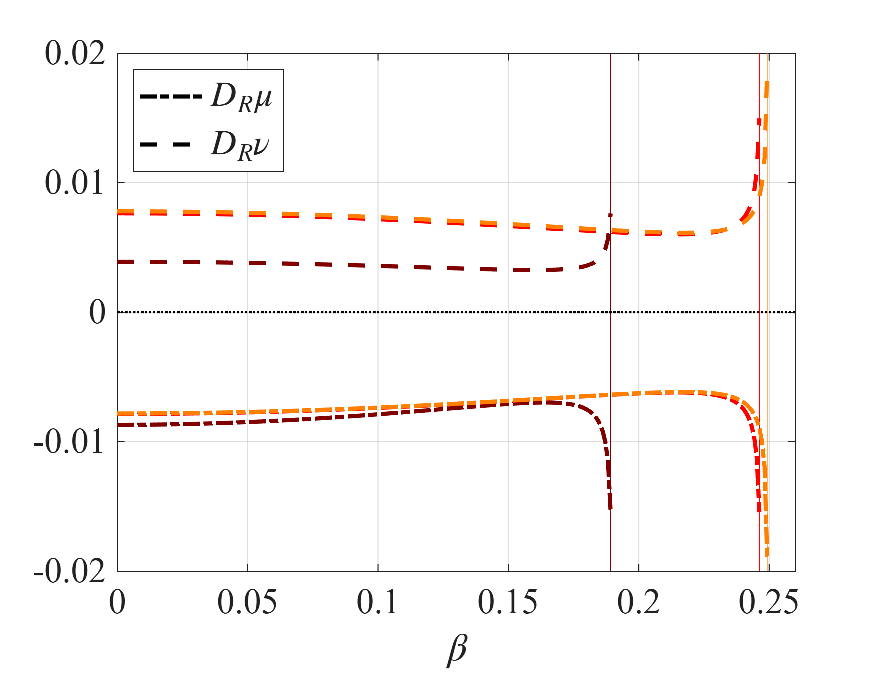}}
             \subcaption{}
             \label{fig:copitch_a}
            \end{subfigure}
            \begin{subfigure}{0.495\textwidth}
            \centering
            \scalebox{0.44}{\includegraphics[trim=0cm 0cm 0cm 0cm]{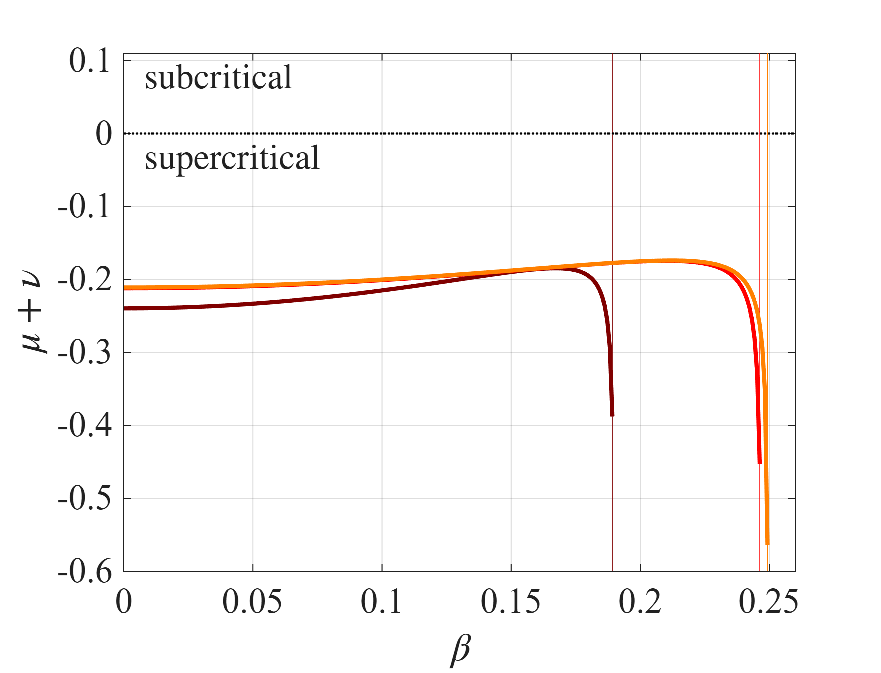}}
             \subcaption{}
             \label{fig:copitch_b}
            \end{subfigure}
     \caption{Weakly nonlinear coefficients $\mu$ (dash-dotted lines) and $\nu$ (dashed lines) as a function of $\bet$, the swimming speed of the particles. Three different values of $D_R \in \set{10^{-4},10^{-3},0.02}$ are considered, each corresponding to a different color (darker shades larger $D_R$).
     Only the range of $\bet$ for which the bifurcation is of pitchfork type (i.e., for which $\om=0$ in figure~\ref{fig:lindr_b}) is considered, and delimited by a thin vertical line (increasing $\bet$ above which makes the bifurcation of Hopf-type). 
     In (a), the coefficients are plotted individually and rescaled by the corresponding $D_R$, whereas their sum (without rescaling) is shown in (b).} \label{fig:copitch}
\end{figure}

Throughout the article, the coefficients are evaluated numerically from the above calculations and are not fitted to DNS data. Recall that, in the pitchfork bifurcation scenario, the coefficients $\mu$ and $\nu$ are real-valued. Their values depend on the chosen normalization of the eigenmodes, and the values reported below are relative to the choice made in (\ref{eq:normal}). 
The weakly nonlinear coefficients are shown individually in figure~\ref{fig:copitch_a}, for three different $D_R$ and rescaled by the corresponding value of $D_R$. Their (non-rescaled) sums are shown in figure~\ref{fig:copitch_b}. 
The curves are plotted as a function of the swimming speed, from $\bet=0$ up to the threshold value at which the bifurcation becomes of the Hopf type. 
For the three $D_R$ considered, the coefficient $\mu$ is always negative and dominates over $\nu$, which is always positive. Consequently, $\He$ always exists, and thus the pitchfork bifurcation is always supercritical. Decreasing $D_R$ even more does not change this fact, as the coefficients appear to have already converged between $D_R = 10^{-3}$ and $10^{-4}$. 
Overall, changing $D_R$ has little effect on the sum $\nu+\mu$, and thus on $\He$, but mostly modifies the threshold at which the bifurcation changes from pitchfork to Hopf (i.e., the vertical lines in figure~\ref{fig:copitch}). Interestingly, as $\bet$ closely approaches this threshold, the coefficients undergo abrupt variations. This manifests as $\He$ becoming suddenly smaller as $\bet$ is increased.  

For the remainder of this section, we set $D_R=0.02$, and consider only the two values of the swimming speed $\bet=0$ and $\bet=0.15$. The corresponding coefficients are given in Table~\ref{tab:coeffpitch}. 
\begin{table}
  \begin{center}
\def~{\hphantom{0}}
  \begin{tabular}{l  C{2.1cm} C{2.4cm} C{1.9cm} C{1.9cm} C{2cm}}
      $\bet$   &   $\mu$  & $\nu$  &$\eta$ & $\kap$ & $\al$\\
      \hline
       $0$ (Pitchfork)    & $-0.4346$  & $0.1949$   & $\varnothing$ & $\varnothing$ & $\sqrt{2/\upi}\approx 0.7979$ \\[3pt]
       $0.15$ (Pitchfork) & $-0.3539$  & $0.1642$   & $\varnothing$ & $\varnothing$ & $1.421$\\[3pt]
       $0.5$ (Hopf)  & $-0.1366 + \newline  \ti \times(-0.04364)$  & $0.01190 + \newline \ti \times(-0.0009648)$ & $-0.001864 + \newline \ti \times 0.04251$ & $-0.03903 +\newline  \ti \times0.003319$ & $\varnothing$
  \end{tabular}
  \caption{For $D_R=0.02$, numerical evaluations of the weakly nonlinear coefficients, relative to the choice of normalization in (\ref{eq:nf}). The prefactor $\al$, only defined for the pitchfork bifurcation, is such that $\al\phi$ is the amplitude of the noise acting on the magnitudes in (\ref{eq:bHsys}). The coefficients $\eta$ and $\kap$ are defined only for the Hopf bifurcation. The phase of $\kap$ (only) remains arbitrary.}
  \label{tab:coeffpitch}
  \end{center}
\end{table}

\begin{figure}
            \begin{subfigure}{0.495\textwidth}
            \centering
            \scalebox{0.44}{\includegraphics[trim=0cm 0cm 0cm 0cm]{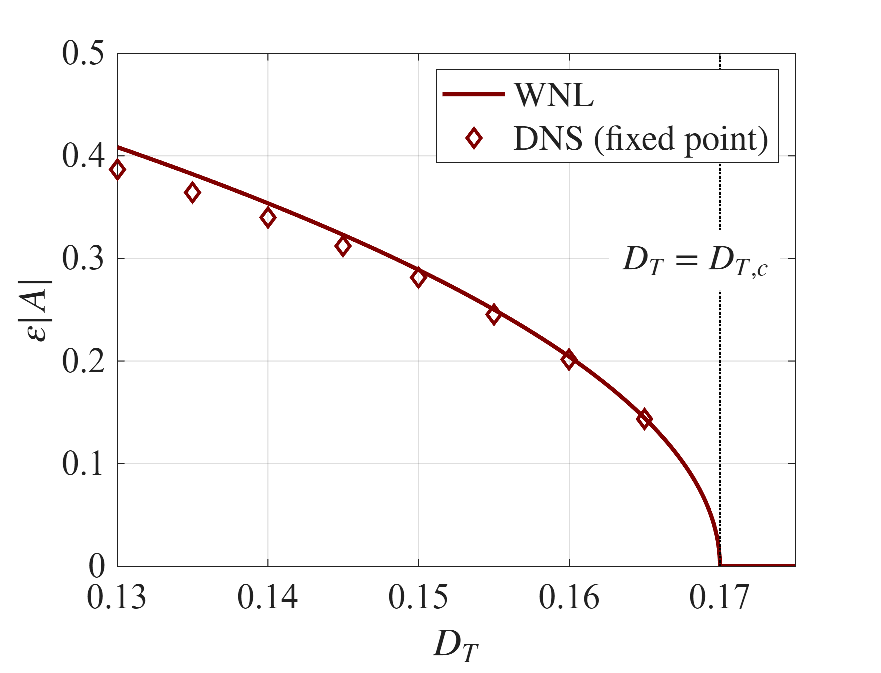}}
             \subcaption{
             $\bet=0$}
              \label{fig:bdp_a}
            \end{subfigure}
            \begin{subfigure}{0.495\textwidth}
            \centering
            \scalebox{0.44}{\includegraphics[trim=0cm 0cm 0cm 0cm]{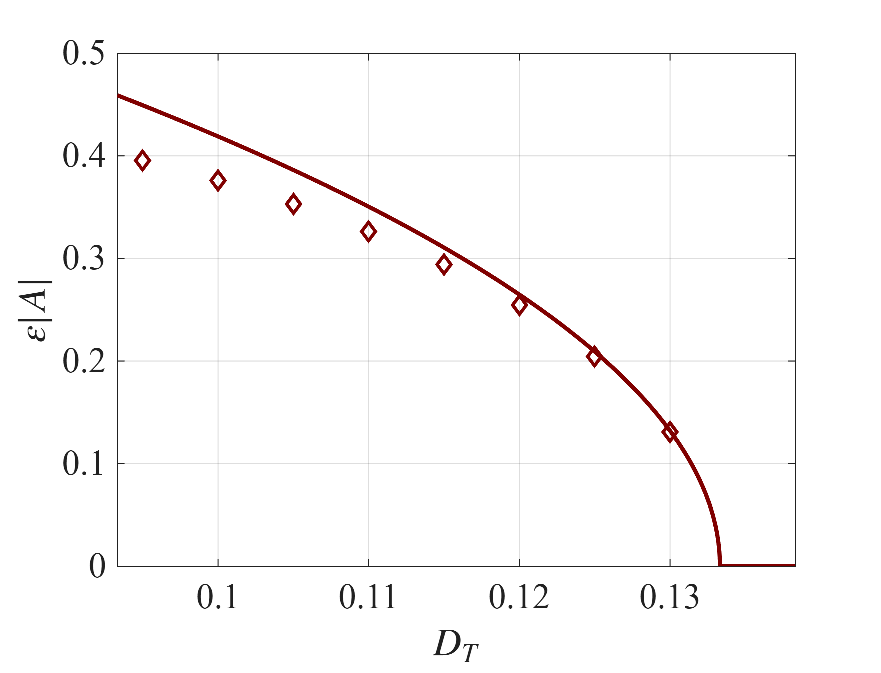}}
             \subcaption{
             $\bet=0.15$}
             \label{fig:bdp_b}
            \end{subfigure}
     \caption{Bifurcation diagrams in $D_T$ of the system (\ref{eq:ec_st}) with $F=0$ (deterministic regime) and $D_R=0.02$. 
     Each frame corresponds to a different 
     $\bet$ in the pitchfork bifurcation region; the corresponding value of $D_{T,c}$ can be deduced from figure~\ref{fig:lindr_a}. 
     %
     %
     Two different approaches are compared. The first approach is fully nonlinear and involves running a DNS of (\ref{eq:ec_st}) (with $F=0$) until an equilibrium state is reached. A diamond marker indicates a fixed point.
     The second approach is the weakly nonlinear expansion (WNL), predicting a fixed point with amplitude $\e H_e = \sqrt{D_{T,c}-D_T}H_e$ (continuous line), $H_e$ being given in (\ref{eq:He_Pi}). } \label{fig:bifdiagpitch}
\end{figure}

All results discussed so far only referred to weakly nonlinear quantities. We show in figure~\ref{fig:bifdiagpitch} the bifurcation diagram of the system (\ref{eq:ec_st}) (with $F=0$), as the uniform base solution is destabilized by decreasing $D_T$ below its critical value. The weakly and fully nonlinear approaches are compared. The measure of comparison, or ``order parameter", is naturally chosen as the equilibrium value of $\e \mA$, or, equivalently, of $\e \mB$.

In the weakly nonlinear approach, at equilibrium $\mA =\mB=\He$. Since $\He$ is a constant, $\e \mA = \sqrt{D_{T,c}-D_T}\He$ evolves as a square root when plotted against $D_T$. In the fully nonlinear method, DNS of the system (\ref{eq:ec_st}) are run from different initial conditions and until reaching a steady state, and the corresponding amplitudes $A$ and $B$ are then extracted following (\ref{eq:Aex}). 

In figure~\ref{fig:bifdiagpitch}, we found a unique stable fixed point in the DNS for all the $D_T$ considered (as predicted by the weakly nonlinear approach). The comparison between the two approaches, for both the immotile case at $\bet = 0$ in figure~\ref{fig:bdp_a}, and the motile case at $\bet=0.15$ in figure~\ref{fig:bdp_b}, yields convincing results. This validates our calculations of the weakly nonlinear coefficients. As expected, the agreement progressively degrades as $D_T$ decreases, since it corresponds to larger $\e$ and thus the neglected terms in the weakly nonlinear expansion become important. 

\begin{figure}
            \begin{subfigure}{0.95\textwidth}
            \centering
            \scalebox{0.4}{\includegraphics[trim={0cm 0cm 0cm 0cm},clip]{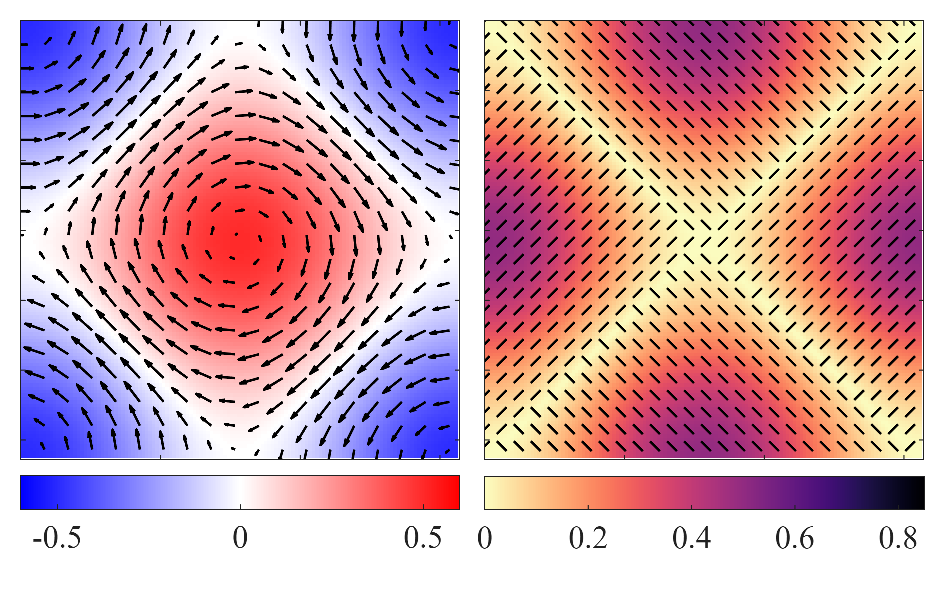}}
             \subcaption{$\bet=0$}
             \label{fig:stimmo_a}
            \end{subfigure}
            \hfill
            \begin{subfigure}{0.95\textwidth}
            \centering
            \scalebox{0.4}{\includegraphics[trim={0cm 0cm 0cm 0cm},clip]{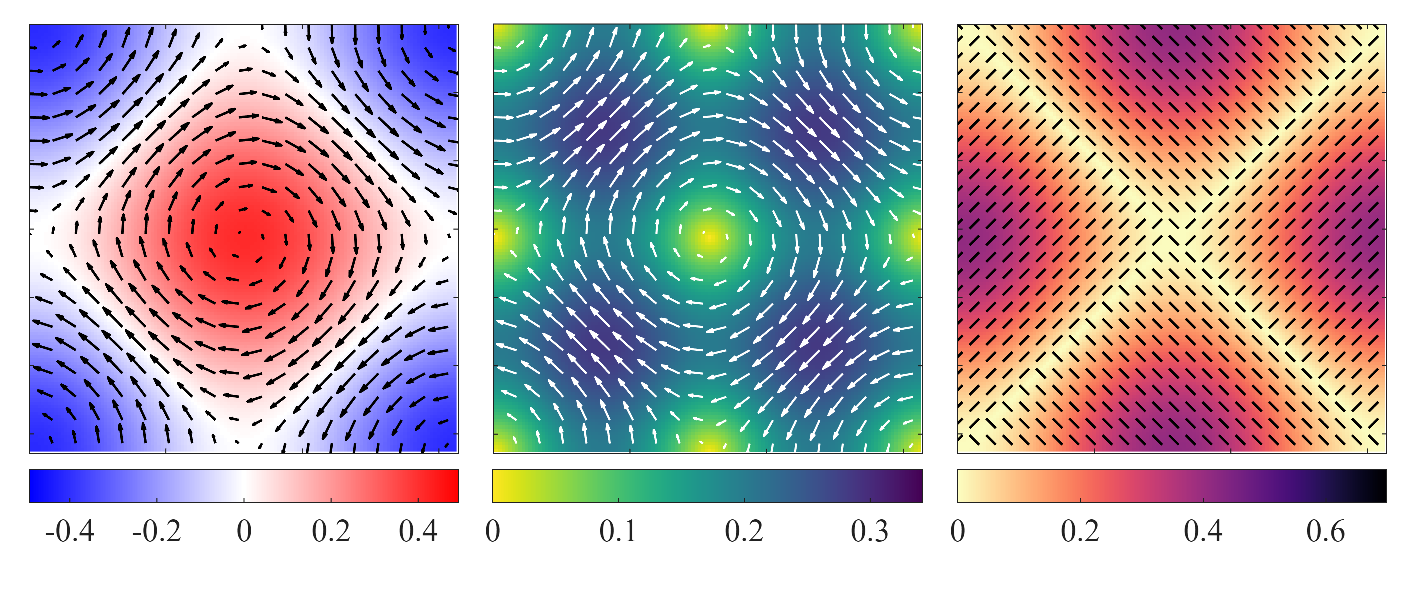}}
             \subcaption{$\bet=0.15$}
             \label{fig:stimmo_b}
            \end{subfigure}
     \caption{For $D_R=0.02$, values of $\bet$ corresponding to the pitchfork bifurcation region, and over the doubly periodic square domain $(x,y) \in [0,2\upi] \times [0,2\upi]$. 
     Leading-order, weakly nonlinear deterministic equilibrium solution of (\ref{eq:ec_st}). Left frame: vorticity (colormap) with arrows for the velocity. Middle frame (only for $\bet\neq 0$): polarization vector field with the colormap for the norm. Right frame: nematic order parameter, given by the leading eigenvalue of $c^{-1}_1 \bQ_1$ (colormap), and representing the strength of particle local alignment; the associated eigenvector is the principal direction of particle alignment (short straight lines). 
     Any uniform rescaling of the fields is also a solution, corresponding to a different value of $D_T \leq D_{T,c}$.} \label{fig:stimmo}
\end{figure}

From the knowledge of $\mA$ and $\mB$, the leading-order weakly nonlinear equilibrium solution can be reconstructed from (\ref{eq:q1c2P}).  The results are shown in figure~\ref{fig:stimmo}.
For the immotile case (top row), only the velocity and the nematic order parameter (i.e., the strength of particle alignment) are nonzero. The superposition of the two bifurcating eigenmodes, one oscillating purely along $x$ and the other purely along $y$, results in a macroscopic velocity vortex in the left frame. On the frame on the right, the nematic order parameter assumes its maximum values, i.e., the particles exhibit a clear preferred orientation, wherever the vorticity is zero. For the motile case at $\bet = 0.15$, the velocity and nematic order parameter are qualitatively similar to those for the immotile case. The motility makes the polarization vector nonzero, and it coincides with the velocity field.


\subsection{Stochastic results ($\phi \neq 0$): steady statistics}
\label{sec:pitch_sts}

We now enable stochastic forcing, i.e., $\phi \neq 0$, and compare the trajectories and probability distributions in a statistically steady regime. 

Since the stochastic forcing applied to the original system is white in space, the processes $\xi_A$ and $\xi_B$, expressed in (\ref{eq:xis_Pi}) and acting on the amplitudes, are complex-valued and uncorrelated with each other (since they refer to distinct wavenumbers). For each process, the real and imaginary parts are also uncorrelated. Furthermore, they are white in time and Gaussian-distributed with zero average. This follows from the fact that all the components in $\hbf_{\bk}$ are white in time. In addition, because components of $\hbf_{\bk}$ are uncorrelated at nonzero time lag (see Table~\ref{tab:cov_f}), the projection of $\hbf_{\bk}$ onto a constant vector must also be white noise. More precisely, if $\bxi \doteq (\xi_A,\xi_B)^T$, then 
\begin{align}
\begin{split}
\ea{\bxi(t)} =\bz, \quad \ea{\bxi(t)\bxi^T(t')} = \bz, \quad \ea{\bxi(t)\bxi^{T,*}(t')} = 2\al^2\de(t-t')\bI, 
\label{eq:noxi}
 \end{split}
\end{align}
where we have defined the scalar $\al$ such that $\al^2$ is the noise intensity of both the real and imaginary parts of both $\xi_A$ and $\xi_B$. We show in Appendix~\ref{app:noise} that the value of $\al$ follows directly from (\ref{eq:xis_Pi}). Numerical values of $\al$ are given in Table.~\ref{tab:coeffpitch}.

Rewriting $A=\mA e^{\ti \psi_A}$ and $B=\mB e^{\ti \psi_B}$, it is easy to show that the equations for the magnitudes, $\mA$ and $\mB$, are decoupled from those for the phases, $\psi_A$ and $\psi_B$, such that (\ref{eq:ampeq_Pi}) can be reduced to a $2$-dof system. Specifically, using It\^{o}'s Lemma, the magnitudes obey the equation
\begin{align}
\begin{split}
&\frac{\di \bH(\tau)}{\di \tau} = -\nab V(\bH(\tau)) + \phi \al \bzet(\tau), \quad \text{with} \\
&\ea{\bzet(\tau)} =\bz, \quad \ea{\bzet(\tau)\bzet^{T}(\tau')} = \de(\tau-\tau')\bI, 
\label{eq:bHsys}
 \end{split}
\end{align}
and where we have defined $\bH(\tau) \coloneq (\mA,\mB)^T(\tau)$. The deterministic part of the right-hand side derives from the potential $V$, defined as
\begin{align}
\begin{split}
V(\bH) \coloneq -\frac{1}{2}\pae{\mA^2 + \mB^2} - \frac{\mu}{4}\pae{\mA^4 + \mB^4} - \frac{\nu}{2}\mA^2\mB^2 - \frac{(\al\phi)^2}{2}\pae{\ln{\mA}+\ln{\mB}}.
\label{eq:pot}
 \end{split}
\end{align}
We note the presence of logarithmic, infinitely high barriers of potential along the axes $\mA=0$ and $\mB=0$, which prevent $\mA$ and $\mB$ from becoming negative. In (\ref{eq:bHsys}), we imply $\nab =(\pa_{\mA},\pa_{\mB})^T$. 

Each magnitude is independent of its associated phase, but the converse is not true. Indeed, the equations for the phases read
\begin{align}
\begin{split}
\frac{\di \psi_A}{\di \tau} = \frac{\phi \al}{\mA}\zeta_{\psi_A}(\tau), \quad \text{and} \quad \frac{\di \psi_B}{\di \tau} = \frac{\phi \al}{\mB} \zeta_{\psi_B}(\tau), 
\label{eq:Phasys}
 \end{split}
\end{align}
where, again, $\zeta_{\psi_\bLac}$ is a Gaussian white noise with zero average and unit intensity (the symbol $\bLac$ denoting either $A$ or $B$ indistinguishably). Importantly, the amplitude of the noise acting on the phase is inversely proportional to the magnitude. This is a simple geometric effect resulting from the fact that a given increment in the trajectory, due to the noise $\zeta_{\psi_\bLac}$, corresponds to a larger increment in the phase if the trajectory is closer to the origin at $|\bLac|=0$. Thus, the effective noise on the phase is proportional to $\zeta_{\psi_\bLac}/|\bLac|$. When the trajectory is exactly at the origin (which occurs with probability zero because the potential diverges there), the phase increment diverges.   

\begin{figure}
            \begin{subfigure}{0.495\textwidth}
            \centering
            \scalebox{0.475}{\includegraphics[trim=0cm 0cm 0cm 0cm]{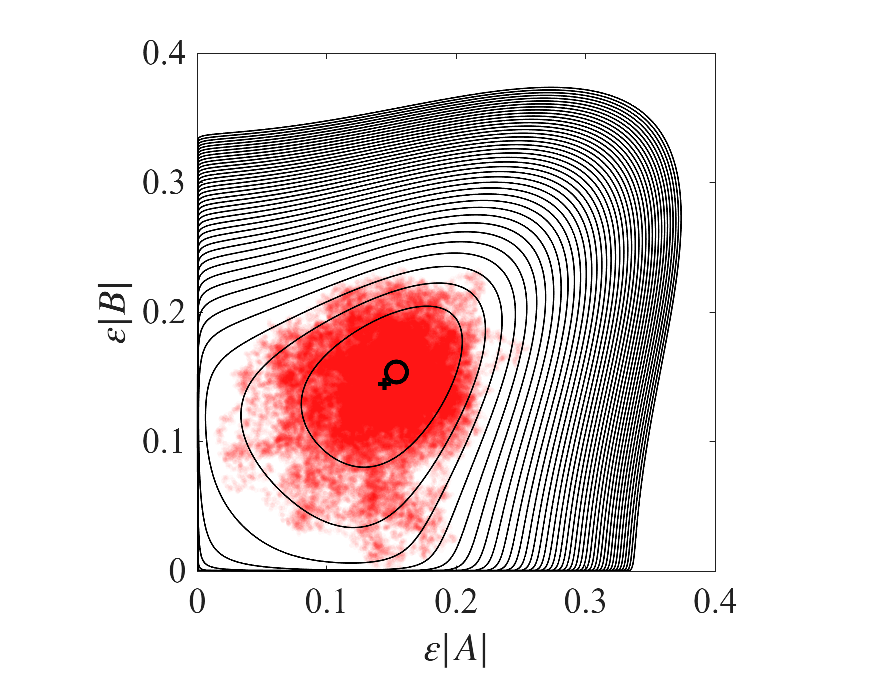}}
             \subcaption{$\bet=0$}
            \end{subfigure}
            \begin{subfigure}{0.495\textwidth}
            \centering
            \scalebox{0.475}{\includegraphics[trim=0cm 0cm 0cm 0cm]{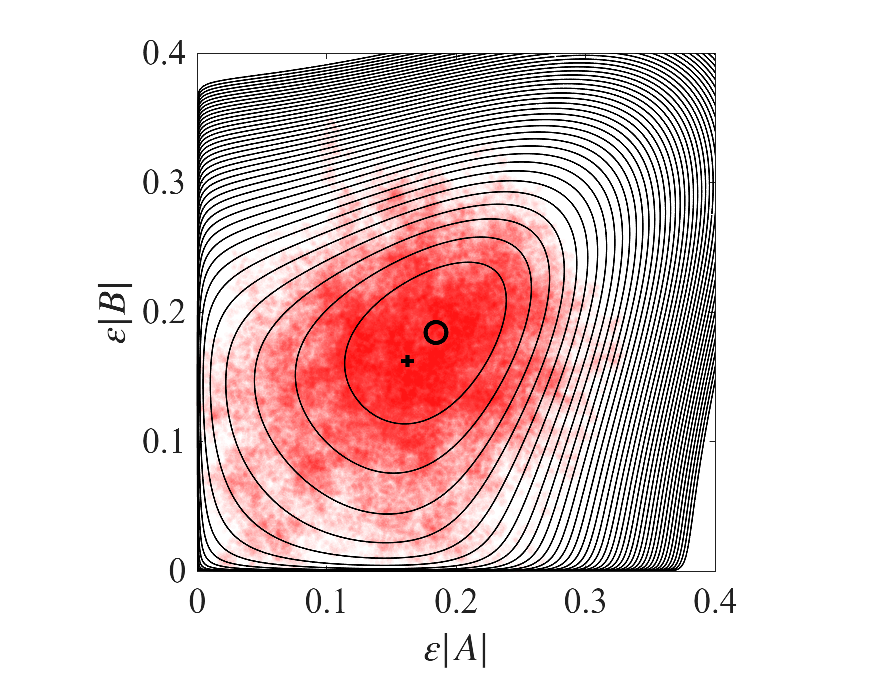}}
             \subcaption{$\bet=0.15$}
            \end{subfigure}
     \caption{For $\e^2 = 0.005$, $\phi=1.41$, and each frame corresponding to different values of $\bet$. 
     A DNS trajectory of length $t=10^4$ ($\e A$ and $\e B$ extracted according to (\ref{eq:Aex})) is shown in the $\e(|A|,|B|)$ phase space with transparent red dots (each corresponds to a different time along the trajectory).
     Forty isolines of the weakly nonlinear potential (\ref{eq:pot}) are also included, starting at the potential minimum (circle marker). 
     The plus marker highlights the deterministic attractor (\ref{eq:He_Pi}). 
     %
     All frames have the same $x$ and $y$ axis range.} \label{fig:inpot}
\end{figure}

Figure~\ref{fig:inpot} shows some fully nonlinear trajectories in the $(\e\mA,\e \mB)$ plane for a fixed $\e^2 = 0.005$.
As explained in Sec.~\ref{sec:dns}, they have been produced by running long DNS of (\ref{eq:ec_st}), and then extracting the component along the eigenmodes $\tbq^A$ and $\tbq^B$ according to (\ref{eq:Aex}). For comparison, we also display some isolines of the WNL potential (\ref{eq:pot}). The latter has a unique global minimum, from which it increases monotonically in all directions. In particular, it increases as both $\mA$ and $\mB$ approach zero, because the uniform base state is unstable. The thin boundary layer where the potential diverges logarithmically at $\mA=0$ and $\mB=0$ is also visible.  

DNS trajectories consistently spend most of their time around the minimum of the WNL potential (where they are initiated). The farther the trajectories depart from the minimum, the rarer the excursions become (i.e., the points are plotted with increasing transparency), especially in the direction where the potential growth is the sharpest.  

Increasing the swimming speed from $\bet=0$ to $\bet=0.15$ between the left and right panels, while keeping $\phi$ fixed, causes the DNS trajectories to explore a larger region of the phase space. In other words, although the amplitude of the externally applied forcing is the same $(=\phi)$ in both cases, the motile configuration appears to experience a greater ``effective'' forcing, or, equivalently, to be more ``receptive'' to the forcing. This is well captured by the weakly nonlinear system (\ref{eq:bHsys}). Here, the amplitude of the stochastic forcing acting directly on the magnitudes is found to be $\al \phi$, which is $\al$ times larger than that of the applied forcing. In Table.~\ref{tab:coeffpitch}, the pre-factor $\al$ is found to be $\approx 1.78$ times larger for $\bet=0.15$ than for $\bet=0$, and thus captures the reported behavior.

We explain the largest receptivity of the system with motile particles, as compared to that for immotile ones, as follows. The polarization field becomes nonzero as soon as $\bet$ does, and direct and adjoint eigenmodes inherently have polarization fields advected in opposite directions by the velocity field. For fixed norms of the direct and adjoint eigenmodes, the projection of the latter onto the former, e.g., $\ssd{\tbq^{A,\da}}{\tbq^A}$, is thus much smaller in the motile case than in the immotile one. That is because of this negative contribution from multiplying direct and adjoint polarization fields. Thereby, enforcing $\ssd{\tbq^{A,\da}}{\tbq^A}=1$ in accordance with (\ref{eq:biorth}), produces an adjoint field of much larger amplitude in the motile case than in the immotile one.
But, according to the expressions (\ref{eq:xis_Pi}) for the noises acting at the amplitude equation level, larger amplitudes of $\tbq^{A,\da}$ and $\tbq^{B,\da}$ must correspond to larger intensities, thus a larger $\al$ from its definition (\ref{eq:noxi}). We insist that this result is not specific to the normalization of the adjoint eigenmodes chosen in (\ref{eq:biorth}). Choosing any another normalization would make appear the prefactors $\ssd{\tbq^{A,\da}}{\tbq^{A}}^{-1}$ and $\ssd{\tbq^{B,\da}}{\tbq^{B}}^{-1}$ in the right-hand sides in (\ref{eq:xis_Pi}), resulting in the same intensities.

Overall, the explanation proposed above is equivalent to noticing that the operators $\bL_{\bka}$ and $\bL_{\bkb}$ have a greater degree of non-normality, in the motile case, as compared to the immotile one. Indeed, the degree of non-normality is increased by the off-diagonal terms in $\bet$ becoming nonzero (i.e., the particles becoming motile) in the operator (\ref{eq:Ldef}). It is well known that a larger degree of non-normality implies poorer spatial support between direct and adjoint eigenmodes, in turn implying a greater receptivity to external forcing \citep{Chomaz05, Giannetti07}. 

\begin{figure}
\centering
\scalebox{0.475}{\includegraphics[trim=1cm 0cm 1cm 0cm]{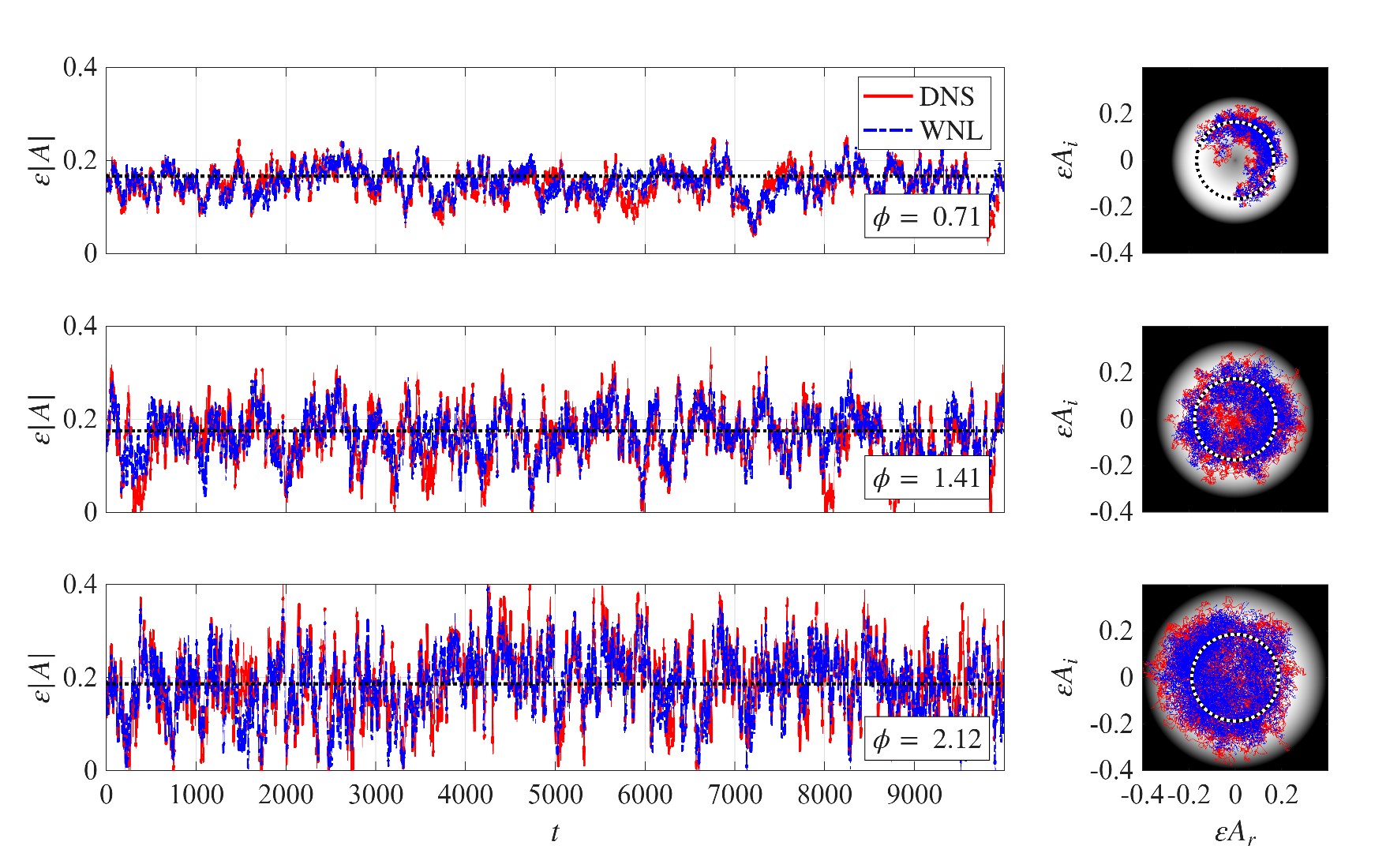}}
     \caption{For $\bet=0.15$ (and $D_R=0.02$, implying $D_{T,c}\approx 0.1333$), corresponding to motile particles, and $\e^2 = 0.005$ (implying $D_T=D_{T,c}-\e^2=0.1283$).
     Left column: temporal evolution of $\e \mA$ over a trajectory for a realization of the stochastic forcing. The dotted line corresponds to the equilibrium value. 
     Right column: $\e A$ shown in the complex plane, with shades of gray corresponding to values of $-\ln(P)/(2\sig^2)$ ranging from the minimum value (lightest shade, highlighted by the dotted circle) to its opposite (darkest shade).
     The fully nonlinear approach (DNS, continuous red line) and the weakly nonlinear one (dash-dotted blue line) are compared. The latter consists of marching (\ref{eq:ampeq_Pi}) in time, with the same noise realization as in the DNS translated into the reduced system according to (\ref{eq:xis_Pi}).
     Three different values of the forcing intensity $\phi \in \set{0.71,1.41,2.12}$ are considered (increasing from top to bottom) 
     }\label{fig:imodstoc}
\end{figure}

For the same parameters as in figure~\ref{fig:inpot}, and including two more values of $\phi$, we compare in figure~\ref{fig:imodstoc} the WNL and DNS trajectories corresponding to $\bet=0.15$.    
In the figure, both approaches use the same noise realization and initial condition for the trajectories. From the spatio-temporal noise generated for the DNS, the components $\hbf_{\bka}(t)$ and $\hbf_{\bkb}(t)$ are extracted at each time step and translated into the processes $\xi_A(t)$ and $\xi_B(t)$, acting on the amplitudes, by applying (\ref{eq:xis_Pi}). The system (\ref{eq:ampeq_Pi}), rewritten over $t=\tau/\e^2$, e.g. $\di_t A = \e^2 (A+\mu A|A|^2+\nu A|B|^2)+ \e \phi \xi_A(t)$, is then marched in time from the initial condition $(A(0),B(0))=\He(1,1)$, which are also imposed in the DNS, and this yields the curves shown for the WNL. The DNS trajectories required several days of CPU time, whereas the WNL ones required a couple of seconds. 

Overall, the two approaches agree well on the magnitude $\e \mA$, though this agreement deteriorates as the forcing amplitude $\phi$ increases. That is because neglected nonlinearities in the WNL expansion are becoming increasingly important. The instantaneous error between the DNS and WNL trajectories, stemming from neglected terms in the expansion, is itself a stochastic process. Although its variance appears low, there are short episodes of time, much shorter than the autocorrelation time $\sim 1/\e^2$, during which the trajectories strongly depart from each other. This occurs more frequently as $\phi$ increases, and one of these episodes is visible just after $t=8000$ and for $\phi =1.41$. That is because, as mentioned, performing integrations by parts in the above development produces terms that can be scaled at higher orders only in the root mean square sense (It\^{o} isometry). This, indeed, does not prevent large instantaneous errors from occurring randomly. In other words, we only expect convergence in the mean of the WNL results to DNS ones as $\phi \rightarrow 0$.    

As expected, increasing $\phi$ from the top to the bottom panels results in a greater variance of $\e \mA$ (around its mean) and thus increasingly frequent excursions toward zero. The trajectories seem more inclined to reach values close to zero than to make an excursion of the same magnitude above the mean. This is consistent with the fact that, in figure~\ref{fig:inpot}, starting from its minimum, the potential increases less sharply in the direction of decreasing values of $\mA$ and $\mB$, than in the direction of increasing ones (except within a thin logarithmic boundary layer at $\mA=0$ and $\mB=0$ where the potential diverges).

The trajectories of $\e A$ plotted in the complex plane in the right panels of figure~\ref{fig:imodstoc}, also reveal the evolution of the phase over time. All trajectories are initiated on the real axis, since $(A(0), B(0))=\He(1,1)$. For the lowest forcing considered, $\phi=0.71$, the magnitude fluctuates only slightly around its mean. Consequently, the phase solving (\ref{eq:Phasys}) is close to simply the integral of white noise, i.e., a Wiener process. Indeed, the trajectory in the right panel resembles a slow random walk around the circle, with the magnitude staying close to equilibrium. For a trajectory of final time $t=10^4$, only about a $3/4$-circle is explored. 

As the forcing is increased, however, the magnitude fluctuates significantly, and the excursions toward zero are associated with an extremely large forcing on the phase in (\ref{eq:Phasys}). Accordingly, within the same time interval, the trajectories explore a much larger region of the complex plane in terms of both phase and magnitude.  

We have performed the same calculation for the immotile case $\bet=0$, and also for $\e^2=0.005$ (not shown). For a given $\phi$, the agreement between the two approaches appears significantly improved with respect to the motile case. This is presumably due to the larger degree of non-normality of the operators for the motile case. This results in (i) a larger effective forcing on the amplitudes for fixed $\phi$ and (ii) a larger magnitude of the neglected higher-order terms, even though they result from the inversion of stable systems, thus making the asymptotic hierarchy less well-posed and reducing the radius of convergence of the series. 

Figure~\ref{fig:imodstoc} compares trajectories (everything else being fixed), but not the associated statistics. The system (\ref{eq:bHsys}) is associated with the Fokker-Planck equation
\begin{align}
\begin{split}
\frac{\pa p }{\pa \tau} = -\nab \bdot \bJ, \ \ \text{with} \ \ \bJ(\bH, \tau) \coloneq  - p(\bH, \tau) \nab V(\bH) - \frac{(\al \phi)^2}{2}\nab p(\bH, \tau),
\label{eq:FP}
 \end{split}
\end{align}
giving the probability density function (pdf) $p(\bH,\tau)$ at a time $\tau$. The probability must be zero at infinity and is naturally also zero at $\mA=0$ and $\mB=0$ because the potential diverges there. In a statistically steady regime, the density function $p_s(\bH) \coloneq \lim_{\tau \rightarrow \infty}p(\bH,\tau) = \mZ^{-1}\exp\pae{-2V(\bH)/(\al \phi)^2}$. The scalar $\mZ$ is a normalization factor, and the subscript $s$ in $p_s$ emphasizes that the pdf relates to a ``steady'' regime. Integrating $p_s$ over one coordinate from zero to infinity gives the steady pdf for the other, such that 
\begin{equation}
\begin{split}
P(\mA) &= \mZ^{-1} \int_{0}^{\infty} p_s(\bH)  \di \mB \\
&= \mZ^{-1} \mA\exp\sae{\frac{\mA^4(\nu^2-\mu^2) + 2\mA^2(-\mu+\nu)}{-2(\al\phi)^2\mu}} \sae{ 1+\erf\pae{\frac{\nu\mA^2+1}{\sqrt{-2(\al\phi)^2\mu}}} }.
\label{eq:bHsys_pdf}
\end{split}    
\end{equation}
Furthermore, since the potential has the symmetry $V(\mA,\mB)=V(\mB,\mA)$ and the two components of the noise $\bzet$ in (\ref{eq:bHsys}) follow the same distribution, it is true that $P(\mB)=P(\mA)$.

\begin{figure}
            \begin{subfigure}{0.495\textwidth}
            \centering
            \scalebox{0.44}{\includegraphics[trim=0cm 0cm 0cm 0cm]{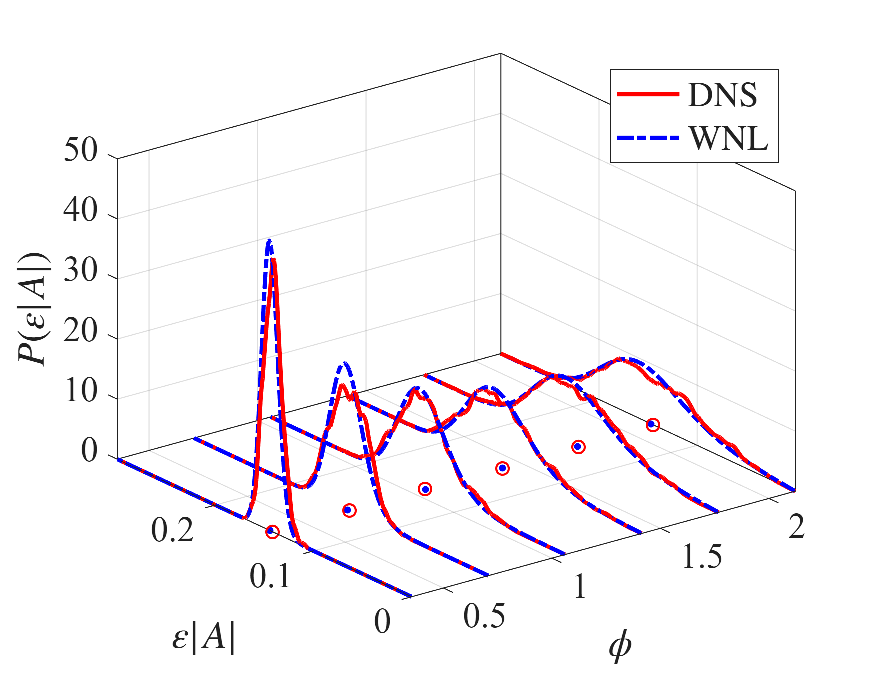}}
             \subcaption{$\bet = 0$}
             \label{fig:modpdf_a}
            \end{subfigure}
            \begin{subfigure}{0.495\textwidth}
            \centering
            \scalebox{0.44}{\includegraphics[trim=0cm 0cm 0cm 0cm]{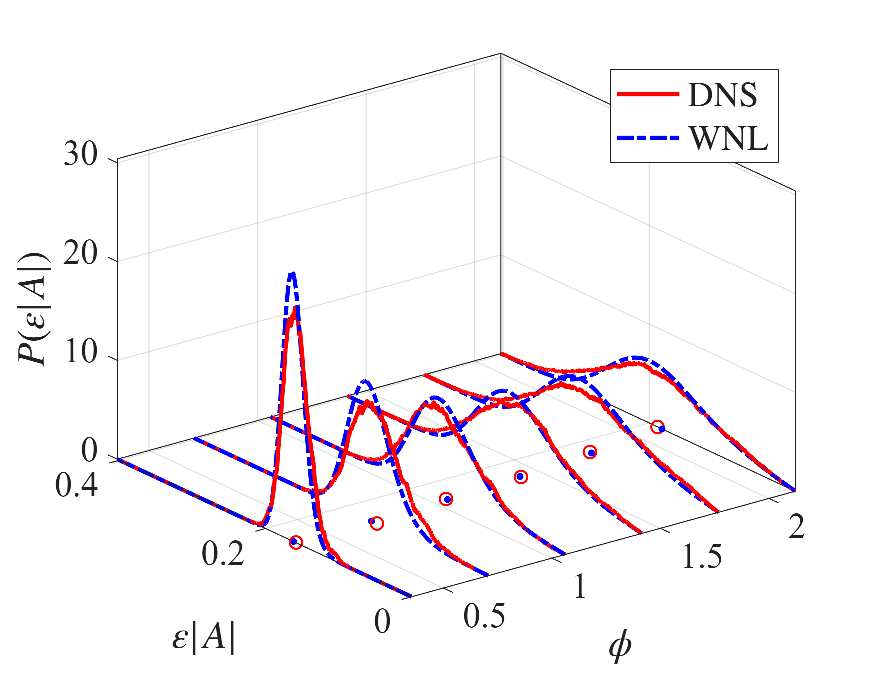}}
             \subcaption{$\bet = 0.15$}
             \label{fig:modpdf_b}
            \end{subfigure}
     \caption{Steady probability density functions (pdf) of $\e \mA$ (also that of $\e \mB$ by symmetry) for six different values of $\phi \in \set{0.35,0.71,1.06,1.41,1.77,2.12}$ and $\e^2=0.005$. 
     The fully nonlinear pdf (DNS, red continuous line) is obtained by postprocessing the data shown in figure~\ref{fig:imodstoc} for $\bet=0.15$ (and for intermediate values of $\phi$), and their equivalent (not shown) for $\bet=0$; for a given $\phi$, a circle marker denotes the barycenter of the pdf, i.e., $\ea{\e \mA}$. 
     The expression of the weakly nonlinear pdf (blue dashed-dotted line) is given in (\ref{eq:bHsys_pdf}), and a dot marker denotes its barycenter.} \label{fig:modpdf}
\end{figure}

Expression (\ref{eq:bHsys_pdf}) for the steady pdf of each weakly nonlinear magnitude is plotted in figure~\ref{fig:modpdf}. Figure~\ref{fig:modpdf_a} corresponds to the immotile case, whereas figure~\ref{fig:modpdf_b} to $\bet=0.15$. 
In both cases, six different values of the forcing are considered. The pdfs corresponding to the DNS trajectories are also reported, obtained by postprocessing the data already presented in figures~\ref{fig:imodstoc} and its equivalent for the immotile case (not shown). 

The weakly and fully nonlinear stationary pdfs are in satisfactory overall agreement. Expression (\ref{eq:bHsys_pdf}) captures well the flattening and widening of the probability distribution as the forcing amplitude increases. It also systematically predicts a correct value for (at least) the first moment of the pdf, i.e. $\ea{\e\mA}=\ea{\e\mB}$.

At least for the parameters considered, the results presented in this section suggest that the WNL expansion accurately predicts the stationary statistics of the leading-order dynamics of the system (\ref{eq:ec_st}) beyond the onset of a bifurcation point. This holds at a considerably lower numerical cost than running DNS. 

In the next section, we will assess whether the amplitude equation system can also predict the statistics and paths of rare, noise-induced structural rearrangements of the system.

\subsection{Stochastic results: dynamical statistics of rare events}
\label{sec:pitch_dys}

\begin{figure}
\centering
\scalebox{0.475}{\includegraphics[trim=1cm 0cm 1cm 0cm]{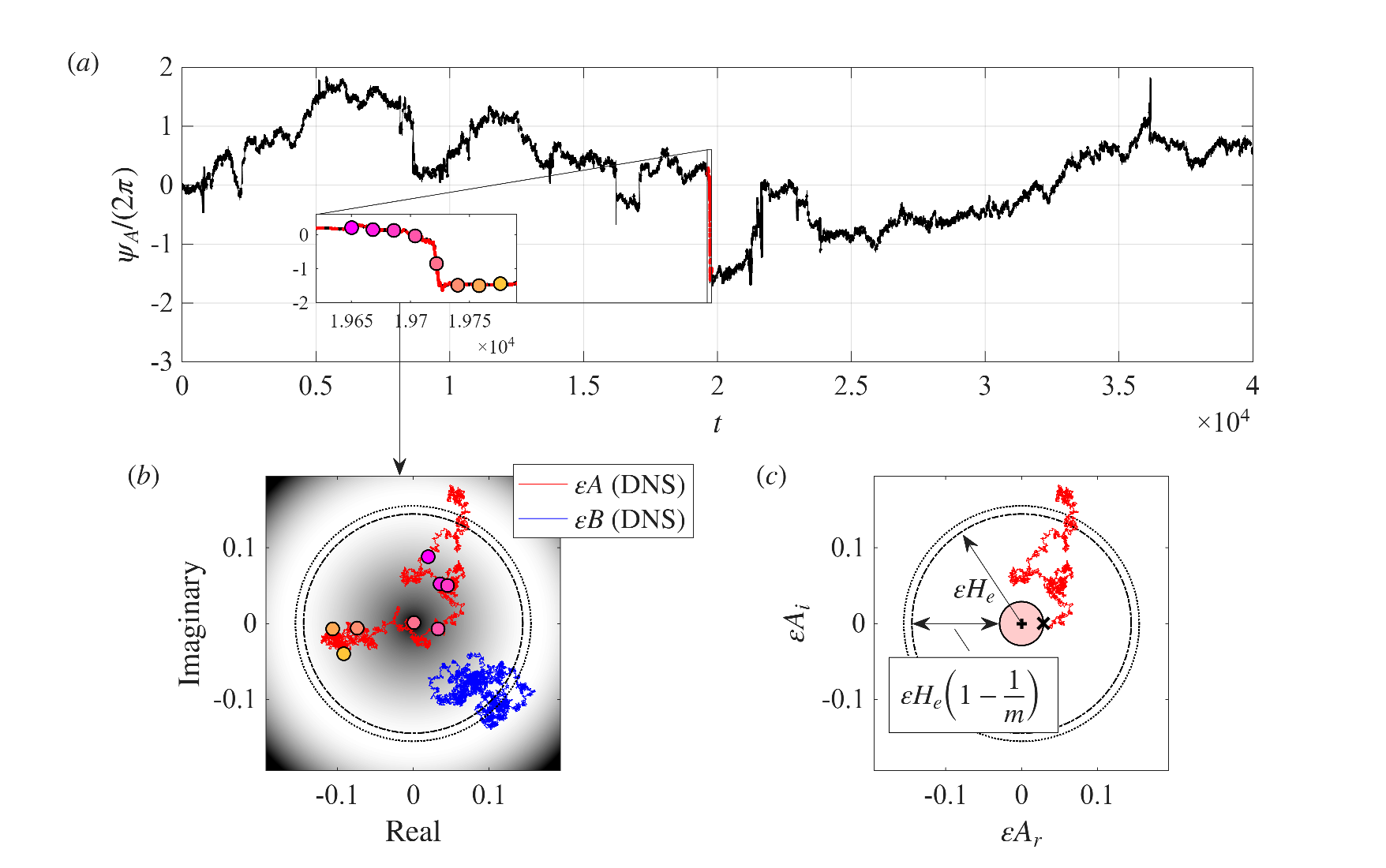}}
     \caption{For $\bet=0$ and $\e^2=0.005$ (same as in figure~\ref{fig:imodstoc}). (a) Evolution over time of the phase of $\psi_A$, the phase of $A$, normalized by $2\pi$ and extracted from a generic DNS realization for $\phi = 1.77$ (black line).
     A ``phase-slip" event, where $\psi_A$ suddenly varies over a time scale much shorter than that of its otherwise slow Brownian drift, is highlighted by a red dotted line and enlarged in the inset. 
     Colored dot markers are located at eight, uniformly spaced times $t$. See figure~\ref{fig:snaptraj} for the corresponding vorticity fields at these times. 
     (b) Within the time frame of this phase-slip event, the corresponding evolutions of both $\e A$ (red line) and $\e B$ (blue line) are shown in the complex plane; along the trajectory of $\e A$, colored dot markers are placed at the same times as those in (a).
     Shades of gray for values of $-\ln(P)/(2\sig^2)$ (depending only on the magnitude, thus appearing axisymmetric on the plot) are also included. The latter function realizes its minimum along the dotted circle, whereas the dash-dotted one has radius $\e H_e$. 
     (c) On the same plane, the event of interest is defined as $\e \mA$ and/or $\e \mB $ becoming smaller than $\e H_e/m$. This is denoted by the black cross on the plot, where the trajectory of $\e A$ (the same as in (b)) enters the red disk of radius $\e H_e/m$.
     } \label{fig:extri}
\end{figure}

For the immotile case and the second-largest forcing considered in figure~\ref{fig:modpdf}, an example of the temporal evolution of the phase $\psi_A$ (rescaled by $2\upi$) of the amplitude $A$ is shown in figure~\ref{fig:extri}a. 
The data relates to a DNS conducted over a very long period of time. The phase evolution resembles a slow, random walk over long time intervals, associated with the patterns shown in figure~\ref{fig:stimmo} stochastically drifting through space. This phenomenon is interspersed with episodes during which the phase changes substantially over a short period of time, much shorter than the otherwise slower random walk.  These abrupt variation episodes occur seemingly at random times and will be referred to as ``phase-slip'' events below.    

One of these phase-slip events is highlighted with a red dashed line in the figure, and magnified in the inset. The corresponding evolution of $\e A$ and $\e B$ is displayed in the complex plane in figure~\ref{fig:extri}b (the colored dots serve as time markers). The weakly nonlinear axisymmetric log-density of the magnitude is also shown. In figure~\ref{fig:extri}b, the trajectory for $\e A$ starts at some random phase but close to the equilibrium magnitude (corresponding to the minimum of the log-density in the WNL perspective). The phase-slip event then corresponds to $\e A$ radially approaching the local maximum of the log-density at the origin, $\mA=0$, very closely, making about one turn and a half around it there (corresponding to a $\approx -3\upi$ increment in the phase), then relaxing to the equilibrium magnitude. Thereby, during the process, the phase has varied considerably in a very short time. That is a consequence of the geometric effect already evoked. 
In particular, if the trajectory is close to the origin, a few time steps suffice to make a full turn around it, each turn corresponding to $\pm 2\upi$ in the phase. 

This is formalized by the WNL equation (\ref{eq:Phasys}) for the phase. Here, the probability distribution of $\di \psi_A $, conditioned to $\mA$, follows a Gaussian with standard deviation $\sqrt{\di \tau}/\mA = \e\sqrt{\di t}/\mA$. Consequently, as $\mA \rightarrow 0$, large phase increments within a time step become increasingly likely (see figure~$5$ in \cite{Rigas15} for an illustration).  

\begin{figure}
\centering
\scalebox{0.44}{\includegraphics[trim=0cm 3.5cm 0cm 0cm]{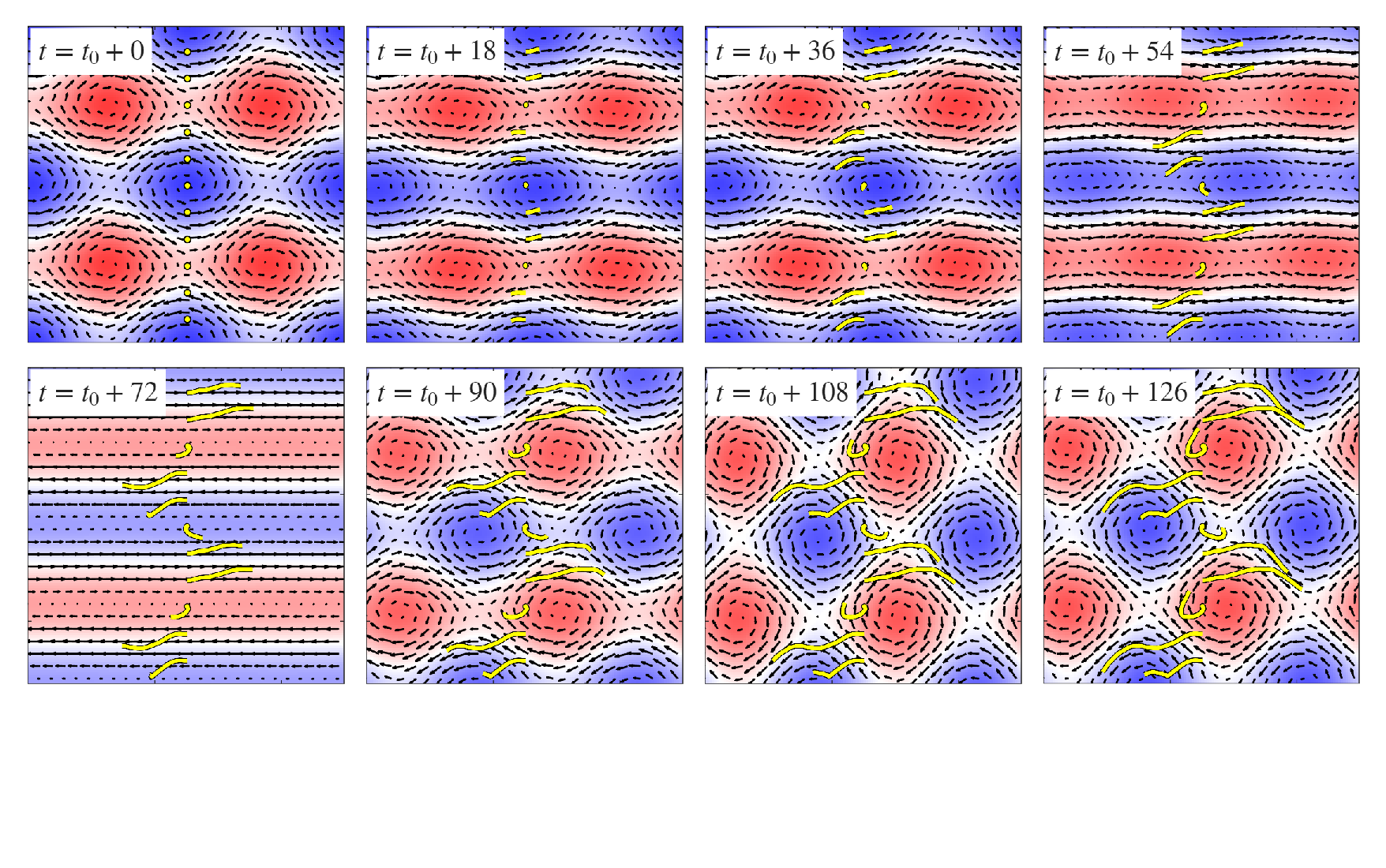}}
     \caption{Eight successive snapshots of the velocity (arrows) and vorticity (colormap) of the flow, during the phase-slip event visible in the inset of figure~\ref{fig:extri}, and at times corresponding to the eight colored dot markers. The shortest time is renamed $t_0$ for the plot. Fields are shown over the doubly periodic square domain $(x,y) \in [0,4\upi] \times [0,4\upi]$ (the pattern being $2\upi$-periodic) and reconstructed from (\ref{eq:q1c2P}) knowing $\e A$ and $\e B$. The yellow lines are passive tracers introduced into the flow at time $t=t_0$ for visualization purposes. } \label{fig:snaptraj}
\end{figure}

We show in figure~\ref{fig:snaptraj} velocity snapshots associated with the trajectories in figure~\ref{fig:extri}b, and thus corresponding to a phase-slip event. 
In figure~\ref{fig:extri}b, at the same time that the amplitude $\e A$ is near the origin of the complex plane, the amplitude $\e B$ remains around the equilibrium magnitude. Thereby, $\e B$, associated with the $\tbq^{B}\exp(\ti y)$ component, dominates over $\e A$, associated with the $\tbq^{A}\exp(\ti x)$ one. This manifests as the solution becoming uniform in the $x$-direction in figure~\ref{fig:snaptraj}. Afterwards, $\e A$ relaxes to its equilibrium magnitude and the pattern of velocity vortices reforms, with a large spatial shift in the $x$-direction (modulo $2\upi$ in the snapshots), relative to the pattern at $t=t_0$. This is precisely because the phase of $\e A$ drifted substantially in the time interval. This is analogous to a noise-induced transition, in the sense that the solution pattern rearranges itself on a much shorter time scale than it would otherwise drift randomly. 

Overall, the phase-slip events in the evolution of $\psi_A$, visible in figure~\ref{fig:extri}a, are enabled by the magnitude $\mA$ randomly getting close to zero. From figure~\ref{fig:extri}b, we also understand that these must remain rare events for the weak forcing considered, as they demand the amplitudes getting near the maximum of the log-density at the origin. A more probable way to achieve a given phase difference is to perform a circular random walk while keeping the magnitude close to its equilibrium value. 

Consequently, the rare event of interest in what follows is formalized as $A$ and/or $B$ entering a disk of small radius $\He/m$ around the origin of the complex plane, with $m \gg 1$ a free parameter (see figure~\ref{fig:extri}c for an illustration). As $m$ increases, $\mA$ and/or $\mB$ approach the origin of the complex plane more closely, making large phase increments increasingly likely. The mean return time between two such events, when $\mA$ and/or $\mB$ decreases below the threshold value $\He/m$, and as measured on the slow time scale $\tau$, is named $r(m)$. In what follows, we determine $r(m)$ for increasingly large values of $m$, and compare the weakly nonlinear and fully nonlinear approaches.

For the WNL counterpart, we exploit the low-dimensionality of the system (\ref{eq:bHsys}) to compute the mean return time $r(m)$ using standard tools from statistical mechanics. For this purpose, the first step is to define the domains in the two-dimensional magnitude phase space, 
\begin{equation}
\begin{split}
\Omega_m \coloneq \set{(\mA,\mB) \in \mathbb{R}_{+}^2 : \mA > \He/m, \ \text{and} \ \mB > \He/m},
\nonumber
\end{split}    
\end{equation}
as well as $\dB \doteq \mathbb{R}_{+}^2 \setminus \Omega_m$. The two domains $\Omega_m$ and $\dB$ (whose union is $\mathbb{R}_{+}^{2}$) are separated by the boundary 
\begin{equation}
\begin{split}
\boB \coloneq  \set{ (\He/m,\mB) : \mB \geq \He/m \} \cup \{ (\mA,\He/m) : \mA \geq \He/m }.
\label{eq:bA}
\end{split}    
\end{equation}
We also define the boundary of $\Omega_m$ at infinity,
\begin{equation}
\begin{split}
\bO_m^{\infty} \coloneq  \set{ (\infty,\mB) : \mB\geq\He/m \} \cup \{  (\mA,\infty) : \mA\geq\He/m },
\nonumber
\end{split}    
\end{equation}
such that $\pa \Omega_m = \bO_m^{\infty} \cup \boB$. The boundary $\boB$ is represented in figure~\ref{fig:imoreac} for $m=10$. For a trajectory initiated anywhere within $\Omega_m$, the rare event of interest corresponds to reaching $\dB$, which can only happen by crossing the boundary $\boB$. From the Fokker-Planck equation, it is then possible to show that $\tau_e(\bH)$, defined as the average time for a trajectory beginning at $\bH \in \Omega_m $ to reach the boundary $\boB$, obeys 
\begin{equation}
\begin{split}
-1 = - \nab \tau_e(\bH) \bdot \nab V(\bH) + \frac{(\al \phi)^2}{2} \Delta \tau_e(\bH), \ \ \text{over $\Omega_m$},
\label{eq:taue}
\end{split}    
\end{equation}
subject to the boundary conditions 
\begin{equation}
\begin{split}
\tau_e &= 0 \quad \text{for $\bH \in \boB$ \ (``absorbing'' b.c.)} \\ 
(\nab \tau_e) \cdot \bn &= 0 \quad \text{for $\bH \in \bO_m^{\infty}$.}
\label{eq:taue_BC}
\end{split}    
\end{equation}
We refer to \cite{Gardiner10}, Chapter $5.4$ for a more detailed derivation. The time $\tau_e(\bH)$ is a function of the starting point $\bH$ of the trajectories, which is unknown but whose density in a statistically steady regime, $p_s(\bH)$, was given above. Thenceforth, the mean return time $r(m)$ is given by
\begin{equation}
\begin{split}
r(m) = \int_{\Omega_m}\tau_e(\bH) p_s(\bH)\di \bH.
\label{eq:rmwnl}
\end{split}    
\end{equation}

\begin{figure}
            \begin{subfigure}{0.495\textwidth}
            \centering
            \scalebox{0.44}{\includegraphics[trim=0cm 0cm 0cm 0cm]{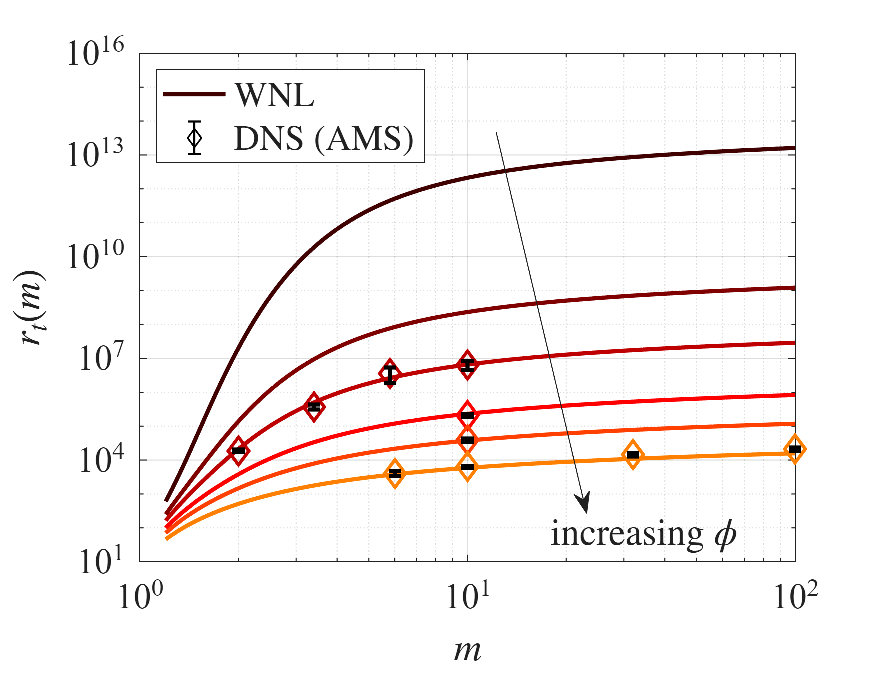}}
             \subcaption{$\bet = 0$}
             \label{fig:AMSpitch_a}
            \end{subfigure}
            \begin{subfigure}{0.495\textwidth}
            \centering
            \scalebox{0.44}{\includegraphics[trim=0cm 0cm 0cm 0cm]{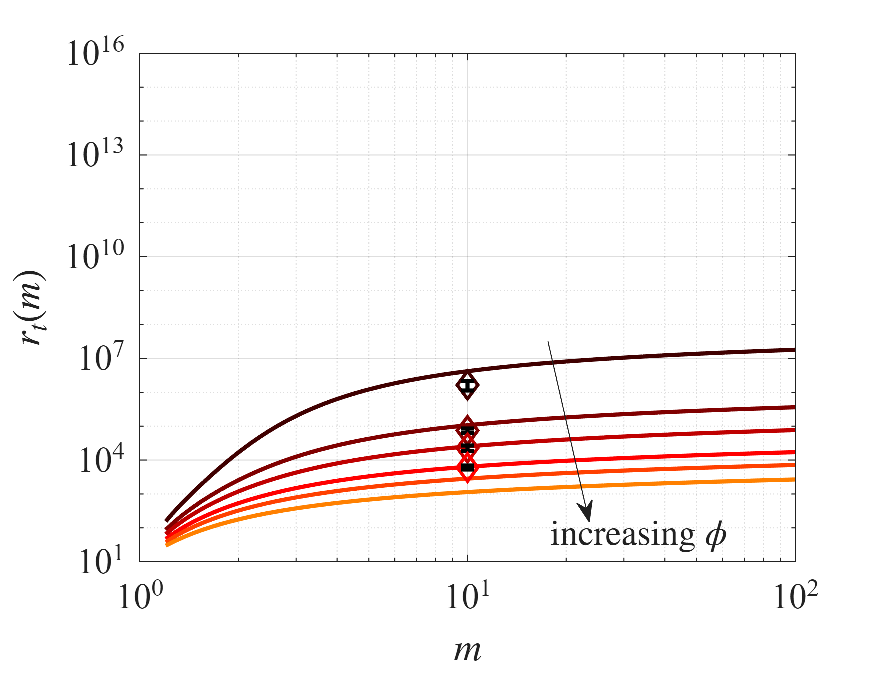}}
             \subcaption{$\bet = 0.15$}
             \label{fig:AMSpitch_b}
            \end{subfigure}
     \caption{For $\e^2 = 0.005$ and $\bet$ different for each frame. Mean return time $r_t(m)$ between two events, where an event is defined as $\mA$ and/or $\mB$ becoming smaller than $H_e/m$. 
     The mean return time is plotted as a function of $m$. 
     Different $\phi \in \set{0.46,0.60,0.71,0.88,1.06,1.41}$ are considered (larger $\phi$ lighter shades). 
     Two approaches are compared. The first uses the Adaptive Multilevel Splitting (AMS) algorithm on the fully nonlinear model (\ref{eq:ec_st}), with the parameters given in Table~\ref{tab:AMSp}. 
     The second approach is weakly nonlinear (continuous line), in which the mean return times can be determined according to (\ref{eq:rmwnl}).} \label{fig:AMSpitch}
\end{figure}

Solving for $\tau_e$ for many different values of $m$ and $\phi$, each time using (\ref{eq:rmwnl}) to extract the corresponding mean return time, yields the curves shown in figure~\ref{fig:AMSpitch}. There, the mean return time is rescaled as $r_t(m) \coloneq r(m)/\e^2$ to correspond to the physical time $t$. This way, it can be compared to fully nonlinear results, all obtained over $t$. The mean return time is plotted as a function of $m$, and corresponds to the same $\e$ and forcing amplitudes $\phi$ considered in figure~\ref{fig:modpdf}.

For the fully nonlinear counterpart, the mean return times are obtained by using the AMS algorithm presented in Sec.~\ref{sec:AMS}, and with the algorithm parameters reported in Table~\ref{tab:AMSp}.
\begin{table}
  \begin{center}
\def~{\hphantom{0}}
  \begin{tabular}{C{3cm} C{1cm} C{0.8cm} C{0.8cm} C{0.8cm} C{0.8cm} C{0.8cm} C{0.8cm} C{0.8cm} C{0.8cm} }
               Version & $\di t$ & $N$ & $h_{\dA}$  & $h_{\dS}$ & $h_{\dB}$ & $t_{\text{max}}$ & $N_t$ & $K$ & $\Nre$ \\
               \hline
               Mean return time & $2\times 10^{-2}$ & $64$ & $\varnothing$ & $\varnothing$ & $\varnothing$ & $5/\e^2$  & $50$  & $1$ & $6$ \\ [3pt]
               Mean transition times & $2\times 10^{-2}$ & $64$ & $-0.95$  &  $-0.8$  &  $0.95$ & $\varnothing$  & $50$  & $1$ & $6$
  \end{tabular}
  \caption{Parameters of the AMS algorithm, using the notations introduced in Sec.~\ref{sec:AMS}. The system is discretized in time with a time step $\di t$, and in space over a uniform grid of $N\times N$ points. The mean return time version of the algorithm is employed in Sec.~\ref{sec:pitch_dys} with cost function (\ref{eq:rcoor_Pi}), whereas the mean transition time version is employed in Sec.~\ref{sec:hopf_dys} with cost function (\ref{eq:rcoor_Ho}).} 
  \label{tab:AMSp}
  \end{center}
\end{table}

To sort the trajectories, the cost function is chosen as  
\begin{equation}
\begin{split}
\varphi(\bH) \coloneq 1 - \log\pae{\mA\frac{m}{\He}}\log\pae{\mB\frac{m}{\He}}, \ \ \text{defined for} \ \ \mA, \mB \geq \frac{\He}{m},
\label{eq:rcoor_Pi}
\end{split}    
\end{equation}
which reaches its maximum $(=1)$ along the boundary $\boB$, and decreases monotonically as $\bH$ departs from the latter. 

Perhaps unsurprisingly, for a given $\phi$, the mean return time in figure~\ref{fig:AMSpitch} is monotonically increasing with $m$. This is because the disk of radius $\He/m$ centered at the origin reduces to a single point of measure zero. By contrast, for a given $m$, the mean return time decreases monotonically with $\phi$ as trajectories explore a larger region of space (visible in figure~\ref{fig:inpot}). 

For $\bet=0$ in figure~\ref{fig:AMSpitch_a}, and for all the $m$ and $\phi$ considered, the weakly and fully nonlinear approaches are in good agreement. This is particularly remarkable given that the two approaches employ distinct techniques to calculate the mean return time from their respective governing systems. Although these techniques should, in principle, yield the same results, many sources of bias could arise in the AMS algorithm \citep{Brehier15, Rolland15}.

The good agreement between the two approaches suggests that the predictions of the WNL reduced system are valid not only for dynamical statistics of rare events, but also for deterministic and statistically steady quantities. This is particularly useful given that the largest return time obtained with the AMS algorithm, of $O(10^7)$, required several days of calculations. For this reason, we did not try the algorithm in a regime where we expected an even larger return time. The WNL curves, however, required in total only a couple of minutes of computation time and could predict return times of (at least) $O(10^{13})$. 

For the motile case at $\bet=0.15$ in figure~\ref{fig:AMSpitch_b}, for a given $\phi$, the mean return times are substantially smaller than for the immotile case. That is again because the noise intensity $\al$, acting on the magnitude, is larger in the motile case. For $m=10$, the agreement between the weakly and fully nonlinear approaches appears to degrade slightly more rapidly as $\phi$ decreases than for the immotile case.    

\begin{figure}
            \begin{subfigure}{0.495\textwidth}
            \centering
            \scalebox{0.44}{\includegraphics[trim=0cm 0cm 0cm 0cm]{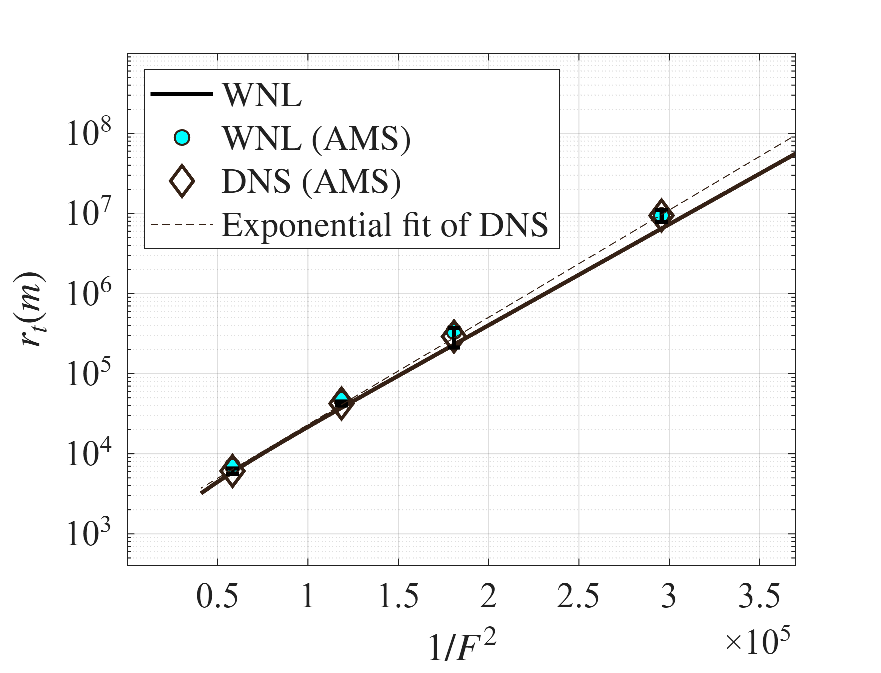}}
             \subcaption{$\e^2 = 0.0025$}
             \label{fig:epseff_a}
            \end{subfigure}
            \begin{subfigure}{0.495\textwidth}
            \centering
            \scalebox{0.44}{\includegraphics[trim=0cm 0cm 0cm 0cm]{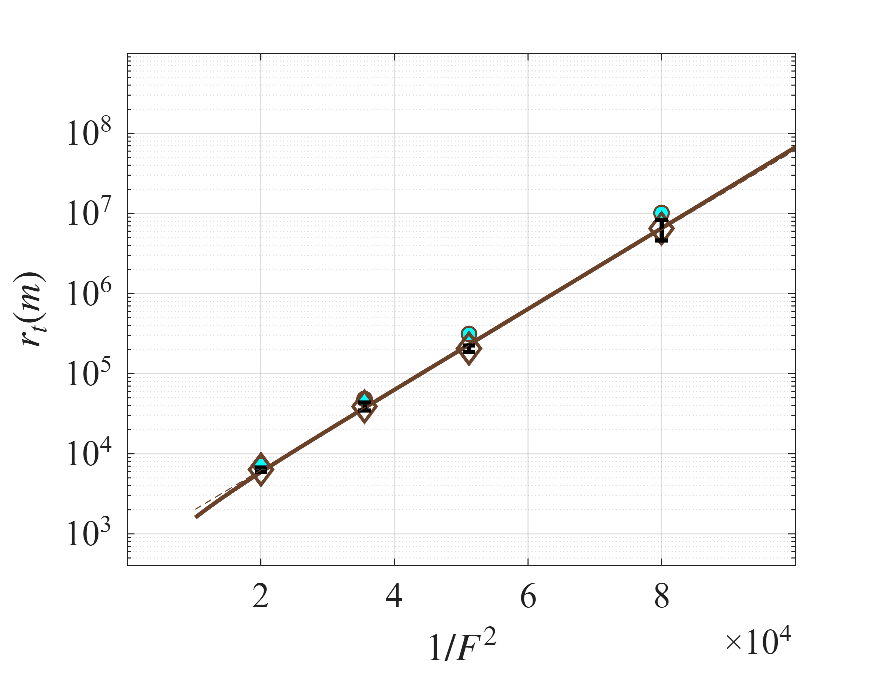}}
             \subcaption{$\e^2 = 0.005$}
             \label{fig:epseff_b}
            \end{subfigure}
            \begin{subfigure}{0.495\textwidth}
            \centering
            \scalebox{0.44}{\includegraphics[trim=0cm 0cm 0cm 0cm]{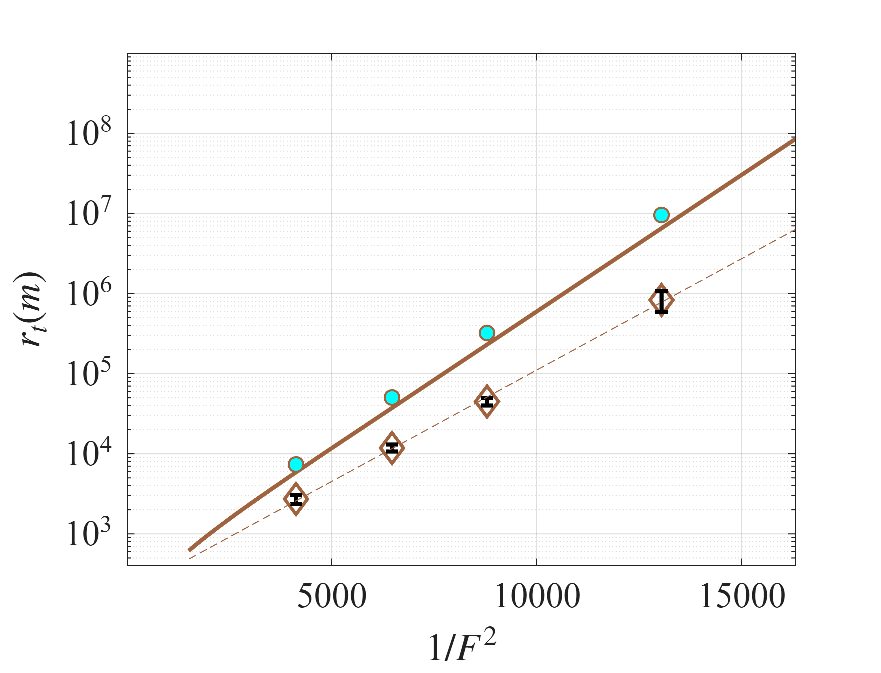}}
             \subcaption{$\e^2 = 0.013$}
             \label{fig:epseff_c}
            \end{subfigure}
            \begin{subfigure}{0.495\textwidth}
            \centering
            \scalebox{0.44}{\includegraphics[trim=0cm 0cm 0cm 0cm]{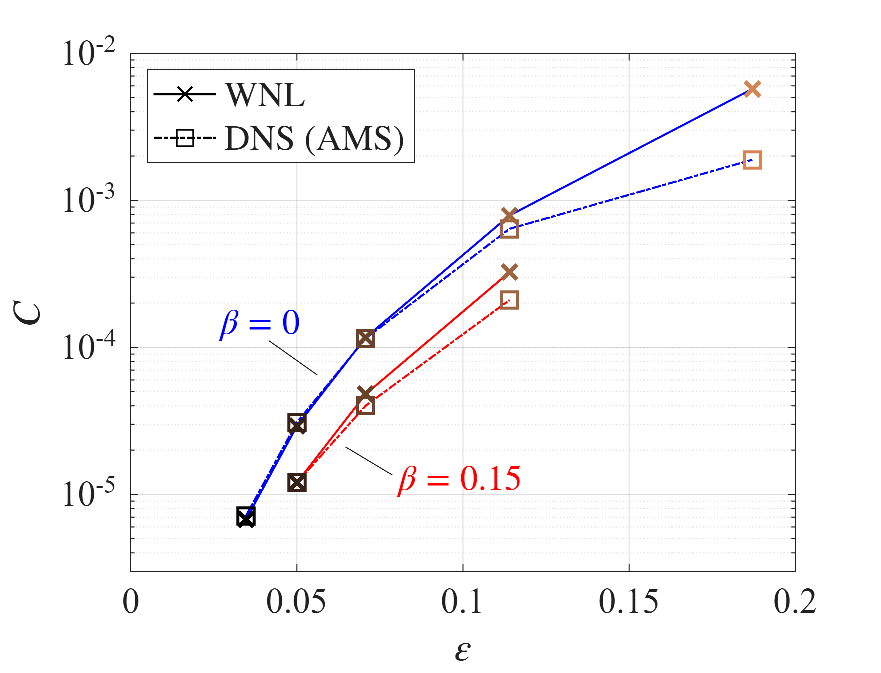}}
             \subcaption{}
             \label{fig:epseff_d}
            \end{subfigure}
     \caption{For $\bet=0$ and different $\e$ for each frames in (a)-(c) (larger $\e$ lighter shades).
     (a)-(c) Mean return time $r_t(m)$ for $m=10$ and as a function of $1/F^2$, where $F=\e^2\phi$ is the forcing intensity. 
     The same two approaches as in figure~\ref{fig:AMSpitch} are compared (diamond marker for the DNS with the AMS algorithm, and a continuous line for the weakly nonlinear estimate).
     Predictions made by using the AMS algorithm directly on the amplitude equations system are also included (cyan dot markers) to reveal potential biases due to the AMS itself.  
     The best exponential fit of the form $r_t \propto \exp(C/F^2)$, consistent with the Arrhenius law, is also computed (shown with a dashed line for the DNS).
     The corresponding rates $C$ are reported in (d) as a function of $\e$ (in blue); the results for $\bet=0.15$ are also included (in red) for comparison.} \label{fig:epseff}
\end{figure}

\begin{figure}
\centering
    \begin{subfigure}{0.495\textwidth}
    \centering
    \scalebox{0.33}{\includegraphics[trim={3cm 0cm 4cm 0cm},clip]{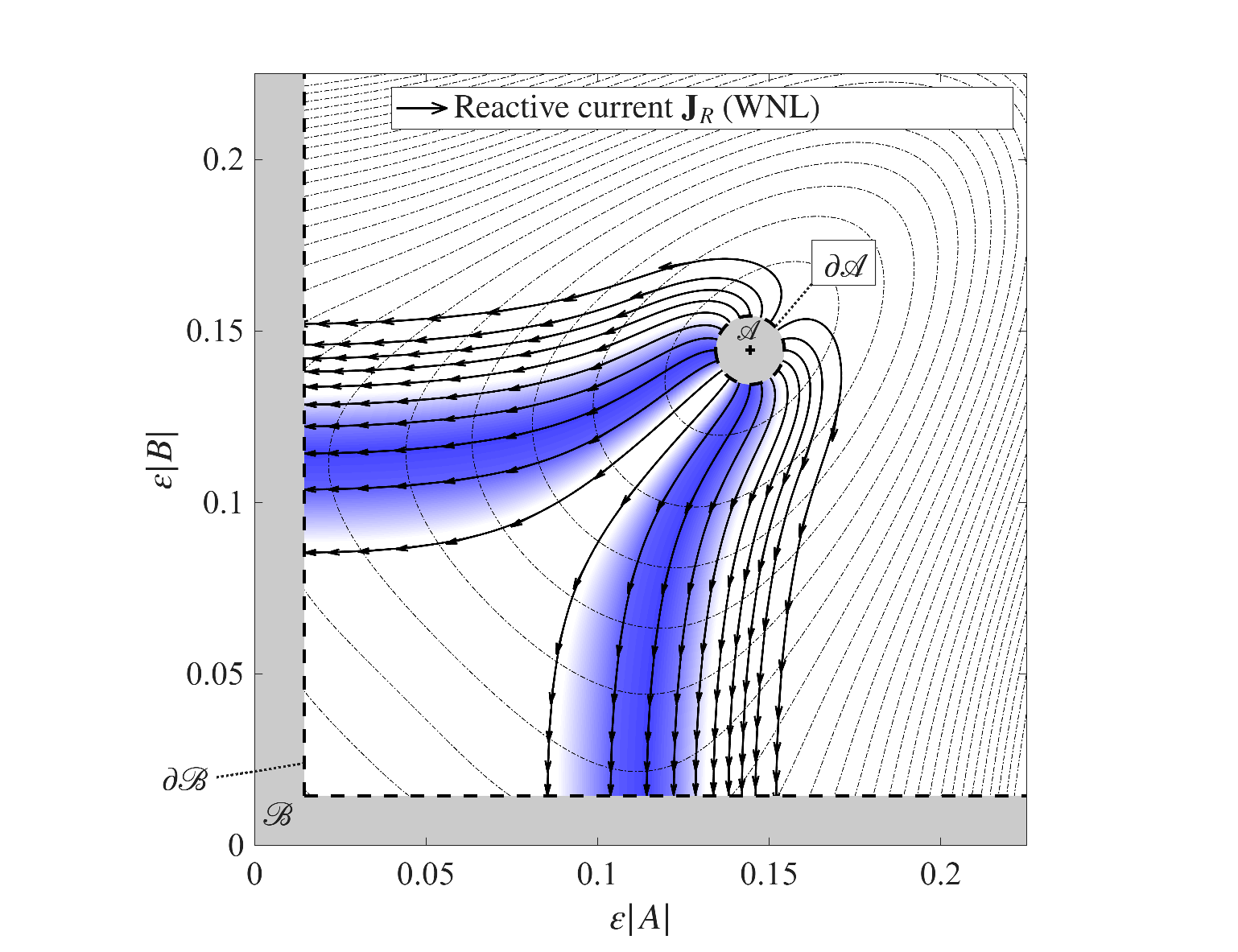}}
     \subcaption{}
     \label{fig:reac_a}
    \end{subfigure}
    \begin{subfigure}{0.495\textwidth}
    \centering
    \scalebox{0.33}{\includegraphics[trim={3cm 0cm 4cm 0cm},clip]{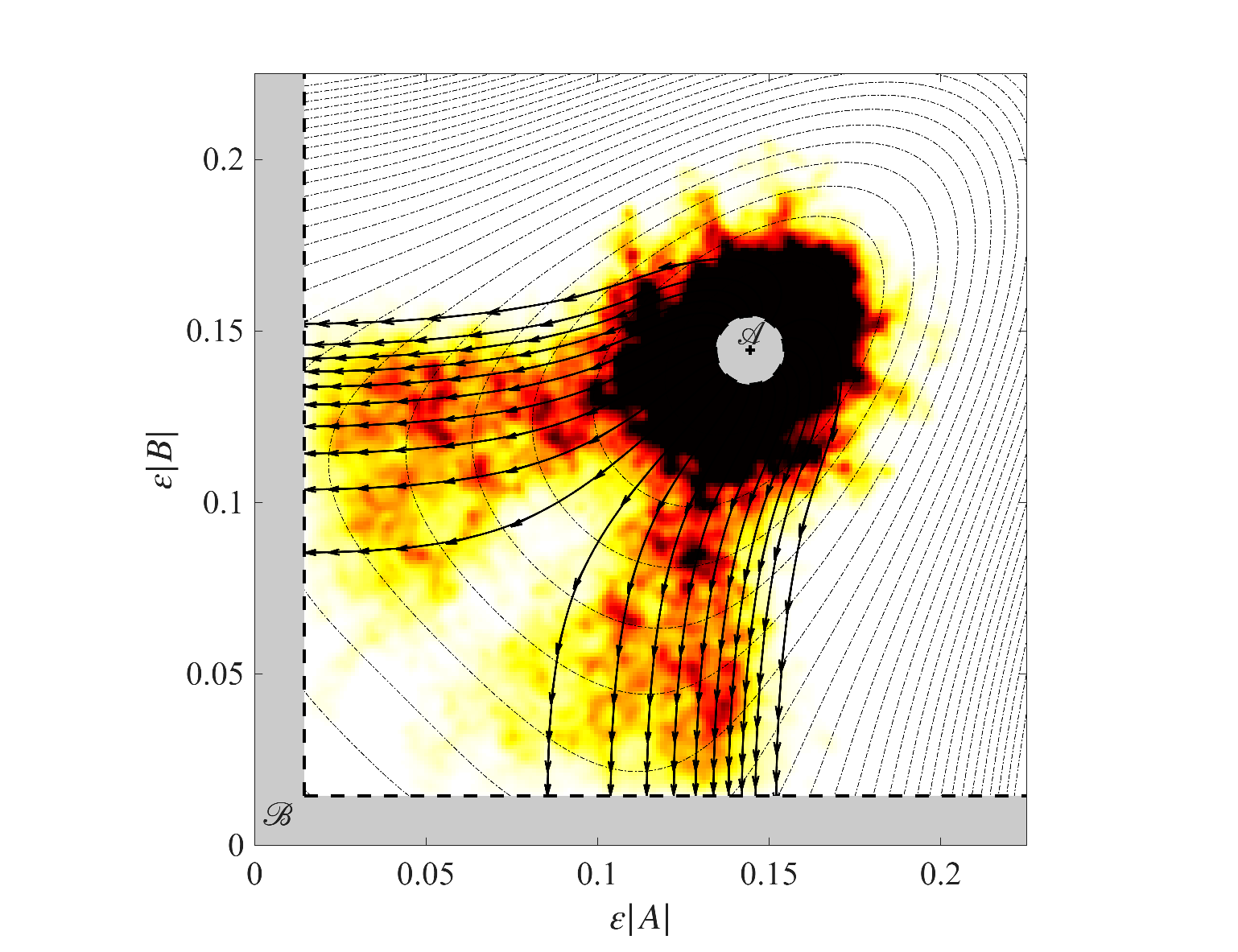}}
     \subcaption{}
     \label{fig:reac_b}
    \end{subfigure}
     \caption{For $\bet=0$, $\e^2=0.005$, $\phi=0.71$ and $m=10$.
     (a) Streamlines of the reactive probability current $\bJ_R$, which characterize how, on average, trajectories go from $\dA$ to $\dB$.
     The current $\bJ_R$ results from the weakly nonlinear approach, and is computed according to (\ref{eq:jr}). 
     Each streamline is colored according to its own weight (the colors are interpolated between the streamlines). This weight is computed as flux of the reactive probability current through the $1/2$-isocontour of the $\qp$ committor function, at the point where the streamline intersects it. The darker the shade, the greater the flux of reactive trajectories passing through the region. 
     Thereby, along the boundary $\boB$, darker shades are naturally observed at the local minima of the weakly nonlinear potential, whose isocontours are shown as dash-dotted lines. 
     %
     (b) The same streamlines as in (a) are shown, together with a density plot of the $\dA \rightarrow \dB$ reactive trajectories, $120$ in total, as obtained from the fully nonlinear approach with the AMS algorithm.
     %
     %
     }  \label{fig:imoreac}
\end{figure}

Setting $m=10$ and focusing on the immotile case, we study in figure~\ref{fig:epseff} the effects of varying the small parameter $\e$ on the mean return times. 
The latter is plotted as a function of $1/F^2$, where we recall that $F=\e^2\phi$ is the non-rescaled forcing amplitude. When plotted this way, the Arrhenius law for an equilibrium system (i.e., deriving from a potential) predicts that the mean return time grows exponentially. The same two weakly and fully nonlinear approaches as in figure~\ref{fig:AMSpitch} are shown and compared. To assess potential bias coming from the AMS algorithm, we also use the latter directly and the WNL system (\ref{eq:bHsys}) (instead of solving (\ref{eq:taue}) then (\ref{eq:rmwnl})). The results are labeled WNL (AMS) in the figure. 

For all values of $\e$ considered, and for both approaches, the mean return time is indeed exponential in the inverse of the forcing intensity. The associated exponential rates are interpreted as the potential barrier. For the smallest $\e$ considered in figures~\ref{fig:epseff_a}, there is a slight mismatch between the rates (i.e., the slopes in the figures in log-lin scale) of the WNL and DNS (with the AMS algorithm) methods. However, this mismatch is corrected by also applying the AMS to the amplitude equation system (the circle markers align well with the diamond ones). Therefore, we conclude that this slight discrepancy is due to algorithm biases and/or discretization errors.

Nonetheless, increasing $\e$ eventually makes the mismatch between weakly and fully nonlinear exponential rates significant, and this is not due to numerical biases. This is perhaps better seen in figures~\ref{fig:epseff_d}, where the rates are reported as a function of $\e$. While the rates converge towards each other in the limit $\e \rightarrow 0$, indicating that the weakly nonlinear expansion is well-posed, they depart from each other above $\e \sim 10^{-1}$. In particular, the fully nonlinear rates and associated mean return times become substantially smaller than the weakly nonlinear predictions. 

This overestimation can be attributed to the neglected, higher-order multiplicative noise terms $\xi^{(2)}_{\bLac_j}$ appearing at $O(\e^2)$ in (\ref{eq:nf}), and detailed in Appendix~\ref{app:xi2}. For increasingly large $\e$, these multiplicative noises presumably lower the potential barrier and greatly reduce the return times. 

We note that, by contrast, for $\e \sim 10^{-1}$ and $\bet=0$, corresponding to $D_T=0.16$ in figure~\ref{fig:bdp_a}, the weakly/fully nonlinear agreement for the deterministic equilibrium magnitude remains excellent. This suggests that the radius of convergence of the weakly nonlinear expansion is much smaller for predicting rare event statistics than for deterministic quantities. This confirms that neglecting the higher-order multiplicative noises is the main source of error in figure~\ref{fig:epseff_d}. The mean return time, due to its exponential dependence on the potential-barrier-to-noise ratio, amplifies enormously any small mistake made on the latter. Nonetheless, $\e \sim 10^{-1}$ remains a reasonably large value, below which the weakly nonlinear expansion accurately predicts the mean return time between two rare events. 

Still in figure~\ref{fig:epseff_d}, the exponential rates corresponding to $\bet=0.15$ yield a lesser radius of convergence, as compared to the immotile case. Again, this is presumably due to the motile dynamics having a greater degree of non-normality and thus being more receptive to forcing. 

We now turn our attention to the associated transition paths in the $(\mA,\mB)$ phase space. For this purpose, we first define a disk $\dA$ of rather small radius $\rA$ centered around $(\He,\He)$, i.e., $\dA \coloneq \{\bH: d(\bH,\bH_e) < \rA \}$, such that $\dA$ denotes the region of the phase space near the deterministic attractor, where trajectories spend most of their time. Therefore, we seek to characterize how, on average, trajectories go from $\dA$ to $\dB$. This amounts to computing the reactive probability current, denoted $\bJ_{R}$ in what follows. For the weakly nonlinear approach, the reactive probability current can be computed directly, as shown in Appendix~\ref{app:reac} and again made possible by (\ref{eq:bHsys}) having only two degrees of freedom. 

For the immotile case, and the forcing intensity corresponding to the largest return time ($\sim 10^7)$ computed by the AMS algorithm in figure~\ref{fig:AMSpitch}, we show the streamlines of the reactive probability current in figure~\ref{fig:imoreac}. 
These streamlines show how, on average, the reactive trajectories transition from $\dA$ to $\dB$. Furthermore, trajectories from $\dA$ to $\dB$ are more likely to follow streamlines with a darker color in figure~\ref{fig:reac_a}. For comparison, we have also included in figure~\ref{fig:reac_b} a density plot of the reactive trajectories produced by the AMS algorithm applied to the fully nonlinear model. Both approaches seem in excellent agreement. Firstly, because the fully nonlinear reactive trajectories appear to, on average, follow the streamlines of the reactive probability current, obtained from the system of amplitude equations. Secondly, because the reactive trajectories indeed seem to concentrate where the flux carried by the streamlines is the largest. 

The reactive trajectories correspond to $\mA$ or $\mB$ decreasing below $\He/m$, while the other remains close to its equilibrium value (as was already visible in figure~\ref{fig:extri}b). Because of the $V(\mA,\mB)=V(\mB,\mA)$ symmetry of the potential, the two scenarios are equally likely. In going from $\dA$ to $\dB$, the trajectories choose a path that minimizes the potential elevation along the way. For example, if the rare event is realized by $\mA$ decreasing below $\He/m$, then, for each value of $\mA$ along the way, the corresponding reactive trajectories concentrate around the minimum of the potential along $\mB$. In the limit $m \rightarrow \infty$, we expect the reactive trajectories to reach $\boB$ through the minimum of $V(\mA,0)$, or, with the same probability, the minimum of $V(0,\mB)$, these two points indeed correspond to saddle points of (\ref{eq:ampeq_Pi}).

That reactive trajectories follow the minimal-energy path on average is a well-known result for systems deriving from a potential. The main result of figure~\ref{fig:imoreac} is precisely to reveal that the fully nonlinear trajectories follow such a minimal energy path, even though the deterministic part of (\ref{eq:ec_st}) does not derive from a potential. However, the weakly nonlinear expansion has shown that, in a certain regime, the system (\ref{eq:ec_st}) can be represented in a reduced set of reaction coordinates whose dynamics do derive from a potential. Here, these reaction coordinates naturally emerge as the modulation amplitudes of the bifurcation eigenmodes.

Overall, the results presented in this section build our confidence that weakly nonlinear expansion techniques are also capable of making quantitative predictions about rare-event dynamical statistics. 
 
\section{Results in the Hopf bifurcation region}
\label{sec:hopf}

Let us now study the dynamics past the onset of the codimension-4 Hopf bifurcation. The results in the deterministic regime are first given in Sec.~\ref{sec:hopf_det}, in particular, the co-existence of two stable periodic orbits. The consequent statistical results of the noise-induced transitions are presented in Sec.~\ref{sec:hopf_dys}. In this section, unlike in the previous one, the system of amplitude equations does not derive from a potential. 

\subsection{Deterministic results}
\label{sec:hopf_det}

The amplitude equations system (\ref{eq:ampeq_H}) for the Hopf bifurcation involves more eigenmodes and thus more nonlinear interactions than for the pitchfork one. This time, all coefficients are generically complex-valued, and $\eta$ and $\kap$ did not exist for the pitchfork case. In the deterministic regime, i.e., for $\phi=0$, the system (\ref{eq:ampeq_H}) can be rewritten for the magnitudes of the amplitudes as
\begin{equation}
\begin{split}
\frac{\di \mAp}{\di \tau} = & \mAp + \mu_r \mAp^3 + \eta_r \mAp\mAm^2  + \nu_r \mAp \nn{\boldsymbol{B}}^2 + \Re(\kap e^{\ti \de}) \mAm \mBp \mBm, \\
\frac{\di \mAm}{\di \tau}=& \mAm + \mu_r \mAm^3 + \eta_r \mAm\mAp^2 + \nu_r \mAm \nn{\boldsymbol{B}}^2 + \Re(\kap e^{\ti \de}) \mAp \mBp \mBm, \\
\frac{\di \mBp}{\di \tau}=& \mBp + \mu_r \mBp^3 + \eta_r \mBp\mBm^2 + \nu_r \mBp \nn{\boldsymbol{A}}^2 - \Re(\kap e^{-\ti \de}) \mAp \mAm \mBm, \\
\frac{\di \mBm}{\di \tau}=& \mBm + \mu_r \mBm^3 + \eta_r \mBm\mBp^2 + \nu_r \mBm \nn{\boldsymbol{A}}^2 - \Re(\kap e^{-\ti \de}) \mAp \mAm \mBp,
\label{eq:sysmodr}
\end{split}    
\end{equation}
where $\nn{\boldsymbol{B}}^2 =\mBp^2+\mBm^2$, $\nn{\boldsymbol{A}}^2 =\mAp^2+\mAm^2$ and both $\Re(\bullet)$ and the subscript ``$r$'' stand for the real part. Remarkably, the evolution of the magnitudes now depends upon
\begin{equation}
\begin{split}
\de \doteq -\psi_{\Ap} + \psi_{\Am} + \psi_{\Bp} - \psi_{\Bm},
\label{eq:dedef}
\end{split}    
\end{equation}
a linear combination of the phases. The evolution equation for $\de$ is
\begin{equation}
\begin{split}
\frac{\di \de}{\di \tau} = & (\mu_i + \eta_i - 2\nu_i)\pae{-\nn{\boldsymbol{A}}^2 + \nn{\boldsymbol{B}}^2} - \Im(\kap e^{\ti\de}) \pae{\frac{\mAm \mBp \mBm}{\mAp} + \frac{\mAp \mBp \mBm}{\mAm}}  \\
&- \Im(\kap e^{-\ti\de}) \pae{\frac{\mAp \mAm \mBm}{\mBp} +  \frac{\mAp \mAm \mBp}{\mBm}}.
\label{eq:eqdel}
\end{split}    
\end{equation}
which couples nonlinearly to the magnitudes. Both $\Im(\bullet)$ and the subscript ``$i$'' denote the imaginary part. The dependence of the magnitudes on the phase combination $\de$ is a consequence of the $3$-wave interaction terms, pre-multiplied by $\kap$ or $\kap^*$ in (\ref{eq:ffka}) and (\ref{eq:ffkb}). These terms indicate that energy can be transferred to a specific eigenmode by nonlinear interactions between the other three eigenmodes. The amount and the direction of energy transferred depend on the phase combination $\de$.

System (\ref{eq:sysmodr})-(\ref{eq:eqdel}) has only two stable fixed points, or, more precisely, two discrete families of stable fixed points $2\upi$-periodic in $\de$. Each family has $\mAp=\mAm=\mBp=\mBm=\He$, where
\begin{equation}
\begin{split}
\He = \frac{1}{\sqrt{- (\mu_r + 2\nu_r + \eta_r) + \sin(\dee) \kap_i}}, \quad \text{and} \quad \sin(\dee) = \pm 1.
\label{eq:eqhe}
\end{split}    
\end{equation}
The symbol $\dee$ designates the equilibrium values of $\de$. The first (family of) stable fixed point(s), which we name ``$\opr$'', corresponds to $\sin(\dee) = 1$ (i.e., $\dee=\upi/2+2\upi n$ with $n \in \mathbb{Z}$). The second (family of) stable fixed point(s) is denoted ``$\ops$" and yields the opposite value, $\sin(\dee) = -1$ (i.e., $\dee=-\upi/2+2\upi n$). We emphasize that, as soon as $\kap_i \neq0$, each family also yields a different value of $\He$ since the latter depends on $\sin(\dee)$. While $\opr$ and $\ops$ are fixed points in the phase space of the amplitude equation system, they are periodic orbits in the phase space of the original system.

Importantly, system (\ref{eq:sysmodr})-(\ref{eq:eqdel}) is invariant under the involution 
\begin{equation}
\begin{split}
\Si : (\mAp, \mAm, \mBp, \mBm, \de) \rightarrow (\mBp, \mBm, \mAp, \mAm, \upi-\de).
\nonumber
\end{split}    
\end{equation}
In particular, $\Si$ maps any representative of the $\opr$ family to another representative of this same family, and similarly for $\ops$. 

Once again, the values of the weakly nonlinear coefficients $\mu$, $\nu$, $\eta$, and $\kap$ depend on the normalization choice of the eigenmodes. Furthermore, a normalized eigenmode is defined only up to a scalar prefactor, with a unit magnitude but an arbitrary phase. The coefficient $\kap$ (only) inherits from this arbitrary phase. 
In other words, it is possible to make $\kap_i$ take any value such that $-|\kap| \leq \kap_i \leq |\kap|$.

The condition for both $\He$ to exist and be finite in (\ref{eq:eqhe}) regardless of the phase of $\kap$, is that 
\begin{equation}
\begin{split}
\mu_r + 2\nu_r + \eta_r < - |\kap|.
\label{eq:supcond}
\end{split}    
\end{equation}
This condition is derived by considering the worst-case scenario in (\ref{eq:eqhe}), where $\sin(\dee)\kap_i=-|\kap|$. If (\ref{eq:supcond}) holds, then the bifurcation is supercritical. If it does not, then it is always possible to adjust the phase of $\kap$ such that one or both $\He$ cease to exist, and the bifurcation becomes subcritical. 

\begin{figure}
\centering
            \begin{subfigure}{0.99\textwidth}
            \centering
            \scalebox{0.44}{\includegraphics[trim=0cm 0cm 0cm 0cm]{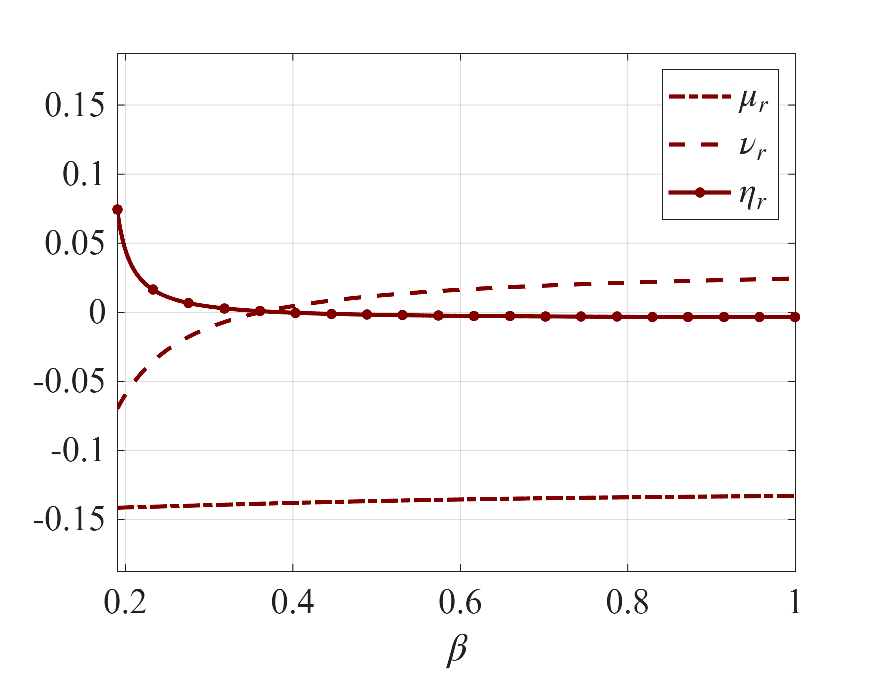}}
            \hfill
            \scalebox{0.44}{\includegraphics[trim=0cm 0cm 0cm 0cm]{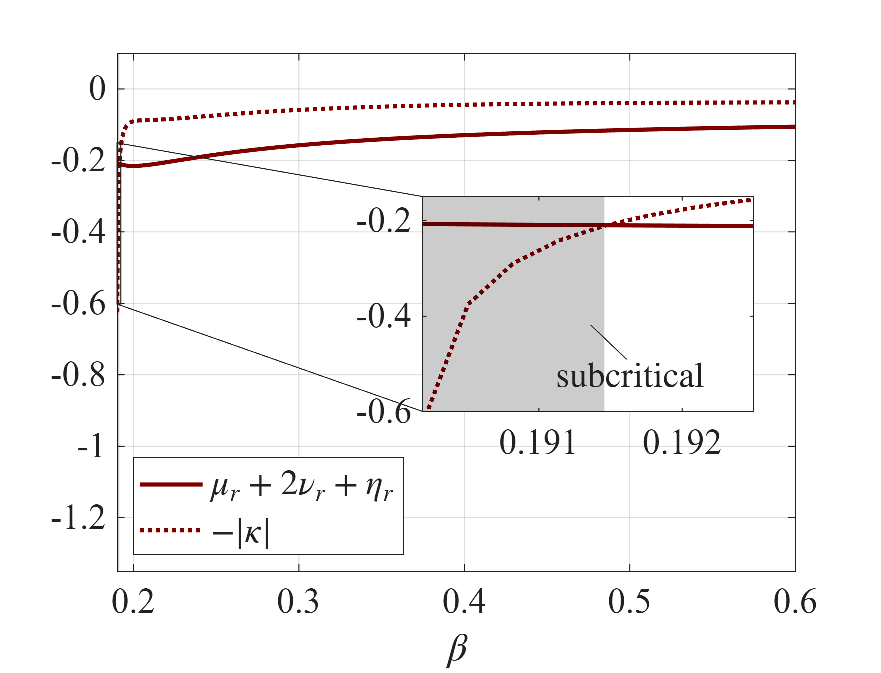}}
            \end{subfigure}
     \caption{Weakly nonlinear coefficients in the range of $\bet$ for which the bifurcation is of Hopf type (i.e., for which $|\om| \neq 0$ in figure~\ref{fig:lindr_b}), and for $D_R=0.02$. 
     The value of $\bet$ at which the horizontal axis begins is that highlighted by a thin vertical line in figure \ref{fig:copitch}.
     Note the appearance of the coefficients $\eta$ and $\kap$, undefined in the pitchfork bifurcation region. 
     The left frames show the real parts of $\mu$, $\nu$ and $\eta$, whereas the right frames show $\mu_r+2\nu_r+\eta_r$ together with $-|\kap|$. According to (\ref{eq:supcond}), whenever the latter is below the former (highlighted by a grayed zone), the phase of the coefficient $\kap$ can be adjusted to make the bifurcation subcritical. Otherwise, the bifurcation is supercritical regardless of the phase of $\kap$.} \label{fig:coeffsHopf_a}
\end{figure}

In figure~\ref{fig:coeffsHopf_a}, for $D_R=0.02$, we show the weakly nonlinear coefficients as a function of the swimming speed $\bet$ (in the range corresponding to the Hopf bifurcation region, and thus the $\bet$-axes begin at the values marked by vertical lines in figure~\ref{fig:copitch}). 
The panel on the right shows the existence of a small interval $\bet \in [0.190,0.192]$ where the condition (\ref{eq:supcond}) is not met. It is highlighted in gray and begins at the threshold value between the pitchfork/Hopf region. Within this interval, the bifurcation can be rendered subcritical by adjusting the phase of $\kap$. From the left panel, we conclude that this subcriticality is enabled by the positive contribution of the coefficient $\eta_r$ to the sum $\mu_r+2\nu_r+\eta_r$ over this interval. The coefficient $\eta_r$ embeds the $2$-wave nonlinear interactions between the pair of eigenmodes with the same wavenumber, but propagating in opposite directions. Furthermore, $|\kap|$ takes particularly large values over this same interval, which also makes the condition (\ref{eq:supcond}) difficult to achieve. Beyond this interval, the bifurcation remains supercritical for all $\bet$ considered. 
We have found that decreasing $D_R$ enlarges the interval of $\bet$ over which the bifurcation is subcritical (not shown). For $D_R=10^{-4}$, this interval is $\bet \in [0.250,0.265]$.

For the rest of this section, we set $D_R=0.02$ (as in the previous section) and $\bet = 0.5$. This corresponds to the supercritical regime in figure~\ref{fig:coeffsHopf_a} (right panel) and thus, past the onset of the bifurcation, both $\opr$ and $\ops$ exist, and the system is bistable. The numerical values of the weakly nonlinear coefficients are reported in Table~\ref{tab:coeffpitch}.   

By deriving from (\ref{eq:ampeq_H}) the evolution equations for the phases of the amplitudes, the WNL calculations predict $\opr$ and $\ops$ to oscillate in time at the frequency $\wnl$ (chosen positive by convention) and such that
\begin{equation}
\begin{split}
\wnl \coloneq  |\om + \underbrace{\e^2 \pae{\mu_i + \eta_i + 2\nu_i + \sin(\dee)\kap_r}\He^2 }_{\text{leading-order nonlinear frequency correction}}|, 
\label{eq:wpo}
\end{split}    
\end{equation}
where, once again, $\sin(\dee)=1$ for $\opr$ and $\sin(\dee)=-1$ for $\ops$ (which also change $\He$), respectively. Consequently, as soon as $\kap_r \neq 0$ (e.g., in Table~\ref{tab:coeffpitch}), the two periodic orbits $\opr$ and $\ops$ have distinct frequencies, with a difference growing $\propto \e^2$.

\begin{figure}
\centering
\scalebox{0.44}{\includegraphics[trim=0cm 3.5cm 0cm 0cm]{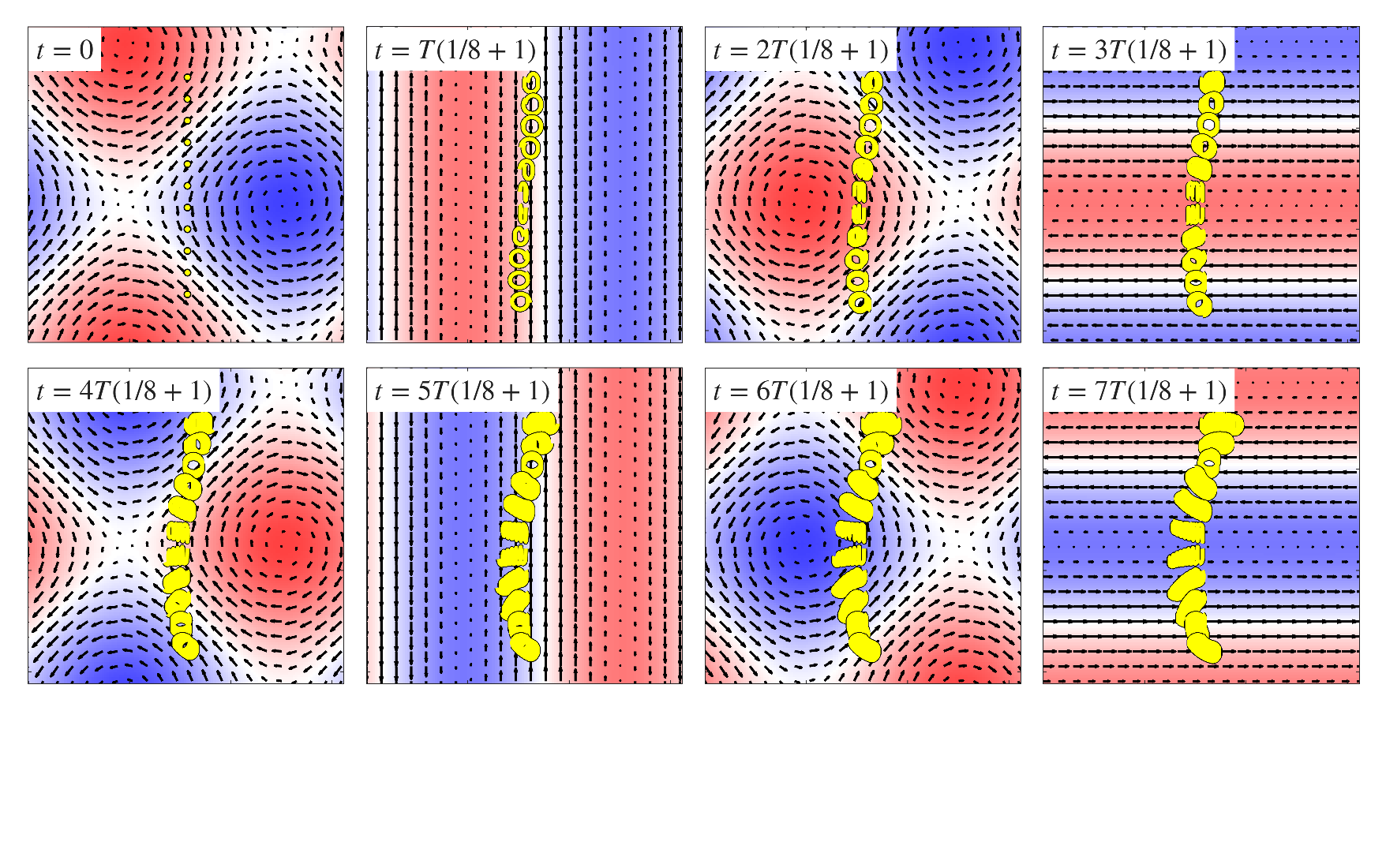}}
     \caption{For $\bet=0.5$ (motile particles, Hopf bifurcation region) and $\e^2 =0.0025$, snapshots of the first stable periodic orbit predicted by the system of amplitude equations (\ref{eq:ampeq_H}) and reconstructed by then evaluating (\ref{eq:q1c4H}). It is named ``$\opr$" to emphasize that the pattern ``reconfigures" over space within a temporal period. 
     The velocity field of the fluid (arrows) and corresponding vorticity (colormap) are shown over the doubly periodic square domain $(x,y) \in [0,2\upi] \times [0,2\upi]$, and the time frame $0 \leq t \leq 7T(1/8+1)$, with $T=2\pi/\wnl$ the (nonlinear) period of the limit cycle shown in figure~\ref{fig:BifHopf}. 
     The snapshots are taken at uniformly spaced times within the time frame, chosen so that the dynamics shown are also those over a single period. Specifically, eight snapshots taken at $t \in \set{(0,1,2,3,4,5,6,7)T/8}$ would yield the same figure for the Eulerian fields. 
     Nonetheless, choosing $7T(1/8+1)$ as the final time permits larger excursions of the Lagrangian trajectories of passive tracers (yellow lines), thus revealing a possible nonlinear drift.} \label{fig:tw1}
\end{figure}

Figure~\ref{fig:tw1} shows velocity snapshots of the periodic orbit $\opr$. Snapshots are spaced uniformly in time, so that the intrinsic dynamics within an oscillation period are visible. They were obtained by simulating the system of amplitude equations (\ref{eq:ampeq_H}) (with $\phi=0$) in the equilibrium regime corresponding to $\opr$, and then reconstructing the first-order solution according to (\ref{eq:q1c4H}).
Passive tracers are also included and reveal some Lagrangian trajectories. In the first snapshot at $t=0$ (an arbitrary reference time in the periodic cycle), the orbit $\opr$ corresponds to macroscopic vortices. Within the next eighth period, it transitions to alternating anti-parallel jets in the vertical direction. Then, within another eighth period, it transitions back to the vortical pattern, although the latter is shifted by $\upi$ in the vertical direction with respect to that at $t = 0$. This process repeats in the horizontal direction within the next quarter period, then a second time in the vertical direction, but with opposite sign (compare the second snapshot to the sixth). It is then repeated a second time in the horizontal direction, but with opposite sign (compare the fourth snapshot to the eighth), which concludes a period.

This explains why we named this periodic orbit ``$\opr$'': to emphasize that the pattern reconfigures over time. The Lagrangian trajectories, evolving over eight periods in the figure, exhibit a clear ``Stokes drift" over time; it is symptomatic of both the ``traveling'' aspect of the pattern and the nonlinearities in the amplitude equation system.   

\begin{figure}
\centering
\scalebox{0.44}{\includegraphics[trim=0cm 3.5cm 0cm 0cm]{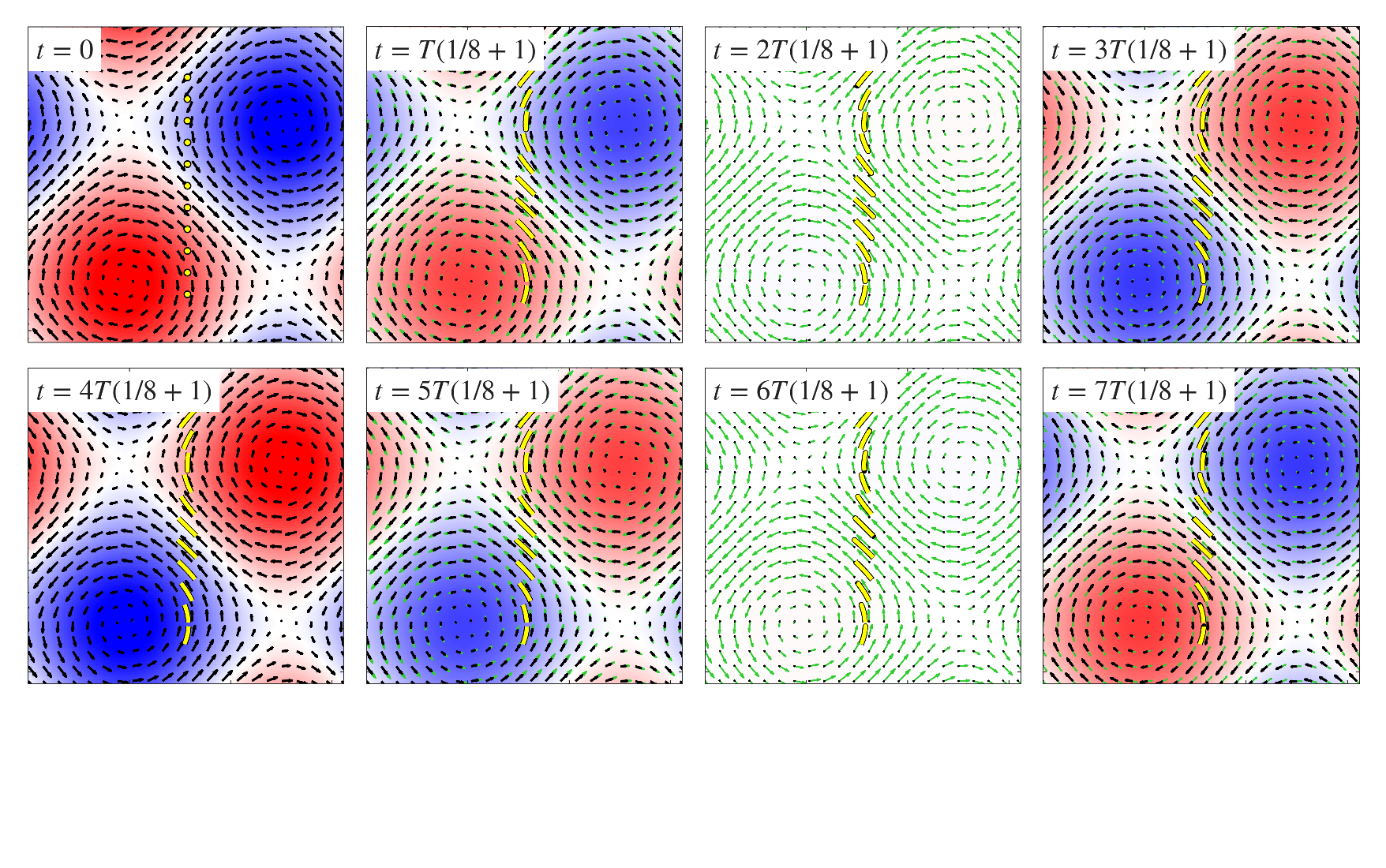}}
     \caption{Same as in figure~\ref{fig:tw1} but for the second co-existing stable periodic orbit predicted by the system of amplitude equations, named ``$\ops$" to emphasize the ``standing" nature of the pattern in space within a temporal period. The green arrows, plotted beneath the black arrows, represent the polarization vector field. } \label{fig:tw2}
\end{figure}

While figure~\ref{fig:tw2} is similar to figure~\ref{fig:tw1}, it displays the second stable periodic orbit, $\ops$. 
Rather than transitioning to anti-parallel jets, the macroscopic vortex structure at the reference time $t=0$ is now preserved. However, the velocity field decays in amplitude until it vanishes at every point in space (at least at the leading order) after a quarter period. Following this, the velocity field regrows in amplitude until, at half period, it becomes the same as at $t=0$ but with vorticity of opposite sign. In other words, within a half period, the flow has experienced a ``vortex-reversal" phenomenon, reminiscent of what was observed in the bacterial suspension experiments of \cite{Nishiguchi25} and \cite{PerezPHD2025} (figure~$5.20$ therein). Within the next half period, the flow reverts to its original state. Strikingly, we observe in the third and seventh snapshots in figure~\ref{fig:tw2} that the polar alignment of the particles reaches its maximum (significant) strength whenever the fluid velocity is uniformly zero. Conversely, the polar alignment becomes uniformly zero when the velocity field reaches its maximal strength. Overall, the velocity and polarization vectors are phase-shifted in time by a quarter period.

Still in figure~\ref{fig:tw2}, the Lagrangian trajectories repeat at each period without drift. In other words, while the pattern is modulated in amplitude, it has a ``standing" nature, hence the name $\ops$. This contrasts with $\opr$, and these two periodic orbits show distinct dynamics.      

\begin{figure}
            \begin{subfigure}{0.495\textwidth}
            \centering
            \scalebox{0.44}{\includegraphics[trim=0cm 0cm 0cm 0cm]{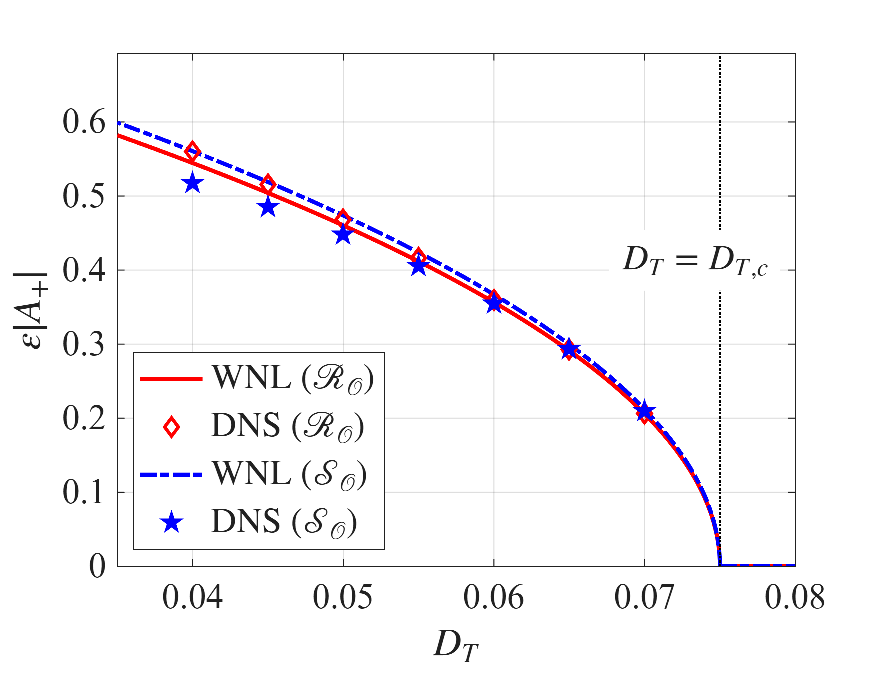}}
             \subcaption{Magnitude}
             \label{fig:BifHopf_a}
            \end{subfigure}
            \begin{subfigure}{0.495\textwidth}
            \centering
            \scalebox{0.44}{\includegraphics[trim=0cm 0cm 0cm 0cm]{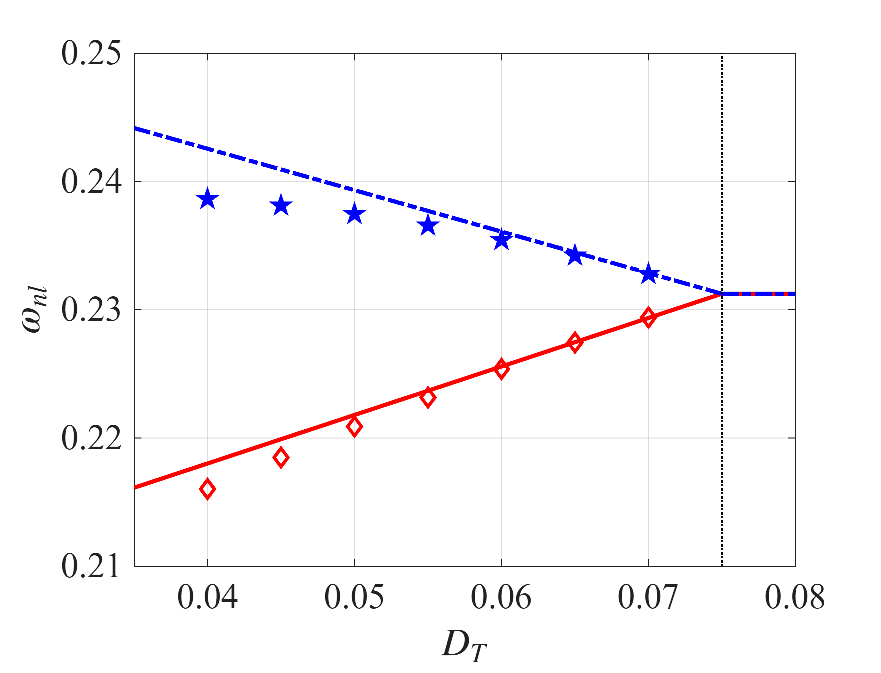}}
             \subcaption{Frequency}
             \label{fig:BifHopf_b}
            \end{subfigure}
     \caption{For $(D_R,\bet)=(0.02,0.5)$, bifurcation diagram of the system (\ref{eq:ec_st}) with $F=0$ (deterministic regime), and by decreasing the translational diffusion parameter $D_T$ below its critical value $D_{T,c}=0.075$. The bifurcation is of Hopf type.
     The fully and weakly nonlinear approaches are compared, both presenting two co-existing stable periodic orbits named ``$\opr$" and ``$\ops$". 
     Their dynamics over a period are distinct, as visible in figure~\ref{fig:tw1} and \ref{fig:tw2}, respectively. 
     Both periodic orbits correspond to slightly different values of the magnitude $H_e$ and the frequency $\wnl$, as given in (\ref{eq:eqhe}) and (\ref{eq:wpo}), respectively. Yet, they are quantitatively distinguished mostly by their corresponding value of $\de$.} \label{fig:BifHopf}
\end{figure}

Deterministic predictions from the amplitude equations are now confronted with their fully nonlinear counterpart. We plot the Hopf bifurcation diagram in figure~\ref{fig:BifHopf}, both for the magnitudes $\e \mAp$ (the equilibrium magnitude is the same for all amplitudes), and the periodic orbit frequencies $\wnl$.

In the fully nonlinear approach, several simulations are run from different initial conditions until steady states are reached. For all values of $D_T$ reported in figure~\ref{fig:BifHopf}, only two stable periodic orbits are found in the DNS, and the system is attracted to one of the two depending on the initial condition; this also reveals that the basins have very different sizes. A Fourier transform in space extracts the $\bka$ and $\bkb$ components, from which we can further extract the amplitudes $\Ap, \Am,$ etc. according to (\ref{eq:Aex}). The corresponding phase combination $\de$ allows us to quantitatively distinguish between $\opr$ and $\ops$.

The DNS and WNL approaches agree well on the magnitudes and frequencies of both periodic orbits. In particular, in the DNS in figure~\ref{fig:BifHopf_b}, the frequencies are indeed found to depart from one another nonlinearly. The two slopes are well predicted by the weakly nonlinear correction in (\ref{eq:wpo}).

We note that the $\e^2$ frequency correction in (\ref{eq:wpo}) appears linear when plotted over $D_T$, due to the scaling $D_T=D_{T,c}-\e^2$. This implies that, for larger $D_T$, the seemingly quadratic behavior of the DNS frequencies can only be captured by correcting (\ref{eq:wpo}) with $O(\e^4)$ terms.   

The coexistence of two stable periodic orbits, captured by the amplitude equations, makes noise-induced transitions possible by activating stochastic forcing. This is presented in the next section. 

\subsection{Stochastic results: dynamical statistics of rare events}
\label{sec:hopf_dys}

Henceforth, we fix the translational diffusivity to $D_T=0.0725$. Since in the Hopf bifurcation region $D_{T,c}=0.075$ (see figure~\ref{fig:lindr_a}), this value of $D_T$ corresponds to $\e=\sqrt{D_{T,c}-D_T}=0.05$ (small indeed). 

\begin{figure}
\centering
\scalebox{0.44}{\includegraphics[trim=2cm 0cm 2cm 0cm]{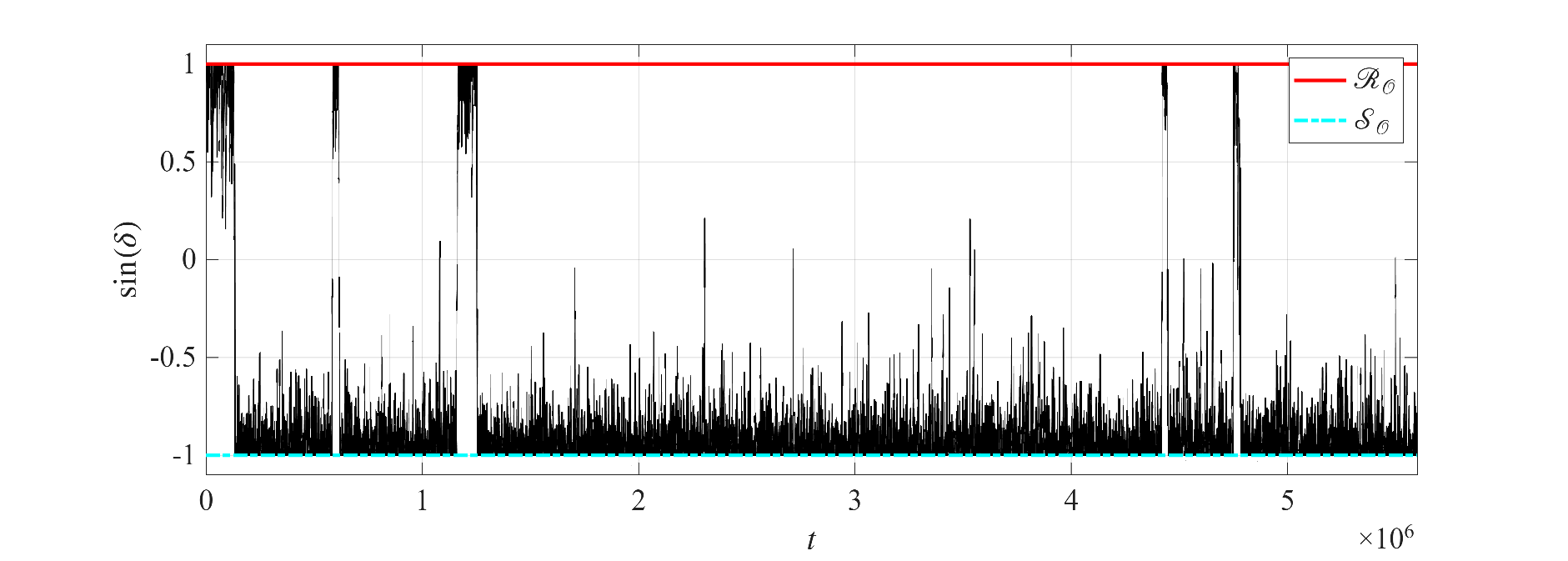}}
     \caption{For $\phi = 0.35$ and $\e^2 = 0.0025$, a weakly nonlinear trajectory (continuous black line) is obtained by simulating (\ref{eq:ampeq_H}) for a realization of the stochastic forcing. Shown is the evolution of $\sin(\de)$ over $t=\tau/\e^2$, exhibiting clear noise-induced transitions between the two attracting values $+1$ (corresponding to $\opr$) and $-1$ (corresponding to $\ops$).} \label{fig:nithopf}
\end{figure}

To illustrate the occurrence of noise-induced transition phenomena between $\opr$ and $\ops$, we simulate the amplitude equation system (\ref{eq:ampeq_H}) for $\phi=0.35$ up to a very large time $t \approx 5.5 \times 10^6$. Running a DNS with the same parameters is now prohibitively expensive. We report the corresponding evolution of $\sin(\delta)$ in figure~\ref{fig:nithopf}. 
Clear transitions of the trajectory between $\opr$ and $\ops$ are visible. These occur at random times, with potentially very long intervals between successive transitions. Strikingly, the trajectory is observed to be around $\opr$ for substantially shorter periods of time than around $\ops$. In other words, the transition $\opr \rightarrow \ops$ is much less rare than the transition $\ops \rightarrow \opr$, indicating that $\opr$ possesses the smaller basin of attraction. 

\begin{figure}
\centering
\scalebox{0.425}{\includegraphics[trim=0cm 0cm 0cm 0cm]{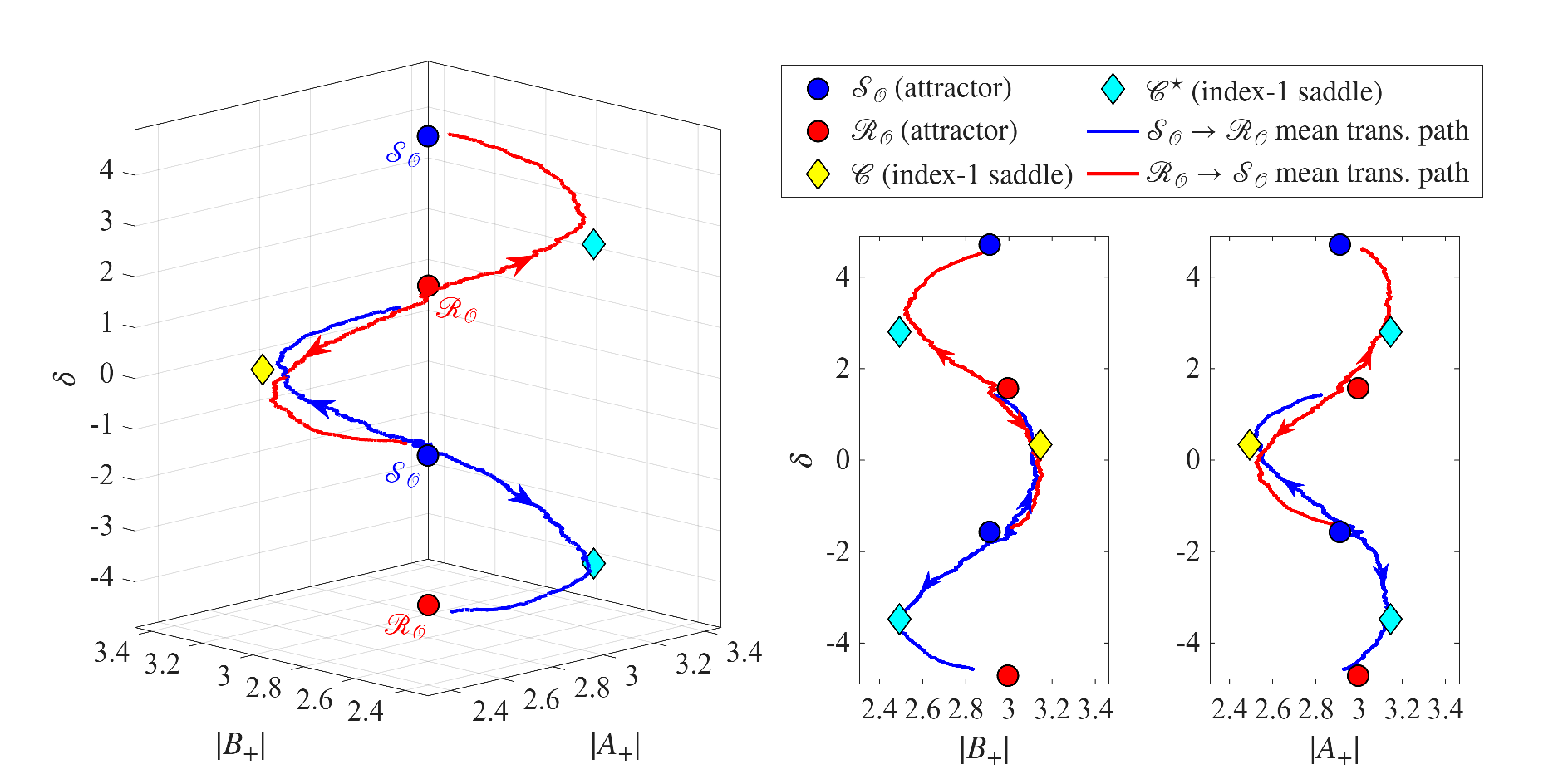}}
     \caption{For $\phi = 0.35$. Weakly nonlinear mean transition paths between the two stable period orbits $\opr$ and $\ops$, and shown in the $(\mAp,\mBp,\de)$ reduced (projected) phase space with $-4.9 \leq \de \leq 4.9$ (left frame); projections in the $(\mBp,\de)$ (middle frame) and the $(\mAp,\de)$ (right frame) are also shown. 
     Dot markers stand for the stable fixed points, whose $z$-coordinates are $\de = \upi/2+2\upi n$ ($n\in \mathbb{Z}$) for $\opr$ (in red) and $\de = -\upi/2+2\upi n$ ($n\in \mathbb{Z}$) for $\ops$ (in blue). 
     The saddle points whose unstable manifold has dimension $1$ (i.e., the ``index-1" saddles) are shown as diamond markers, yellow for $\msC$ and cyan $\msC^{\star} = \Si(\msC)$ (both also having replicas every $2\upi$ in $\de$). 
     The mean transition paths are obtained by applying the AMS algorithm to the system of amplitude equations (\ref{eq:ampeq_H}), then averaging the resulting reactive trajectories after interpolating them onto a common arclength grid. 
     Small arrows indicate the direction of the transition.} \label{fig:mlphopf}
\end{figure}

In figure~\ref{fig:mlphopf}, we study the corresponding average transition paths in the $(\mAp,\mBp,\de)$-projected phase space, for the two transitions  $\opr \rightarrow \ops$ and $\ops \rightarrow \opr$.
These mean transition paths are obtained by applying the AMS algorithm directly to the system (\ref{eq:ampeq_H}) for the amplitudes $\Ap$, $\Am$, $\Bp$, and $\Bm$. The algorithm produces $N_t$ reactive trajectories as an output, each initialized at $\opr$ (resp. $\ops$) and transitioning to the neighborhood of $\ops$ (resp. $\opr$). More precisely, the initial data for the AMS algorithm are such that the four amplitudes have their common equilibrium magnitude $\He$, and three of them are assigned random phases. The initial phase of the fourth amplitude is adjusted for the phase combination $\de$, given in (\ref{eq:dedef}), to be such that $\sin(\de(t=0))=\pm 1$, one of the two equilibrium values; in particular, we assign $\sin(\de(0))=1$ for the trajectories to start at $\opr$ and thus for the algorithm to characterize the $\opr \rightarrow \ops$ transition, and we assign $\sin(\de(0))=-1$ for the algorithm to characterize the $\ops \rightarrow \opr$ transition. In both cases, 
\begin{align}
\varphi[(\Ap,\Am,\Bp,\Bm)](t)\coloneq - \sin(\de(t))\sin(\dee) 
\label{eq:rcoor_Ho}
\end{align}
is chosen as the cost function. This way, the trajectories systematically have $\varphi=-1$ at the initial time, whether they begin in $\ops$ or $\opr$; further, $\varphi$ increases monotonically until the trajectories reach the other orbit, corresponding to $\varphi=1$. The algorithmic parameters are given in Table~\ref{tab:AMSp}, though the number of trajectories is increased to $N_t=10^3$ to obtain a smoother mean path in figure~\ref{fig:mlphopf}. 

In this figure, the stable fixed points $\opr$ and $\ops$ are found every $2\upi$ in $\de$. Importantly, the deterministic system (\ref{eq:sysmodr})-(\ref{eq:eqdel}) also possesses many unstable equilibria. In particular, it possesses the saddle point $\msC$ whose unstable manifold has dimension one (``index-1" saddle), and which corresponds to $\mAp=\mAm=\Hea$, $\mBp=\mBm=\Heb\neq\Hea$ and $\de=\deea$ (plus multiples of $2\upi$). Because the deterministic system is invariant under the involution $\Si$, then $\msC^{\star} \doteq \Si(\msC)$ is also an index-1 saddle point, and by construction has $\mAp=\mAm=\Heb$, $\mBp=\mBm=\Hea$ and $\de=\upi-\deea$ (plus multiples of $2\upi$). We include these two (discrete $2\upi$-periodic families of) saddle points in figure~\ref{fig:mlphopf}.

We note that, in this stochastically forced regime, the system (\ref{eq:sysmodr})-(\ref{eq:eqdel}) for the magnitudes and the phase combination $\de$ should incorporate terms in $\phi^2$, such as in (\ref{eq:bHsys}), and resulting from the It\^{o} stochastic integration of multiplicative noise. This will slightly shift the locations of the stable/saddle points. We believe this difference to be unimportant in the comments below, because $\phi$ is chosen to be small. 

Starting from $\ops$ at $\de=-\upi/2$, the trajectories on average transition to $\opr$ either via an upward path to $\de=\upi/2$, or via a downward path to $\de=-3\upi/2$. In the first scenario, the mean transition path crosses the saddle point $\msC$, and, in the second, it crosses $\msC^{\star}$. We observe that these two possible transition paths are equiprobable. This reflects the fact that they are also observed to be images of each other under the involution $\Si$ (minus $2\upi$), a symmetry of the system. 

By considering the reciprocal transition, starting from $\opr$ at $\de=\upi/2$, the two transition paths to $\ops$ at $\de=-\upi/2$ or $\de=3\upi/2$ also crosses the saddle points $\msC$ or $\msC^{\star}$, respectively. These two mean paths are also found to be equally likely and to be images of each other under the involution $\Si$. However, the saddle-crossing is perhaps less clear than for the $\ops \rightarrow \opr$ transition. 

That the mean transition between two attractors apparently intersects the saddle point whose dimension-one unstable manifolds connect them is consistent with the Freidlin-Wentzell large-deviation theory \citep{Freidlin98} (or even the Arrhenius law for an equilibrium system). The Freidlin-Wentzell theory is valid in the limit of infinitesimally small forcing, and it is not obvious that it should apply here. Indeed, recent examples in the literature show that for small but non-infinitesimal forcing (as here), the saddle points can be bypassed by the mean path \citep{Rolland24, Borner24}. The fact that the saddle points seem slightly closer to the $\ops \rightarrow \opr$ transition path than to its reciprocal could be linked to the greater rarity of the former. 

Another striking feature in figure~\ref{fig:mlphopf} is that the mean transition path of the $\ops \rightarrow \opr$ transition is not the time-reversal of that for the $\opr \rightarrow \ops$ transition. This is particularly evident in the ($\mAp$,$\de$)-projected space (right frame), where the paths diverge significantly in the vicinity of $\msC$. This non-reciprocity of the paths is made possible by the fact that the deterministic system (\ref{eq:sysmodr})-(\ref{eq:eqdel}) does not derive from a potential. The impossibility to construct a potential is due to the three-wave interaction terms in (\ref{eq:sysmodr}). These terms represent energy exchanges between the eigenmodes, the sign of which depends on the phases via $\de$. For instance, when $\de=0$, then $\Ap$ and $\Am$ provide energy to $\Bp$ and $\Bm$, whereas when $\de=\upi$, the opposite occurs. In the language of statistical mechanics, the absence of a potential now makes the dynamics out of equilibrium, and it is expected that the mean paths of reciprocal transitions do not coincide. 

Thus far, the results presented in figures~\ref{fig:nithopf}-\ref{fig:mlphopf} are predictions obtained from the weakly nonlinear method. We now compare these predictions with results obtained by applying the AMS algorithm to the fully nonlinear system, with the parameters provided in Table~\ref{tab:AMSp}. The weakly nonlinear initial conditions on the amplitudes are extended to the full-dimensional space by evaluating (\ref{eq:q1c4H}). 

\begin{figure}
            \begin{subfigure}{0.495\textwidth}
            \centering
            \scalebox{0.44}{\includegraphics[trim=0cm 0cm 0cm 0cm]{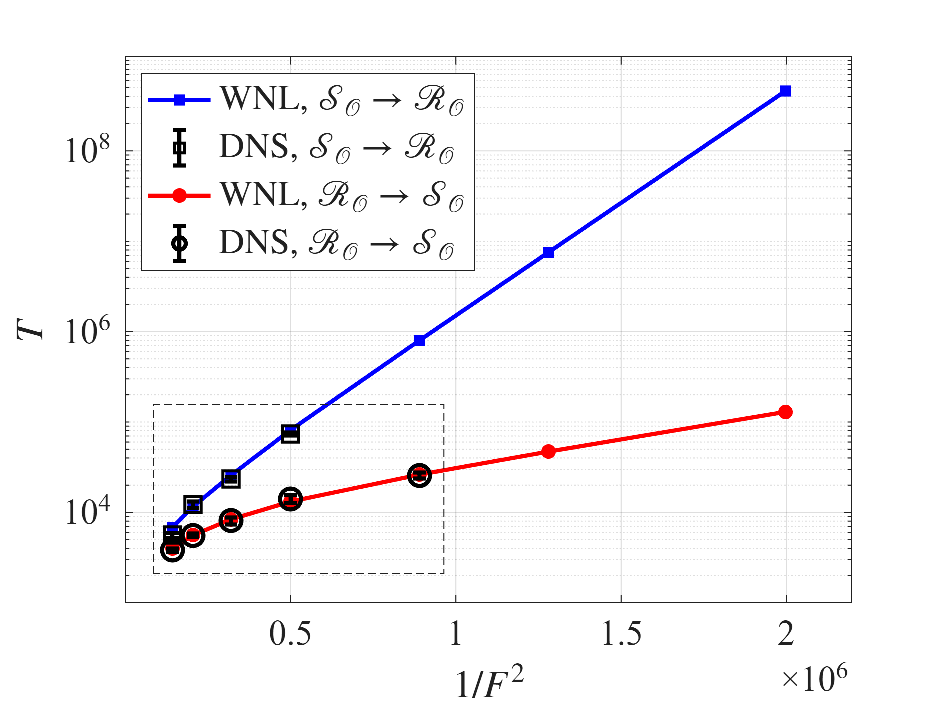}}
             \subcaption{log-lin}
              \label{fig:retHopf_a}
            \end{subfigure}
            \begin{subfigure}{0.495\textwidth}
            \centering
            \scalebox{0.44}{\includegraphics[trim=0cm 0cm 0cm 0cm]{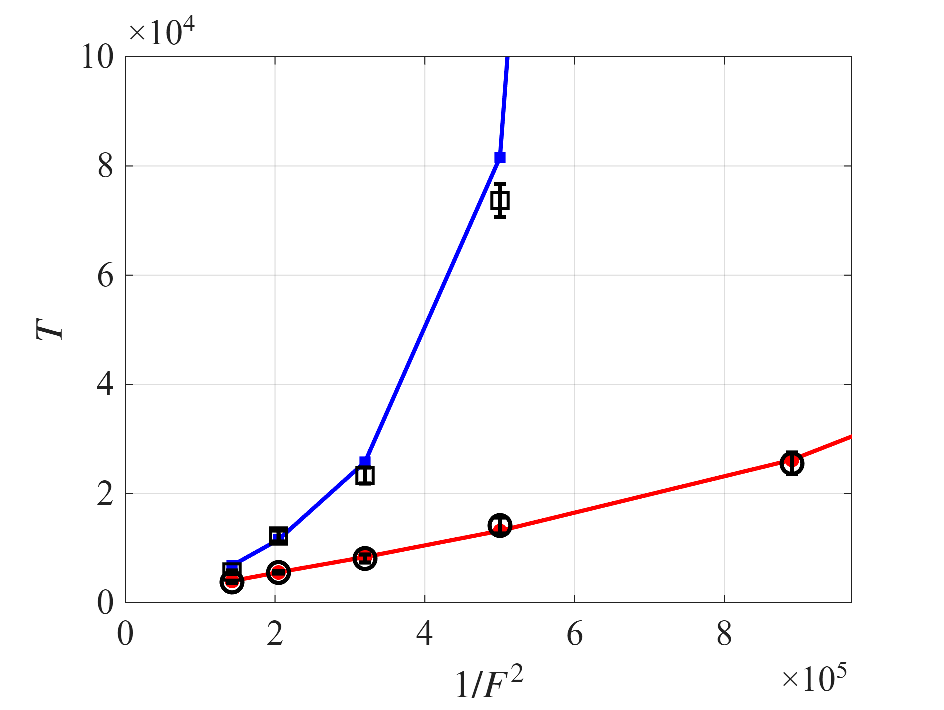}}
             \subcaption{zoomed, lin-lin}
             \label{fig:retHopf_b}
            \end{subfigure}
     \caption{For $\e^2=0.0025$. Mean transition times $T$ of the noise-induced transitions between the two stable periodic orbits $\opr$ and $\ops$.
     The AMS algorithm was used for both the fully and the weakly nonlinear approaches. 
     The two transition directions $\opr \rightarrow \ops$ (plain square for the WNL, empty square with error-bar for the DNS) and $\ops \rightarrow \opr$ (plain circle for the WNL, empty circle with errorbar for the DNS) are considered.
     Panel (b) proposes a zoomed view on lin-lin scale of the dashed frame in (a).} \label{fig:retHopf}
\end{figure}

The mean transition times associated with the two noise-induced transitions, $\ops \rightarrow \opr$ and $\opr \rightarrow \ops$, are reported in figure~\ref{fig:retHopf}.
Recall that the mean transition time $T$ is the average time required for a trajectory that starts in the vicinity of the first orbit to reach the vicinity of the second. For both transitions, the WNL approach predicts that the mean return times are exponential when plotted against $1/F^2$ and in the limit of $F \rightarrow 0$. This is consistent with the large deviation theory.

The mean transition time for the $\ops \rightarrow \opr$ transition is systematically larger than for $\opr \rightarrow \ops$. Thereby, if both periodic orbits are possible states in which to find the system, it is always more likely to find it in $\ops$ (the ``standing'', vortex-reversal periodic orbit) than in $\opr$; furthermore, as $F\to 0$, it is more likely by an exponential factor.

The mean transition times computed from the fully nonlinear model agree very well with the WNL predictions. Each estimate of these fully nonlinear mean return times required about a week of CPU time, and we could not afford to seek times larger than $\sim 10^5$. Plotting the same data in a lin-lin scale in figure~\ref{fig:retHopf_b} highlights slight discrepancies between the two approaches, particularly for the $\ops \rightarrow \opr$ transition and small $F$, corresponding to the largest mean transition times. It is plausible that these mismatches have a numerical/algorithmic origin and disappear as $\di t$ decreases and/or $N_t$ increases, but testing these hypotheses would be numerically too costly. It is also possible that these small mismatches originate from the asymptotic procedure itself, in particular from the neglected higher-order multiplicative noises derived in (\ref{eq:multi2}), similarly as in figure~\ref{fig:epseff}. In the latter scenario, decreasing $\e$ is the only way to improve the agreement. 

\begin{figure}
            \begin{subfigure}{0.495\textwidth}
            \centering
            \scalebox{0.44}{\includegraphics[trim=0cm 0cm 0cm 0cm]{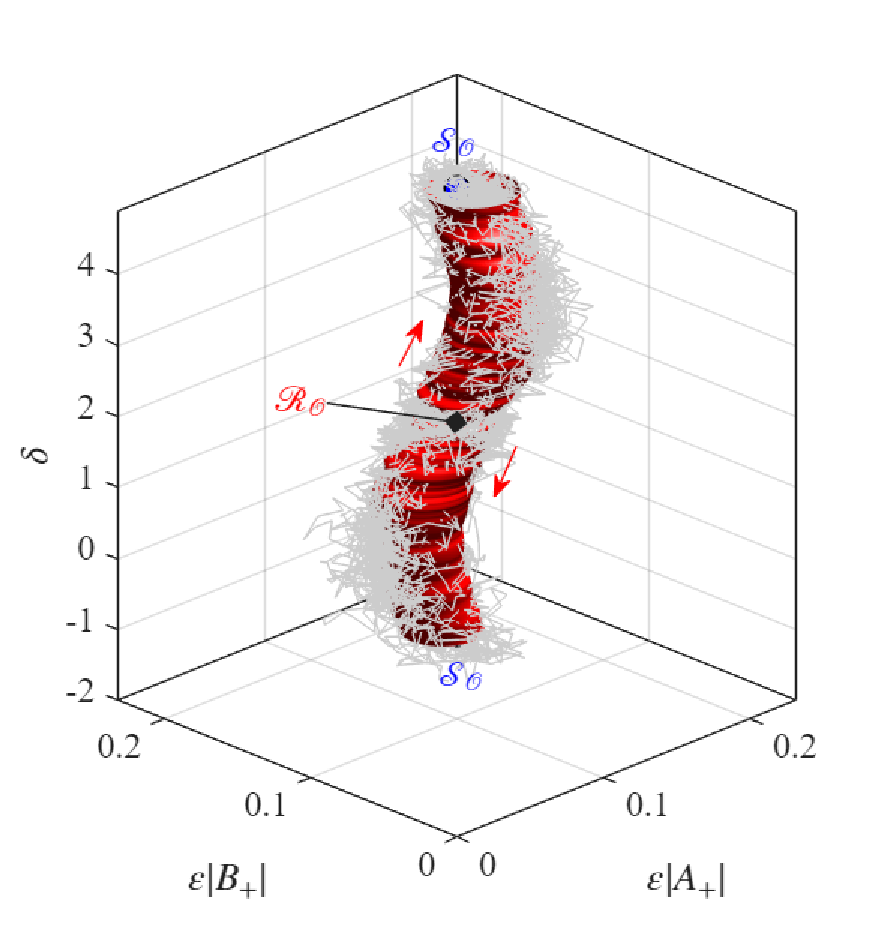}}
             \subcaption{Weakly nonlinear}
            \end{subfigure}
            \begin{subfigure}{0.495\textwidth}
            \centering
            \scalebox{0.44}{\includegraphics[trim=0cm 0cm 0cm 0cm]{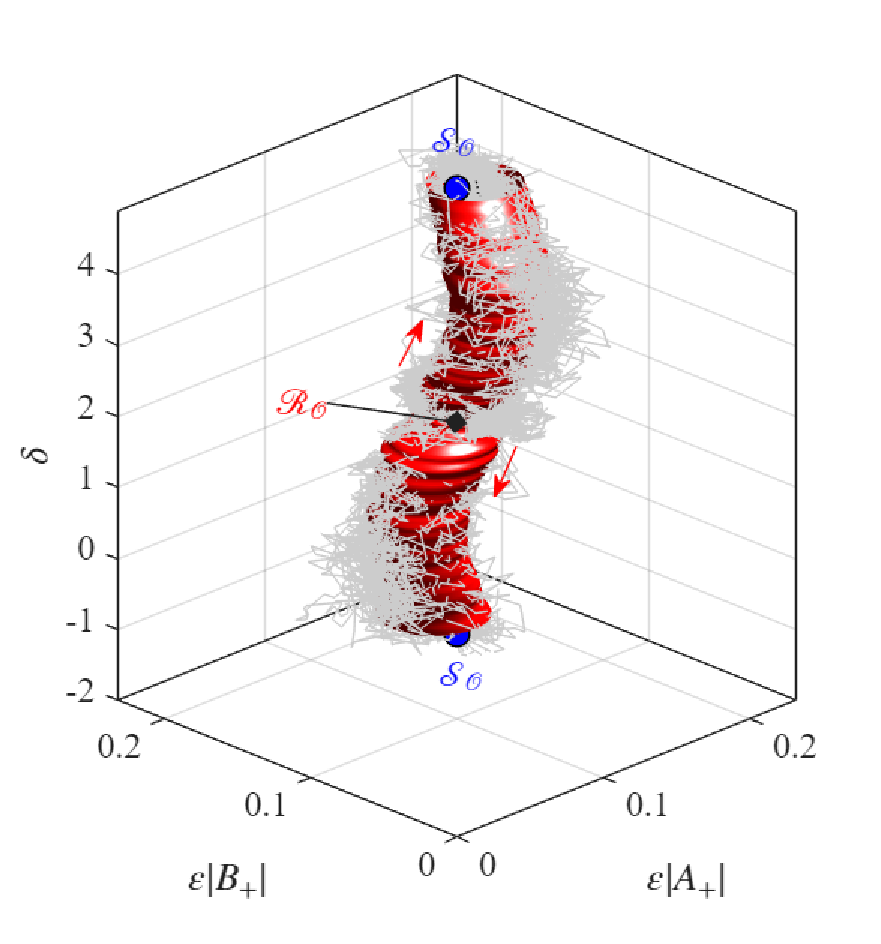}}
             \subcaption{Fully nonlinear}
            \end{subfigure}
     \caption{For $\e^2=0.0025$ and $\phi = 0.42$.
     In the $(\e\mAp,\e\mBp,\de)$ projected phase space, are shown the $80\%$-percentile probability tube in the amplitudes $(\e\mAp,\e\mBp)$ (red surface) of the $\opr \rightarrow \ops$ reactive trajectories.
     In particular, starting at $\opr$ at $\de = \upi/2$, two equiprobable transition paths lead to $\ops$ either at $\de=3\upi/2$ or at $\de = -\upi/2$. 
      By definition, at each resampled arc‑length position $s\in[0,1]$ ($s=0$ corresponding to $\opr$ and $s=1$ to $\ops$), the tube radius is such that $80\%$ of the trajectories lie within that radius when projected onto the $(\e\mAp,\e\mBp)$ coordinates.
     These trajectories have been obtained by using the AMS algorithm on both (a) the fully nonlinear and (b) weakly nonlinear approaches, and $100$ of them are shown as the thin gray lines. 
 } \label{fig:tub1}
\end{figure}

\begin{figure}
            \begin{subfigure}{0.495\textwidth}
            \centering
            \scalebox{0.44}{\includegraphics[trim=0cm 0cm 0cm 0cm]{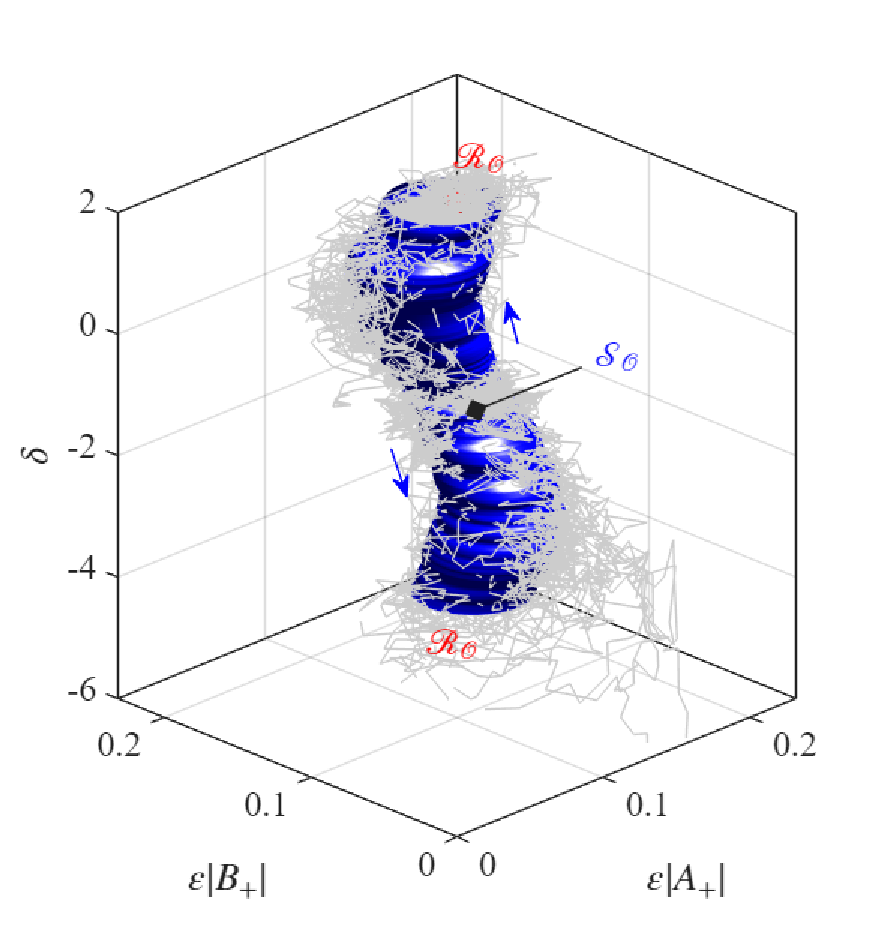}}
             \subcaption{Weakly nonlinear}
            \end{subfigure}
            \begin{subfigure}{0.495\textwidth}
            \centering
            \scalebox{0.44}{\includegraphics[trim=0cm 0cm 0cm 0cm]{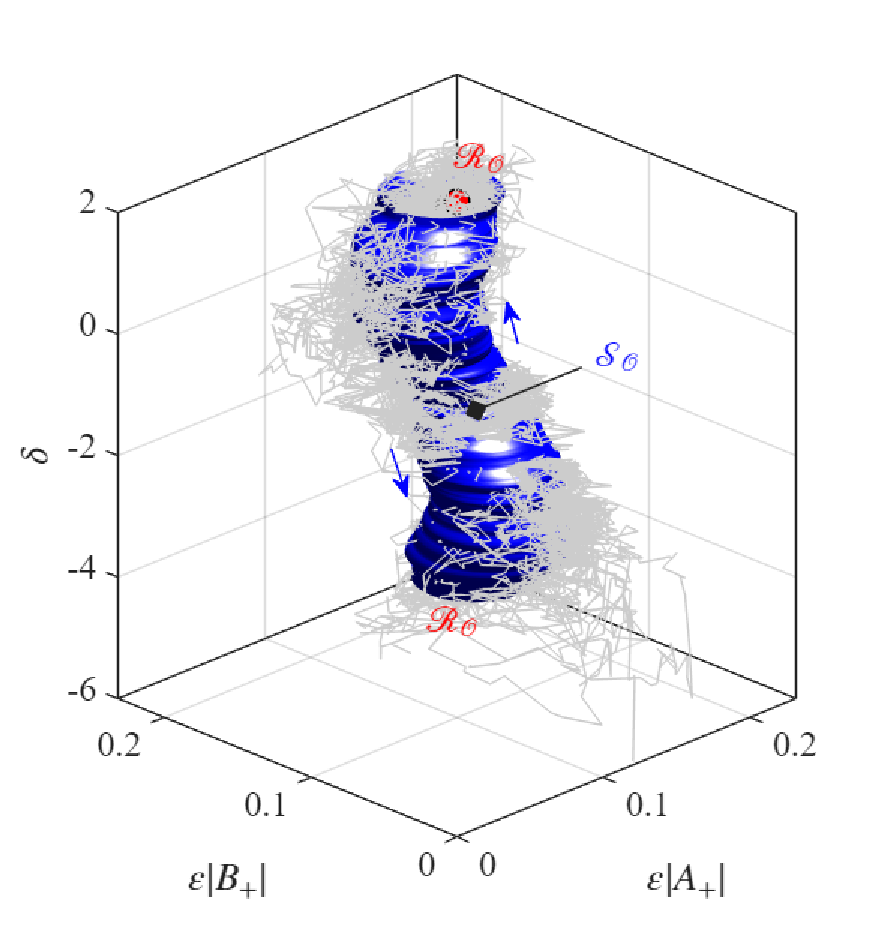}}
             \subcaption{Fully nonlinear}
            \end{subfigure}
     \caption{Same as figure~\ref{fig:tub1} but for the $\ops \rightarrow \opr$ transition at $\phi = 0.56$. Starting at $\ops$ at $\de = -\upi/2$, two equiprobable transition paths lead to $\opr$ either at $\de = \upi/2$ or at $\de= -3\upi/2$. } \label{fig:tub2}
\end{figure}

We further pursue our comparison by showing in figures~\ref{fig:tub1} and \ref{fig:tub2} the $80\%$-percentile probability tube of the transition paths for the $\opr \rightarrow \ops$ and $\ops \rightarrow \opr$ transitions, respectively. Some reactive trajectories are also included. The same $(\mAp,\mBp,\de)$ projected phase space as in figure~\ref{fig:mlphopf} is considered. The corresponding forcing amplitude differs between each figure, as it is the one that yields the largest mean transition time reported in figure~\ref{fig:retHopf} for the fully nonlinear system. 
The probability tubes obtained from both approaches are in excellent agreement, and for both transition directions. From figure~\ref{fig:mlphopf} we understand the probability tubes are curved because, on average, reactive trajectories cross the index-$1$ saddle points between two attractors, and these saddles have $\mAp \neq \mBp$. 

This good agreement, together with figure~\ref{fig:retHopf}, suggests that, at least for the $\e$ considered, the system of amplitude equations can accurately reconstruct the entire flow map of the leading order nonlinear dynamics. Indeed, the reduced system not only correctly locates the stable fixed points, but also the saddle/unstable points, and how their manifolds connect to bring about noise-induced transitions.

\section{Summary and perspectives}
\label{sec:conc}

\subsection{Summary}

We consider the coarse-grained DSS model for a dilute suspension of swimming, rod-like particles, subject to additive white-noise forcing. The model is described by the first three orientational moments of the particle density, i.e., the particle concentration field, the polarity vector field, and the second moment tensor field. The evolution equation for each of these fields is further coupled with the Stokes equations for the fluid velocity. This coarse-grained model, based on the generalized Bingham closure, is thermodynamically consistent with respect to the DSS model \citep{Weady22}.

The uniform, isotropic solution is shown to become unstable by decreasing the strength of the transitional diffusivity, the bifurcation being of either pitchfork or Hopf type depending on the swimming velocity of the particles. In either way, the weakly nonlinear, post-bifurcation dynamics are spanned by a handful of eigenmodes with $\nn{\bk}=1$. Therefore, we reduce the system there to its low-dimensional slow manifold. This reduction persists in the presence of stochastic forcing, provided that the latter is sufficiently weak. 

This results in a system of simple ODEs for the amplitudes of the bifurcating eigenmodes. The deterministic part of the system is derived classically by canceling secular terms $\propto t$ emerging in the expansion. The stochastic part is derived analogously by canceling terms whose standard deviation grows as $\propto \sqrt{t}$. In this way, the additive and multiplicative stochastic forcing acting on the reduced system do not result from \textit{ad hoc} considerations, but rather from a formal treatment. 


The deterministic steady states predicted by the amplitude equations are in excellent quantitative agreement with those computed directly from the coarse-grained equations. That holds for both the codimension-$2$ pitchfork and codimension-$4$ Hopf bifurcations. While the pitchfork bifurcation is always supercritical, the Hopf bifurcation is subcritical over a particular range of particle swimming speeds. 

The weakly nonlinear approach also yields the correct steady density for the magnitudes of the amplitudes over stochastically forced trajectories. In the pitchfork case, the leading-order dynamics are found to derive from a potential. This is not apparent from the original full system, but is revealed by the weakly nonlinear expansion. The fact that an original active matter system can still exhibit equilibrium dynamics, when represented in a few judicious reaction coordinates, is in line with \cite{Nardini23}. By contrast, in the Hopf bifurcation region, no such potential can be constructed, and the weakly nonlinear dynamics are fundamentally out of equilibrium. 

Having established the utility of the amplitude equations in the deterministic regime, we then use them to study rare events. In the pitchfork case, the rare phenomenon of interest is a sudden, large drift of the coherent spatial pattern in a short amount of time (``phase-slip" event). This event is enabled by the trajectory of at least one of the two amplitudes entering a small disk around the origin of the complex plane, where the potential reaches a local maximum (hence the rarity of the event). Because of the low dimensionality of the amplitude equation system, it is possible to compute the mean return time of such an event directly from the associated Fokker-Planck equation. For comparison with the fully nonlinear coarse-grained model, we use the AMS rare event algorithm. These respective approaches, applied to their respective systems, are found in excellent agreement. 


Our analysis also makes clear that motile particles experience a larger effect of stochastic forcing than their immotile counterparts. That is because the linearized operator is more non-normal in the motile case, as motility activates off-diagonal terms by coupling each moment to the higher one. 
This considerably reduces the mean return time for a given forcing intensity. This is in line with the conclusions of \cite{Troude25}, who also find that a greater degree of non-normality is associated with a much reduced transition time via a renormalization of the noise intensity.  

In the case of supercritical Hopf bifurcation, the nonlinear coupling between four co-bifurcating eigenmodes results in two stable coexisting orbits. Each orbit corresponds to a distinct collective state with its own intra-period dynamics. One of the orbits exhibits a time-periodic alternation of macroscopic vortices and anti-parallel jets. The second orbit exhibits a vortex-reversal phenomenon, in which the macroscopic circulation switches periodically between clockwise and counterclockwise directions; between these two states, the system achieves a state of zero (leading-order) velocity. The presence of stochastic forcing enables noise-induced transitions between these two stable collective states, with the predominance of time spent in the second one. 

In comparison with the full coarse-grained DSS model, the amplitude equations give accurate predictions regarding the mean transition times between these two states, at a considerably lower numerical cost. Furthermore, the relative simplicity of the amplitude equation description allows us to understand the geometry of the transition and the importance of saddle states.


\subsection{Perspectives}

We believe that our approach offers perspectives, some of which are proposed below. 

In this article, we consider a two-dimensional periodic box, which quantizes the admissible wavenumber vectors. In the analogous three-dimensional setting, the Hopf bifurcation is of codimension-$6$, which requires including more waves in the weakly nonlinear expansion. Moreover, as the spatial domain becomes very large, an increasing number of eigenmodes at $\nn{\bk}=1$ need to be included in the weakly nonlinear developments, all of them interacting nonlinearly, as their damping rates become asymptotically small. 

Obtaining a system of amplitude equations of higher dimension could capture even richer dynamics. In particular, it could propose a theoretical framework for wave turbulence in active matter systems, which has been reported in many experimental studies \citep{Wensink12}. A rich, hysteretic dynamics could also be found by expanding the amplitude equations to higher orders in the Hopf subcritical regime. 

It is challenging to do by hand the weakly nonlinear calculations that aim to include more eigenmodes and/or higher-order terms. However, the deterministic parts of the amplitude equations can be deduced from symmetry considerations alone. It is the calculation of the coefficients pre-multiplying each monomial that made the developments above complicated. To illustrate this, let us consider the Hopf bifurcation and determine all possible nonlinear combinations of the eighth waves, $\Ap$, $\Am$, $\Bp$, $\Bm$ and complex conjugates, that feed back on the wave $\Ap$. These combinations solve the simple system
\begin{equation}
\begin{split}
\begin{bmatrix}
\bka & \bka & \bkb & \bkb & -\bka & -\bka & -\bkb & -\bkb \\
 \om & -\om & \om & -\om  & -\om &  \om & -\om & \om \\
\end{bmatrix} \boldsymbol{c} = \begin{bmatrix}
 \bka   \\
 \om
\end{bmatrix}
\label{eq:wcomb}
\end{split}    
\end{equation}
where each $c_j$, with $j=1,...,8$, is a positive or null integer. Each solution of the linear matrix-vector system (\ref{eq:wcomb}), and there are an infinite number of them, is associated with a monomial
$\prod_{j=1}^{8} \bLac_j^{c_j}$
to be included in $f_{\Ap}$. In particular, solving under the constraint $\sum_j c_j = n$ yields all the monomials to be included at order $n$. By choosing $n=4$, for instance, system (\ref{eq:wcomb}) has no solutions, from which we deduce that $f_{\Ap}$ in (\ref{eq:ffka}) should not be completed with fourth-order terms. Choosing $n=5$, we found $18$ solutions to (\ref{eq:wcomb}), for example, $\boldsymbol{c}=(0,1,2,0,0,0,1,1)$, corresponding to $\Am\Bp|\Bp|^2\Bm^*$.

System (\ref{eq:wcomb}) can easily be generalized to include more eigenmodes, yielding the form of the amplitude equations for an arbitrary number of modes and up to an arbitrary order (although the shape of higher-order noise terms would also need to be deduced in a systematic manner). From here, the coefficients could be fitted to fully nonlinear trajectories or experimental data, for instance, using the maximum likelihood estimation technique.


Having a systematic way to construct amplitude equations from kinetic (e.g., DSS) or coarse-grained theories, and to infer the coefficients from pre-existing data, could also shed light on the collective nonlinear dynamics in confined (non-periodic) experimental geometries \citep{Lushi14, Nishiguchi25, PerezEstay25}. Among them, the dilute bacterial suspension experiments in a cylindrical cavity performed in \cite{PerezEstay25, PerezPHD2025} exhibit a very rich phenomenology. 

Past the onset of the bifurcation threshold, the authors report the coexistence of a quasi-two-dimensional, coherent vortex state whose size is set by cylinder radius (the largest scale available), with a three-dimensional chaotic state whose characteristic size is that of the smaller, vertical confinement (see \cite{PerezPHD2025}, figure~5.7.b therein). Furthermore, the bacterial system exhibits sporadic transitions between these two states, after very long and possibly random times (see \cite{PerezPHD2025}, figures~5.17 and 5.18 therein). 

It would be experimentally very challenging to produce the statistics of these large transition times, which may motivate a theoretical, reduced-order approach such as that adopted here. Indeed, it is plausible that the coexistence of these two states is captured by the coarse-grained DSS model (\ref{eq:ec_st}) in a three-dimensional, confined cylindrical geometry. Encouragingly, the authors of \cite{PerezPHD2025} report a time-periodic, time-reversal state in figure~5.20, achieved by increasing viscosity. This experimental state is evocative of the state $\ops$ predicted by our calculations (in dimensional units, the vortex size of the latter is also proportional to the domain size).   

Identifying the handful of dominant post-bifurcation eigenmodes from the dispersion relation of the coarse-grained DSS model in a cylindrical geometry, and then applying the above symmetry-based reasoning, will result in a system of nonlinearly coupled amplitudes for these modes. The related coefficients and stochastic forcing intensity can then be inferred from experimental data. By the means advanced in this article, such a system can be used to compute statistics of rare transitions between the coherent vortex and the chaotic attracting states, and possibly many more metastable ones not observed over the short experimental time scales.

\

\begin{bmhead}[Supplementary data]
Supplementary movies are available online.
\end{bmhead}

\begin{bmhead}[Acknowledgements]
YMD is grateful to Scott Weady for sharing his code, adapted for the direct numerical simulation of the coarse-grained model with Bingham closure. YMD is also grateful to Edouard Boujo and Fran{\c c}ois Gallaire for insightful discussions. The computations in this work were performed at facilities supported by the Scientific Computing Core at the Flatiron Institute, a division of the Simons Foundation.
\end{bmhead}

\begin{bmhead}[Funding.]
YMD acknowledges support from the Swiss National Science Foundation under Grant No. 225429 (Postdoc.Mobility scheme). MJS acknowledges support from the Simons Foundation. 
\end{bmhead}

\begin{bmhead}[Declaration of interests]
The authors report no conflict of interest.
\end{bmhead}

\begin{appen}

\section{Practical implementation of the linear stability problem}
\label{app:lind}

While solving the generalized eigenvalue problem (\ref{eq:lin1}), the trace and symmetry constraints must be enforced. This can be done using a projection operator, say $\bP_c$. The latter is such that $\bP_c \hbq$ satisfies the trace and symmetry equations even if an arbitrary $\hbq$ does not. For instance, for $d=2$, the projector $\bP_c$ reads
\begin{align}
\bP_c = \bI_{10} - \bee_6\bee_6^T - \bee_7\bee_7^T + \bee_6\bee_5^T + \bee_7(\bee_1-\bee_4)^T \Rightarrow \bP_c \hbq = (\hc,\hn_x,\hn_y,\hQ_{xx},\hQ_{xy},\hQ_{xy},\hc-\hQ_{xx},\hu_x,\hu_y,\hr)^T,
\nonumber
\end{align}
where $\bee_i$ is the $i$th basis vector. Thereupon, if $\bZz_c$ denotes an orthonormal basis matrix for the range of $\bP_c$, the eigenvalue problem (\ref{eq:lin1}) is better written in the reduced subspace as
\begin{eqnarray}
\sigma \bZz_c^H \bD \bZz_c\hat{\overline{\bq}}_1 = \bZz_c^H \bL_{\bk} \bZz_c \hat{\overline{\bq}}_1,
\label{eq:lin1r}
\end{eqnarray}
where $\hat{\overline{\bq}}_1$ is a reduced state vector free of the $(d-1)d/2+1$ component corresponding to dependent fields (e.g., free of the $Q_{yx}$ and $Q_{yy}$ component in the example above). The eigenvector solutions of (\ref{eq:lin1r}) can then be transformed back in the full-dimensional space by application of $\bZz_c$, the trace and symmetry constraints being thus automatically satisfied. 

Due to its very low dimension, the eigenproblem (\ref{eq:lin1r}) can easily be solved numerically even without an explicit expression for $\bL_{\bk}$. In practice, we solve (\ref{eq:lin1r}) using the \texttt{Matlab} pre-implemented function \texttt{eig}, where the operator $\bL_{\bk}$ is constructed explicitly by successive application over basis vectors. 

Importantly, the special case $\bk=0$ requires a special treatment. Indeed, the action of the linear operator in (\ref{eq:Ldef}) becomes 
\begin{align}
\begin{split}
&\bL_{\bz} \hbq_1 = \pae{0, -(d-1)D_R \hbn_1, \vect{-2dD_R \hbQ_1+ 2D_R \bI \hc_1}, \bz, 0}^T.
\label{eq:Ldef_k0}
\end{split}
\end{align}
The equations for velocity and pressure are reduced to trivial equalities and can be removed. It is then easy to show that the system possesses $d$ real eigenvalues equal to $\sig=-(d-1)D_R$, associated with eigenvectors with nonzero component only in $\hbn_1$. It also possesses $d^2$ real eigenvalues equal to $\sig=-2dD_R$, associated with eigenvectors nonzero only in $\hbQ_1$. Eventually, there exists a null eigenvalue $\sig=0$ whose associated eigenvector has a nonzero component in both $\hc_1$ and $\hbQ_1$. However, the latter linearly neutral eigenvector must be ignored by virtue of the conservation of the total number of particles 
\begin{eqnarray}
\int_{\Omega} c^{\e}\di \bx = \int_{\Omega} c_0 \di \bx  + \e \int_{\Omega} c_1 \di \bx + O(\e^2) =  (2\upi)^d, \quad \forall t. \nonumber
\end{eqnarray}
Since $\int_{\Omega} c_0 \di \bx=(2\upi)^d$, collecting powers of $\e$ implies that $\int_{\Omega} c_j \di \bx=0$ for each order $j=1,2,...$. Accordingly, at each order, the concentration field associated with $\bk=0$ must be zero. This implies that the wavenumber vector $\bk=0$ is strictly stable as soon as $D_R>0$, since then the real parts of all physically admissible eigenvalues are strictly negative. 

Let us now comment on the construction of the adjoint eigenbasis. Since the inner product of interest is the Hermitian dot product, then $\bL^{\da}_{\bk} = \bL^{H}_{\bk}$, and the adjoint eigenmodes are solutions of
\begin{eqnarray}
\sig_j^* \bZz_c^H \bD \bZz_c \tilde{\overline{\bq}}^{\da}_j = \bZz_c^H \bL^H_{\bk} \bZz_c \tilde{\overline{\bq}}^{\da}_j, 
\nonumber
\end{eqnarray}
in the reduced (constraints-free) space. Again, multiplying a reduced adjoint eigenmode by $\bZz_c$ transforms it back into the full space while automatically enforcing the trace and symmetry constraints. 

\section{Showing that higher-order terms have a bounded variance}
\label{App:hotn}

Let $\bww$ denote any of the temporal integrals at $O(\e)$ in (\ref{eq:o2ka_3}), then 
\begin{align}
\begin{split}
\bww(t) &\coloneq \int_{0}^{t} \xi_{\bLac}(s) \bi(s) \di s = \bPhi_{2\bka}(t,0)\int_{0}^{t} \xi_{\bLac}(s) \int_{0}^{s}\bPhi_{2\bka}(0,x) \hbd^{\Ap\Am} \di x \di s \\
&\myeq \bPhi_{2\bka}(t,0)\int_{0}^{t} \pae{ W_{\bLac}(t) -  W_{\bLac}(s) } \bPhi_{2\bka}(0,s) \hbd^{\Ap\Am} \di s, \nonumber
\end{split}
\end{align}
where $\di W_{\bLac}/\di t \coloneq \xi_{\bLac}$. Applying $\bPhi_{2\bka}(0,t)$ to the equation above then differentiating with respect to $t$ leads to 
\begin{align}
\begin{split}
&\di_t\pae{\bPhi_{2\bka}(0,t)\bww(t)} =  \xi_{\bLac}(t) \int_{0}^{t}\bPhi_{2\bka}(0,s) \hbd^{\Ap\Am} \di s \Rightarrow \\
& \bPhi_{2\bka}(t,0) \di_t\pae{\bPhi_{2\bka}(0,t)\bww(t)} =  \xi_{\bLac}(t) \bi(t) \Rightarrow (\bD\di_t-\bL_{2\bka})\bww(t) =  \xi_{\bLac}(t) \bi(t), \nonumber
\end{split}
\end{align}
where we have used (\ref{eq:prpp}). In other words, $\bww(t)$ is the linear response to $\xi_{\bLac}(t) \bi(t)$. Since $\bi(t)$ is bounded, and again by virtue of the strict stability of the linear system $2\bka$, we expect $\bww(t)$ to also have a bounded root mean square. The integral terms at $O(\e)$ in (\ref{eq:o2ka_3}) thus do not threaten the asymptotic hierarchy, at least in the root mean square sense. We postulate that all the integral terms absorbed at $O(\e^2)$ in (\ref{eq:o2ka_3}) are also bounded.

\section{Computing the reactive probability current}
\label{app:reac}

The first step is to solve for the committor functions $\qp(\bH)$ and $\qm(\bH)$. The former, $\qp$, is the probability that, starting from $\bH$, the trajectory reaches $\dB$ before it reaches $\dA$; the latter, $\qm$, is the probability that, on the contrary, the trajectory reaches $\dA$ before it reaches $\dB$. As can be found, for example, in \cite{Vanden10} (equations $(18)$ and $(19)$ therein), among others, the committor function $\qp$ is solution of
\begin{equation}
- \nab \qp \bdot \nab V + \frac{(\al \phi)^2}{2} \Delta \qp = 0, \ \ \text{subject to} \ \
 \begin{cases}
\qp = 0  & \text{on}\;  \bA \\
\qp = 1  & \text{on}\;  \boB. \\
\end{cases}
\nonumber
\end{equation}
Since the drift term in (\ref{eq:bHsys}) derives from a potential, the other committor function is immediately given by $\qm = 1-\qp$. 
From the knowledge of $\qp$, $\qm$, and the probability current $\bJ$ appearing in the Fokker-Planck equation (\ref{eq:FP}), the reactive probability current is given by
\begin{equation}
\begin{split}
\bJ_R \coloneq \qp\qm\bJ + \frac{(\al \phi)^2}{2}p_s\pae{\qm \nab \qp - \qp \nab \qm},
\label{eq:jr}
\end{split}    
\end{equation}
which can be shown to be divergence-free. This expression can be found in \cite{Vanden10}, equation $(26)$ therein.

\section{Deriving the same amplitude equations with the multiple-scale method}\label{app:mms}

The same systems of amplitude equations as (\ref{eq:ampeq_H}) and (\ref{eq:ampeq_Pi}), at leading order, with the same expressions for the coefficients and noise processes, could be derived by using the multiple-scale method. This requires introducing the slow time scale $\tau = \e^2 t$ from the beginning, and making the amplitudes depend on $\tau$ only, e.g., $\Ap = \Ap(\tau)$. It is not necessary to postulate (\ref{eq:nf}) from the beginning, because it arises naturally from the calculations. Then, as proposed in \cite{McMullen24} (or in \cite{DucimetiereTH24}, Chapter 6.2 therein), the external stochastic forcing is decomposed into the sum of two contributions, (i) and (ii), according to  
\begin{align}
\begin{split}
\bff(\bx,t) = & \underbrace{ \pae{\zeta_{\Ap}(t)\tbq^{\Ap} + \zeta_{\Am}(t)\tbq^{\Am}}e^{\ti \bka \cdot \bx} + \pae{\zeta_{\Bp}(t)\tbq^{\Bp} + \zeta_{\Bm}(t)\tbq^{\Bm}}e^{\ti \bkb \cdot \bx} +\cc }_{\text{(i): triggers a response linearly fully contained in the slow manifold.}}\\
& +\underbrace{\pae{\bPoa \hbf_{\bka}(t)e^{\ti \bka \cdot \bx} + \bPob \hbf_{\bkb}(t)e^{\ti \bkb \cdot \bx} + \cc }+ \sum_{\bk\notin\set{\bz,\pm \bka,\pm \bkb} }\hbf_{\bk}(t) e^{\ti \bk \bdot \bx} }_{\text{(ii): triggers a response linearly fully contained in the fast manifold.}},
\end{split}
\label{eq:fdec}
\end{align}
where
\begin{align}
\begin{split}
&\zeta_{\bLac}(t) \doteq \ssd{\tbq^{\bLac,\da}}{\hbf_{\bk}(t)}, \quad  \text{and} \quad  \begin{cases}
\bk = \bka & \text{if}\; \bLac \in \set{\Ap,\Am} \\
\bk = \bkb & \text{if}\; \bLac \in \set{\Bp,\Bm}
\end{cases}.
\nonumber
\end{split}
\end{align}
The operators $\bPoa$ and $\bPob$ are oblique projectors onto the eigen-subspaces spanned by the strictly stable eigenmodes for $\bka$ and $\bkb$, respectively, complement of the neutral subspaces. For a given $\bk$, the full expression of the projector is given by (\ref{eq:bpo}). 

Part (i) of the decomposition excites only the neutral eigenmodes, i.e., yields a response fully contained within the slow manifold. This holds only at the linear level, but only linear problems are solved in the asymptotic expansion. Crucially, since the response into the slow manifold is described as a function of $\tau$ (through the amplitudes), the forcing that triggers it must be as well. This can easily be done by using the scaling invariance of white noise, $\zeta_{\bLac}(t) = \e \zeta_{\bLac}(\e^2 t)$. 

Part (ii) of the decomposition excites only the strictly stable eigenmodes, whether these are in the strictly stable eigen-subspace at $\bka$ and $\bkb$, or in the full eigenspace at all the other $\bk$. In other words, it yields a response fully contained within the fast manifold. For that reason, its temporal dependency is kept as is, since, in the spirit of the multiple-scale method, ``$t$'' is meant to capture the fast variations. The forcing becomes
\begin{align}
\begin{split}
\bff(\bx,t) = & \e\sae{\pae{\zeta_{\Ap}(\tau)\tbq^{\Ap} + \zeta_{\Am}(\tau)\tbq^{\Am}}e^{\ti \bka \cdot \bx} + \pae{\zeta_{\Bp}(\tau)\tbq^{\Bp} + \zeta_{\Bm}(\tau)\tbq^{\Bm}}e^{\ti \bkb \cdot \bx} + \cc} \\
& + \pae{\bPoa \hbf_{\bka}(t)e^{\ti \bka \cdot \bx} + \bPob \hbf_{\bkb}(t)e^{\ti \bkb \cdot \bx} +\cc} + \sum_{\bk\notin\set{\bz,\pm \bka,\pm \bkb} }\hbf_{\bk}(t) e^{\ti \bk \bdot \bx}.
\end{split}
\nonumber
\end{align}
In the asymptotic expansion, $\bff$ is pre-multiplied by $\phi \e^2$ and thus the slow component of the forcing acts at $O(\e^3)$, while the fast component acts at $O(\e^2)$. By then following the multiple-scale formalism, the same systems (\ref{eq:ampeq_H}) and (\ref{eq:ampeq_Pi}) are obtained. Since it is unnecessary to write the expansion under integral form (the integrations by parts are implied in the multiple-scale procedure), the calculations are lighter than those proposed in the main text.

\section{Deriving higher-order noise corrections acting on the amplitudes \label{app:xi2}}

The detailed expression of the multiplicative noise processes $\hbpa$, oscillating at $\bka$, and resulting from quadratic interactions of first-order terms with the second-order stochastic responses $\phi \hbq_{2,\bk \in \Ks}^{\phi}(t)$, is given by
\begin{align}
\begin{split}
\hbpa(\bLa,t) =& \pae{\Am^*\hbg^{\phi}_1(t) + \Bp \hbg^{\phi}_2(t) + \Bm^*\hbg^{\phi}_3(t) }e^{\ti \om t} + \pae{\Ap^*\hbg^{\phi}_4(t) + \Bp^*\hbg^{\phi}_5(t) + \Bm\hbg^{\phi}_6(t) }e^{-\ti \om t},
\end{split}
\nonumber
\end{align}
with the additive stochastic processes
\begin{align}
\begin{cases}
 \hbg^{\phi}_1(t) \doteq \bNc{\tbq^{\Am^*},\hbq^{\phi}_{2\bka}(t)} + \bet \hbr^{\Am^*\phi_{\set{2\bka}}}(t),  \quad&\hbg^{\phi}_2(t) \doteq \bNc{\tbq^{\Bp},\hbq^{\phi}_{-\bkb+\bka}(t)} + \bet \hbr^{\Bp \phi_{\set{-\bkb+\bka}}}(t), \\
\hbg^{\phi}_3(t) \doteq \bNc{\tbq^{\Bm^*},\hbq^{\phi}_{\bkb+\bka}(t)} + \bet \hbr^{\Bm^* \phi_{\set{\bkb+\bka}}}(t), \quad&\hbg^{\phi}_4(t) \doteq \bNc{\tbq^{\Ap^*},\hbq^{\phi}_{2\bka}(t)} + \bet \hbr^{\Ap^* \phi_{\set{2\bka}}}(t),\\
\hbg^{\phi}_5(t) \doteq \bNc{\tbq^{\Bp^*},\hbq^{\phi}_{\bkb+\bka}(t)} + \bet \hbr^{\Bp^* \phi_{\set{\bkb+\bka}}}(t),  \quad &\hbg^{\phi}_6(t) \doteq \bNc{\tbq^{\Bm},\hbq^{\phi}_{-\bkb+\bka}(t)} + \bet \hbr^{\Bm \phi_{\set{-\bkb+\bka}}}(t),
\end{cases}
\nonumber
\end{align}
where $\hbr^{\bLac,\phi_{\set{\bk}}}$ results from the moments of $\hch^{\bLac \phi_{\set{\bk}}} \doteq \hps^{\bLac}\hps^{\phi}_{\bk}$. The stochastic processes $\hbg^{\phi}_j(t)$ are history-dependent, as they involve linear responses to the externally applied white noise via the (strictly stable) operators associated with the wavenumbers $\bk \in K$; for that same reason, they are not generically white, but rather colored (low-pass filtered). Furthermore, recall that $\hbq_{\bz}^{\phi}(t)=\bz$ because the spatial mean is unforced, and thus does not contribute to the term above.  From here, the term $\hqpa(\bLa,t)$ appearing in (\ref{eq:o3ka_oe3}), resulting from inverting and integrating by parts, reads
\begin{align}
\begin{split}
&\hqpa(\bLa,t) = \Am^*\int_{0}^{t}\bPhi(t,s)\hbg_{1}^{\phi}(s)e^{\ti \om s} \di s + \Bp\int_{0}^{t}\bPhi(t,s)\hbg_{2}^{\phi}(s)e^{\ti \om s} \di s + \Bm^*\int_{0}^{t}\bPhi(t,s)\hbg_{3}^{\phi}(s)e^{\ti \om s} \di s \\
& + \Ap^*\int_{0}^{t}\bPhi(t,s)\hbg_{4}^{\phi}(s)e^{-\ti \om s} \di s + \Bp^*\int_{0}^{t}\bPhi(t,s)\hbg_{5}^{\phi}(s)e^{-\ti \om s} \di s + \Bm\int_{0}^{t}\bPhi(t,s)\hbg_{6}^{\phi}(s)e^{-\ti \om s} \di s \\
&- \sum_{j=1}^{8} \bLac_j \int_{0}^{t} \bPhi(t,s)\bD\tbq^{\Ap}\xi^{(2,j)}_{\Ap}(s) e^{\ti \om s}\di s - \sum_{j=1}^{8} \bLac_j  \int_{0}^{t} \bPhi(t,s)\bD\tbq^{\Am}\xi^{(2,j)}_{\Am}(s) e^{-\ti \om s}\di s,
\end{split}
\label{eq:hbqa}
\end{align}
where we have postulated a polynomial expansion for the multiplicative noise corrections, according to $\xi^{(2)}_{\bLac_i}(\bLa,t)= \sum_{j=1}^{8} \bLac_j \xi^{(2,j)}_{\bLac_i}(t)$ (other terms could have been included and then shown to be zero). 

Again, because $\bPhi(t,s)$ possesses neutral eigenmodes, the temporal integrals in (\ref{eq:hbqa}) diverge in the root-mean-square. For instance, 
\begin{align}
\begin{split}
\int_{0}^{t}\bPhi(t,s)\hbg_{1}^{\phi}(s)e^{\ti \om s} \di s =&  e^{\ti \om t}\tbq^{\Ap}\underbrace{\int_{0}^{t} \spap{\hbg_{1}^{\phi}(s)} \di s}_{\text{standard deviation is $\propto \sqrt{t}$}.}  + e^{-\ti \om t}\tbq^{\Am} \underbrace{\int_{0}^{t}\spam{\hbg_{1}^{\phi}(s)} e^{2 \ti \om s} \di s}_{\text{standard deviation is $\propto \sqrt{t}$}.} \\
& +\underbrace{\int_{0}^{t} \bPhi(t,s) \bPo \hbg_{1}^{\phi}(s) e^{\ti \om s} \di s.}_{\text{in the strictly stable manifold thus remains bounded}.}  \nonumber
\end{split}
\end{align}
In the equation above, the integral terms multiplying $\tbq^{\Ap}$ and $\tbq^{\Am}$, respectively, each have a standard deviation going as $\propto \sqrt{t}$ (even though the integrand are generally not white noises), and thus they appear at $O(\e^2)$ after a time $t\sim 1/\e^2$ and at $O(\e)$ after a time $t\sim 1/\e^4$. This is asymptotically inconsistent with the fact that these terms were collected at $O(\e^3)$. Therefore, we use the freedom afforded by the $\xi^{(2,j)}$ to cancel all integrand yields that cause such divergence in (\ref{eq:hbqa}), which yields
\begin{align}
\begin{split}
\xi^{(2)}_{\Ap}(\bLa,t) =& \Am^* \spap{\hbg_{1}^{\phi}(t)} + \Bp\spap{\hbg_{2}^{\phi}(t)} + \Bm^*\spap{\hbg_{3}^{\phi}(t)} \\
& + \pae{\Ap^* \spap{\hbg_{4}^{\phi}(t)}+ \Bp^*\spap{\hbg_{5}^{\phi}(t)} + \Bm\spap{\hbg_{6}^{\phi}(t)}}e^{-2\ti \om t}, \\ 
\xi^{(2)}_{\Am}(\bLa,t) =& \pae{\Am^* \spam{\hbg_{1}^{\phi}(t)} + \Bp\spam{\hbg_{2}^{\phi}(t)} + \Bm^*\spam{\hbg_{3}^{\phi}(t)}}e^{2\ti \om t}\\
& + \Ap^* \spam{\hbg_{4}^{\phi}(t)}+ \Bp^*\spam{\hbg_{5}^{\phi}(t)} + \Bm\spam{\hbg_{6}^{\phi}(t)}.
\label{eq:multi2}
\end{split}
\end{align}
Furthermore, in this way, the higher-order term $\hqpa(\bLa,t)$ is entirely contained within the strictly stable eigen-subspace.

The noise corrections $\xi^{(2)}_{\Bp}$ and $\xi^{(2)}_{\Bm}$ can be derived by proceeding similarly.

\section{Full expressions of the nonlinearly-induced forcing terms at third-order \label{app:o3t}}

We provide below the detailed expressions of the nonlinear forcing terms, collected at $O(\e^3)$, and involved in computing the weakly nonlinear deterministic coefficients. In the equation for $\Ap$, we need 
\begin{align}
\begin{cases}
&\dt{1,1}{2} \coloneq \hbd^{\Ap,\Ap \Ap^*} + \hbd^{\Ap^*,\Ap \Ap} + \bet\hbr^{\Ap \Ap \Ap^*} ,  \\
&\dt{1,2}{2}  \coloneq \hbd^{\Ap,\Am\Am^*} + \hbd^{\Am,\Ap\Am^*} + \hbd^{\Am^*,\Am\Ap} + \bet\hbr^{\Ap\Am\Am^*},  \\
&\dt{1,3}{2}  \coloneq \hbd^{\Ap,\Bp\Bp^*} + \hbd^{\Bp,\Ap \Bp^*} + \hbd^{\Bp^*,\Bp\Ap} + \bet\hbr^{\Ap\Bp\Bp^*}, \\
&\dt{1,4}{2}  \coloneq \hbd^{\Ap,\Bm\Bm^*} + \hbd^{\Bm,\Ap\Bm^*} + \hbd^{\Bm^*,\Bm\Ap} + \bet\hbr^{\Ap\Bm\Bm^*}, \\
&\dt{1}{1}   \coloneq \hbd^{\Am,\Bp\Bm^*} + \hbd^{\Bp,\Am\Bm^*} + \hbd^{\Bm^*,\Bp\Am} + \bet\hbr^{\Am\Bp\Bm^*},
\end{cases}
\label{eq:rty}
\end{align}
where we have defined
\begin{align}
\begin{split}
\hbd^{X,YZ} \coloneq \bNc{\tbq^{X},\hbq^{YZ}} + \bet\hbr^{X,YZ}. \nonumber
\end{split}
\end{align}
Note the comma separating $X$ from $YZ$ in the superscript of $\hbd^{X,YZ}$ and $\hbr^{X,YZ}$. While the order of $Y$ and $Z$ doesn't matter, i.e., $\hbd^{X,YZ}=\hbd^{X,ZY}$, the comma emphasizes that $X$ and $YZ$ cannot be interchanged. That is, $\hbd^{X,YZ} \neq \hbd^{YZ,X}$. We have also defined the contribution $\hbr^{X,YZ}$, arising from the moments of the quadratic contributions $\hch^{X,YZ}$ to the third-order forcing terms in density, given in (\ref{eq:chi_j}), such that
\begin{align}
\begin{split}
\hch^{X,YZ}(\bp) = \hps^{X}(\bp)\hps^{YZ}(\bp).
\nonumber
\end{split}
\end{align}
In (\ref{eq:rty}), there are also terms in $\hbr^{XYZ}$. The absence of a comma in the superscript implies that the order of appearance of $X$, $Y$, and $Z$ there does not matter. These terms arise from the moments of the cubic contributions to the third-order forcing terms in density in (\ref{eq:chi_j}), given by
\begin{align}
\begin{split}
\hch^{XYZ}(\bp) = \gamma \hps^{X}(\bp)\hps^{Y}(\bp)\hps^{Z}(\bp), \quad \text{with} \quad \gamma = \begin{cases}
-1 & \text{if} \; \text{$Y=X$ and $Z=X^*$, or $Y=X^*$ and $Z=X$}. \\
-2 & \text{otherwise.} \;
\end{cases} 
\nonumber
\end{split}
\end{align}

Computing the coefficients involved in the equation for $\Am$ requires
\begin{align}
\begin{cases}
&\dt{2,1}{2} \coloneq \hbd^{\Am,\Ap\Ap^*} + \hbd^{\Ap,\Am\Ap^*} + \hbd^{\Ap^*,\Ap\Am} + \bet\hbr^{\Am\Ap\Ap^*} ,  \\
&\dt{2,2}{2} \coloneq \hbd^{\Am, \Am \Am^*} + \hbd^{\Am^*, \Am \Am} + \bet\hbr^{\Am \Am \Am^*}, \\
&\dt{2,3}{2} \coloneq \hbd^{\Am,\Bp\Bp^*} + \hbd^{\Bp,\Am\Bp^*} + \hbd^{\Bp^*,\Bp\Am} + \bet\hbr^{\Am\Bp\Bp^*}, \\
&\dt{2,4}{2} \coloneq \hbd^{\Am,\Bm\Bm^*} + \hbd^{\Bm,\Am\Bm^*} + \hbd^{\Bm^*,\Bm\Am} + \bet\hbr^{\Am\Bm\Bm^*}, \\
&\dt{2}{1} \coloneq \hbd^{\Ap,\Bm\Bp^*} + \hbd^{\Bm,\Ap\Bp^*} + \hbd^{\Bp^*,\Ap\Bm} + \bet\hbr^{\Ap\Bm\Bp^*}.
\end{cases}
\nonumber
\end{align}
The coefficients in the equation for $\Bp$ are based on 
\begin{align}
\begin{cases}
&\dt{3,1}{2} \coloneq \hbd^{\Bp,\Ap\Ap^*} + \hbd^{\Ap,\Bp\Ap^*} + \hbd^{\Ap^*,\Ap\Bp} + \bet\hbr^{\Bp\Ap\Ap^*},  \\
&\dt{3,2}{2} \coloneq \hbd^{\Bp,\Am\Am^*} + \hbd^{\Am,\Bp\Am^*} + \hbd^{\Am^*,\Bp\Am} + \bet\hbr^{\Bp\Am\Am^*},  \\
&\dt{3,3}{2} \coloneq \hbd^{\Bp, \Bp \Bp^*} + \hbd^{\Bp^*, \Bp \Bp} + \bet\hbr^{\Bp \Bp \Bp^*}, \\
&\dt{3,4}{2} \coloneq \hbd^{\Bp,\Bm\Bm^*} + \hbd^{\Bm,\Bp\Bm^*} + \hbd^{\Bm^*,\Bm\Bp} + \bet\hbr^{\Bp\Bm\Bm^*}, \\
&\dt{3}{1} \coloneq \hbd^{\Ap,\Bm\Am^*} + \hbd^{\Am^*,\Ap\Bm} + \hbd^{\Bm,\Ap\Am^*} + \bet\hbr^{\Ap\Bm\Am^*}.
\end{cases}
\nonumber
\end{align}
Eventually, coefficients in the equation for $\Bm$ stem from
\begin{align}
\begin{cases}
&\dt{4,1}{2} \coloneq \hbd^{\Bm,\Ap\Ap^*} + \hbd^{\Ap,\Bm\Ap^*} + \hbd^{\Ap^*,\Ap\Bm} + \bet\hbr^{\Bm\Ap\Ap^*},  \\
&\dt{4,2}{2} \coloneq \hbd^{\Bm,\Am\Am^*} + \hbd^{\Am,\Bm\Am^*} + \hbd^{\Am^*,\Am\Bm} + \bet\hbr^{\Bm\Am\Am^*} ,  \\
&\dt{4,3}{2} \coloneq \hbd^{\Bm,\Bp\Bp^*} + \hbd^{\Bp,\Bm\Bp^*} + \hbd^{\Bp^*,\Bp\Bm} + \bet\hbr^{\Bm\Bp\Bp^*}, \\
&\dt{4,4}{2} \coloneq \hbd^{\Bm,\Bm\Bm^*} + \hbd^{\Bm^*,\Bm\Bm} + \bet\hbr^{\Bm\Bm\Bm^*}, \\
&\dt{4}{1} \coloneq \hbd^{\Am,\Bp\Ap^*} + \hbd^{\Ap^*,\Am\Bp} + \hbd^{\Bp,\Am\Ap^*} + \bet\hbr^{\Am\Bp\Ap^*}.
\end{cases}
\nonumber
\end{align}

\section{Autocorrelation of the noise acting on the amplitudes}\label{app:noise}

In this Appendix, we derive the autocorrelation law for the noise processes appearing in the amplitude equations. The first step is to compute the autocorrelation law of the $\bk$-Fourier component of forcing, given by $\hbf_{\bk}(t) \coloneq (4\upi^2)^{-1} \int_{\Omega} \bff(\bx,t) e^{- \ti \bk \bdot \bx } \di \bx$, where we recall that $|\Omega|=4\upi^2$ and $\bff$ is detailed in (\ref{eq:fcgd}). This gives
\begin{align}
\begin{split}
\ea{\hf_{\bk,i}(t)\hf^*_{\bk,j}(t')} = \frac{1}{(4\upi^2)^2} \iint_{\Omega} \ea{f_i(x,t)f_j(x',t')} e^{- \ti \bk \bdot (\bx-\bx') } \di \bx \di \bx'.
\nonumber
\end{split}
\end{align}
where the scalar subscripts $i$ or $j$ denote the component of the forcing vector, e.g., $f_1=f_c$, $f_2=f_{n_x}$, $f_3=f_{n_y}$, etc. It is also true that $\ea{f_i(x,t)f_j(x',t')} = \al_{ij} \de(\bx-\bx')\de(t-t')$, where the scalar $\al_{ij}$ correspond to the values reported in Table.~\ref{tab:cov_f}. Therefore, $\ea{\hf_{\bk,i}(t)\hf^*_{\bk,j}(t')} =  (4\upi^2)^{-1}\al_{ij}\de(t-t')$, from which it follows from the definitions (\ref{eq:xika}) and (\ref{eq:xikb}) that
\begin{align}
\begin{split}
\ea{\xi_{\bLac}(t)\xi^*_{\bLac}(t')} =& \ea{(\tbq^{\bLac,\da})^H\bD\hbf_{\bk}(t) \hbf_{\bk}(t')^H\bD \tbq^{\bLac,\da}}=\de(t-t')\underbrace{\frac{1}{4\upi^2} \sum_{i=1}^5\sum_{j=1}^5 \al_{ij} (\tq_i^{\bLac,\da})^*\tq_j^{\bLac,\da}}_{=\text{intensity of $\xi_{\bLac}(t)$}},
\nonumber
\end{split}
\end{align}
where $\bk=\bka$ (resp. $\bk=\bkb$) if $\bLac \in \set{A,\Ap,\Am}$ (resp. $\bLac \in \set{B,\Bp,\Bm}$).

\end{appen}\clearpage

\bibliographystyle{jfm}
\bibliography{jfm}

\end{document}